\RequirePackage{lineno}\newdimen\linenumbersep\linenumbersep=2pt
\documentclass[aps,prd,twocolumn,superscriptaddress,amsmath,amssymb,floatfix,nofootinbib,showkeys]{revtex4}
\usepackage{xcolor,graphicx,verbatim}
\usepackage[colorlinks=true,citecolor=blue,filecolor=blue,linkcolor=blue,urlcolor=blue,pdftex]{hyperref}
\newcommand{\CENS}{$\mathrm{CE}\nu\mathrm{NS}$}
\newcommand{\Er}{E_{\mathrm{R}}}
\newcommand{\Ed}{E_{\mathrm{R}}}
\newcommand{\Sone}{\mathrm{S1}}
\newcommand{\Stwo}{\mathrm{S2}}
\newcommand{\mN}{m_{\mathrm{N}}}
\newcommand{\tpb}{t_{\mathrm{pb}}}
\newcommand{\Msun}{\mathrm{M}_{\odot}}

\newcommand{\amsterdam}{\affiliation{GRAPPA Centre of Excellence, University of Amsterdam, Amsterdam, The Netherlands}}
\newcommand{\bologna}{\affiliation{INFN-Bologna, Bologna, Italy}}
\newcommand{\copenhagen}{\affiliation{Niels Bohr International Academy, Niels Bohr Institute, Copenhagen, Denmark}}
\newcommand{\purdue}{\affiliation{Department of Physics and Astronomy, Purdue University, West Lafayette, Indiana, USA}}

\begin{document}

\title{Supernova neutrino physics with xenon dark matter detectors: \\ A timely perspective}

\author{Rafael~F.~Lang}\email{rafael@purdue.edu}\purdue
\author{Christopher McCabe}\email{c.mccabe@uva.nl}\amsterdam
\author{Shayne Reichard}\email{sreichar@purdue.edu}\purdue
\author{Marco Selvi}\email{selvi@bo.infn.it}\bologna
\author{Irene Tamborra}\email{tamborra@nbi.ku.dk}\copenhagen

\begin{abstract}
Dark matter detectors that utilize liquid xenon have now achieved tonne-scale targets, giving them sensitivity to all flavours of supernova neutrinos via coherent elastic neutrino-nucleus scattering. Considering for the first time a realistic detector model, we simulate the expected supernova neutrino signal for different progenitor masses and nuclear equations of state in existing and upcoming dual-phase liquid xenon experiments. We show that the proportional scintillation signal~(S2) of a dual-phase detector allows for a clear observation of the neutrino signal and guarantees a particularly low energy threshold, while the backgrounds are rendered negligible during the supernova burst. XENON1T (XENONnT and LZ; DARWIN) experiments will be sensitive to a supernova burst up to~25~(35; 65)~\!kpc from Earth at a significance of more than $5\sigma$, observing approximately~35~(123; 704) events from a 27~\!$\Msun$ supernova progenitor at 10~\!kpc. Moreover, it will be possible to measure the average neutrino energy of all flavours, to constrain the total explosion energy, and to reconstruct the supernova neutrino light curve. Our results suggest that a large xenon detector such as DARWIN will be competitive with dedicated neutrino telescopes, while providing complementary information that is not otherwise accessible.
\end{abstract}


\maketitle

\section{Introduction}

Core-collapse supernovae are among the most energetic transients that occur in the Universe, originating from the death of very massive stars~\cite{Janka:2016fox,Janka:2012wk}. Despite the remarkable progress we have seen in our understanding of the core-collapse physics over the last decade, we are still far from fully grasping the physical processes that underlie the supernova (SN) engine and, in particular, the role that neutrinos play in powering it~\cite{Janka:2016fox,Mirizzi:2015eza,Chakraborty:2016yeg}. A high-statistics detection of neutrinos from the next Galactic SN explosion in detectors that operate with different technologies will shed light on both the stellar engine and the properties of neutrinos. Neutrino flavour discrimination will be crucial to investigate neutrino oscillation physics and scenarios with non-standard neutrino properties~\cite{Mirizzi:2015eza,Esmaili:2014gya,Wu:2014kaa,EstebanPretel:2007yu,Stapleford:2016jgz,Hidaka:2007se}.
On the other hand, the detection of all six neutrino flavours will be essential to reconstruct global emission properties, such as the total explosion energy emitted into neutrinos~\cite{Drukier:1983gj,Beacom:2002hs,Horowitz:2003cz}. 

At present, several neutrino detectors are ready for the next Galactic SN explosion, while others are under construction or being planned~\cite{Mirizzi:2015eza,Scholberg:2012id}. Among these experiments, the most promising technologies include Cherenkov telescopes and liquid scintillators, as used in or proposed for IceCube~\cite{Abbasi:2011ss}, Super-Kamiokande~\cite{Abe:2010hy,Abe:2016waf}, IceCube-Gen2~\cite{Aartsen:2014njl}, Hyper-Kamiokande~\cite{Abe:2011ts}, LVD~\cite{Agafonova:2014leu}, Borexino~\cite{Cadonati:2000kq}, JUNO~\cite{An:2015jdp}, RENO-50~\cite{Kim:2014rfa}, and KamLAND~\cite{Barger:2000hy,Asakura:2015bga}. Both of these technologies will be able to probe $\bar{\nu}_e$ neutrinos with high accuracy. In contrast, the planned liquid argon detector within the DUNE facility~\cite{Acciarri:2015uup} will accurately probe the $\nu_e$ channel. There are also proposals to study the $\nu_e$ properties with Cherenkov telescopes or liquid scintillators~\cite{Laha:2014yua,Laha:2013hva,Jia-Shu2016} or with experiments that use lead or iron targets~\cite{Kolbe:2000np,Volpe:2001gy,Shantz:2010th}. Together, these experiments will accurately measure the $\bar{\nu}_e$ and $\nu_e$ fluxes from the next Galactic SN explosion~\cite{Scholberg:2012id}.

The elastic scattering of neutrinos on protons~\cite{Beacom:2002hs,Dasgupta:2011wg} and on nuclei~\cite{Drukier:1983gj} are alternative tools to detect astrophysical neutrinos. Neutrino-nucleus scattering is especially attractive because, at low energies, the scattering cross-section is coherently enhanced by the square of the nucleus's neutron number~\cite{Freedman:1977xn}. Supernova neutrinos with energies of $\mathcal{O}(10)$~MeV induce $\mathcal{O}(1)$~keV nuclear recoils through coherent elastic neutrino-nucleus scattering (\CENS). Although recoils in this energy range are too small to be detected by conventional neutrino detectors, it is precisely this energy range for which direct detection dark matter experiments are optimized~\cite{Undagoitia:2015gya}. The primary purpose of these experiments is to search for nuclear recoils induced by Galactic dark matter particles. Yet, sufficiently large experiments ($\gtrsim$ tonne of target material) are also sensitive to \CENS\ from SN neutrinos~\cite{Horowitz:2003cz,Monroe:2007xp,Strigari:2009bq,XMASS:2016cmy}.

Mediated by $Z$-boson exchange, \CENS\ is especially intriguing because it is equally sensitive to all neutrino flavours. Detectors that observe \CENS\ are therefore sensitive to the $\bar{\nu}_{\mu}$, $\nu_{\mu}$, $\bar{\nu}_{\tau}$ and $\nu_{\tau}$ (otherwise dubbed~$\nu_x$) neutrinos within their main detection channel, in addition to the $\bar{\nu}_e$ and $\nu_e$ neutrinos~\cite{Drukier:1983gj}. This feature carries numerous implications for experiments that detect \CENS. For instance, the neutrino light curve could be reconstructed without the uncertainties that arise from neutrino oscillation in the stellar envelope~\cite{Chakraborty:2013zua}, the total energy emitted into all neutrino species could be measured, or, by assuming adequate reconstruction of the $\bar{\nu}_e$ and $\nu_e$ emission properties with other detectors,
 \CENS\  detectors provide a way to reconstruct the $\nu_x$ emission properties. 

In this paper, we revisit the possibility of detecting \CENS\ from SN neutrinos in the context of XENON1T~\cite{Aprile:2015uzo} and larger forthcoming direct detection dark matter experiments that employ a xenon target, such as XENONnT~\cite{Aprile:2015uzo}, LZ~\cite{Akerib:2015cja}, and DARWIN~\cite{Aalbers:2016jon}. Among the various technologies used in direct detection experiments, dual-phase xenon experiments have many advantages: the large neutron number of the xenon nucleus enhances the \CENS\ rate compared to nuclei used in other direct detection experiments; they are sensitive to sub-keV nuclear recoils; the deployment of XENON1T heralds the era of tonne-scale experiments, which are relatively straightforward to scale to even larger masses; despite their large size, the background rates are very low; and, finally, they have excellent timing resolution, $\mathcal{O}(100)$~\!$\mu$s, in the data analysis mode discussed here. As we demonstrate in section~\ref{sec:results}, these factors mean that XENON1T is already able to detect neutrinos from a SN up to 25~kpc at a significance of more than~$5\sigma$.

To forecast the signal that is expected from SN neutrinos in forthcoming xenon detectors, we adopt inputs of four hydrodynamical SN simulations from the Garching group~\cite{Mirizzi:2015eza,Huedepohl:2013} that differ in the progenitor's mass and nuclear equation of state in such a way as to provide a reasonable estimate of the signal band. The neutrino properties for the adopted progenitor models are introduced in section~\ref{sec:SNmodels}. In section~\ref{sec:xenondet}, for the first time, we accurately simulate the expected signal in terms of the measured quantities in dual-phase xenon experiments (scintillation photons and ionization electrons). In section~\ref{sec:S2only}, we discuss the advantages of a dual-phase xenon detector in the observation of SN neutrinos (the low-energy sensitivity of the proportional-scintillation-signal analysis mode) as well as the expected backgrounds and achievable threshold. Section~\ref{sec:results} contains our main physics results. We discuss the detection significance of SN neutrinos with future-generation xenon detectors, the reconstruction of the SN neutrino light curve as well as the average neutrino energy and the total explosion energy. Uncertainties related to the detector modeling are outlined in section~\ref{sec:discussion}. Finally, in section~\ref{sec:conclusions}, we present our conclusions.

\section{Supernova neutrino emission} \label{sec:SNmodels}

In order to forecast the expected recoil signal in a xenon detector, we require the differential flux of the $\nu_e$, $\bar{\nu}_e$ and $\nu_x$ neutrinos as a function of time and energy. The differential flux for each neutrino flavour~$\nu_{\beta}$ at a time~$t_{\rm pb}$ after the SN core bounce for a SN at a distance~$d$ is parametrized by
\begin{equation}
\label{eq:nuflux}
f_{\nu_\beta}^0(E,t_{\rm pb})= \frac{L_{\nu_\beta}(t_{\rm pb})}{4 \pi d^2}\,\frac{\varphi_{\nu_\beta}(E,t_{\rm pb})}{\langle E_{\nu_\beta}(t_{\rm pb}) \rangle}\ ,
\end{equation}
where $L_{\nu_\beta}(t_{\rm pb})$ is the $\nu_\beta$ luminosity, $\langle E_{\nu_\beta}(t_{\rm pb}) \rangle$ is the mean energy, and $\varphi_{\nu_\beta}(E,t_{\rm pb})$ is the neutrino energy distribution. The neutrino energy distribution is defined in~\cite{Keil:2002in,Tamborra:2012ac} as
\begin{equation}
\begin{split}
\label{alphafit}
\varphi_{\nu_\beta}(E,t_{\rm pb})&=\xi_\beta(t_{\rm pb}) \left(\frac{E}{\langle E_{\nu_\beta}(t_{\rm pb}) \rangle}\right)^{\alpha_\beta(t_{\rm pb})}\\
&\quad\times \exp\left[-\frac{(\alpha_\beta(t_{\rm pb})+1) E}{\langle E_{\nu_\beta}(t_{\rm pb}) \rangle}\right] .
\end{split}
\end{equation}
The fit parameter $\alpha_\beta(t_{\rm pb})$ satisfies the relation
\begin{equation}
\label{eq:earelate}
\frac{\langle E_{\nu_\beta}(t_{\rm pb})^2 \rangle}{\langle E_{\nu_\beta}(t_{\rm pb}) \rangle^2} =
\frac{2+\alpha_\beta(t_{\rm pb})}{1+\alpha_\beta(t_{\rm pb})}\ ,
\end{equation}
while $\xi_\beta(t_{\rm pb})$ is a normalization factor defined such that $\int dE \,\varphi_{\nu_\beta}(E,t_{\rm pb})=1$. In the following, we show results for a benchmark distance of $d=10$~\!kpc for the Galactic~SN.

\subsection{Supernova neutrino emission properties}

The neutrino emission properties that we adopt are from the one-dimensional (1D) spherically symmetric SN hydrodynamical simulations by the Garching group~\cite{Mirizzi:2015eza,Huedepohl:2013,snarchive}. More recent three-dimensional (3D) SN simulations exhibit hydrodynamical instabilities such as large-scale convective overturns and the standing accretion shock instability (SASI) that are responsible for characteristic modulations in the neutrino signal not observable in 1D SN simulations~\cite{Lund:2010kh,Tamborra:2013laa,Tamborra:2014hga,Tamborra:2014aua,Janka:2016fox}. However, as in this paper we are interested in the general qualitative behaviour of the SN neutrino event rate in a xenon detector, we can safely neglect these effects and adopt the outputs from 1D spherically symmetric SN simulations.

To investigate the variability of the expected signal as a function of the progenitor mass, we use the neutrino emission properties from the hydrodynamical simulations of two SN progenitors with masses of $11.2~\!\Msun$ and $27~\!\Msun$. We also consider the dependence of the expected event rates on the nuclear equation of state (EoS) by adopting, for each SN progenitor, simulations obtained from the Lattimer and Swesty EoS~\cite{Lattimer:1991nc} with a nuclear incompressibility modulus of $K = 220$~\!MeV (LS220 EoS) and the Shen EoS~\cite{Shen:1998gq}. These four progenitors provide a gauge of the astrophysical variability of the expected recoil signal.

\begin{figure*}[!htbp]
\centering
\includegraphics[width=1.9\columnwidth]{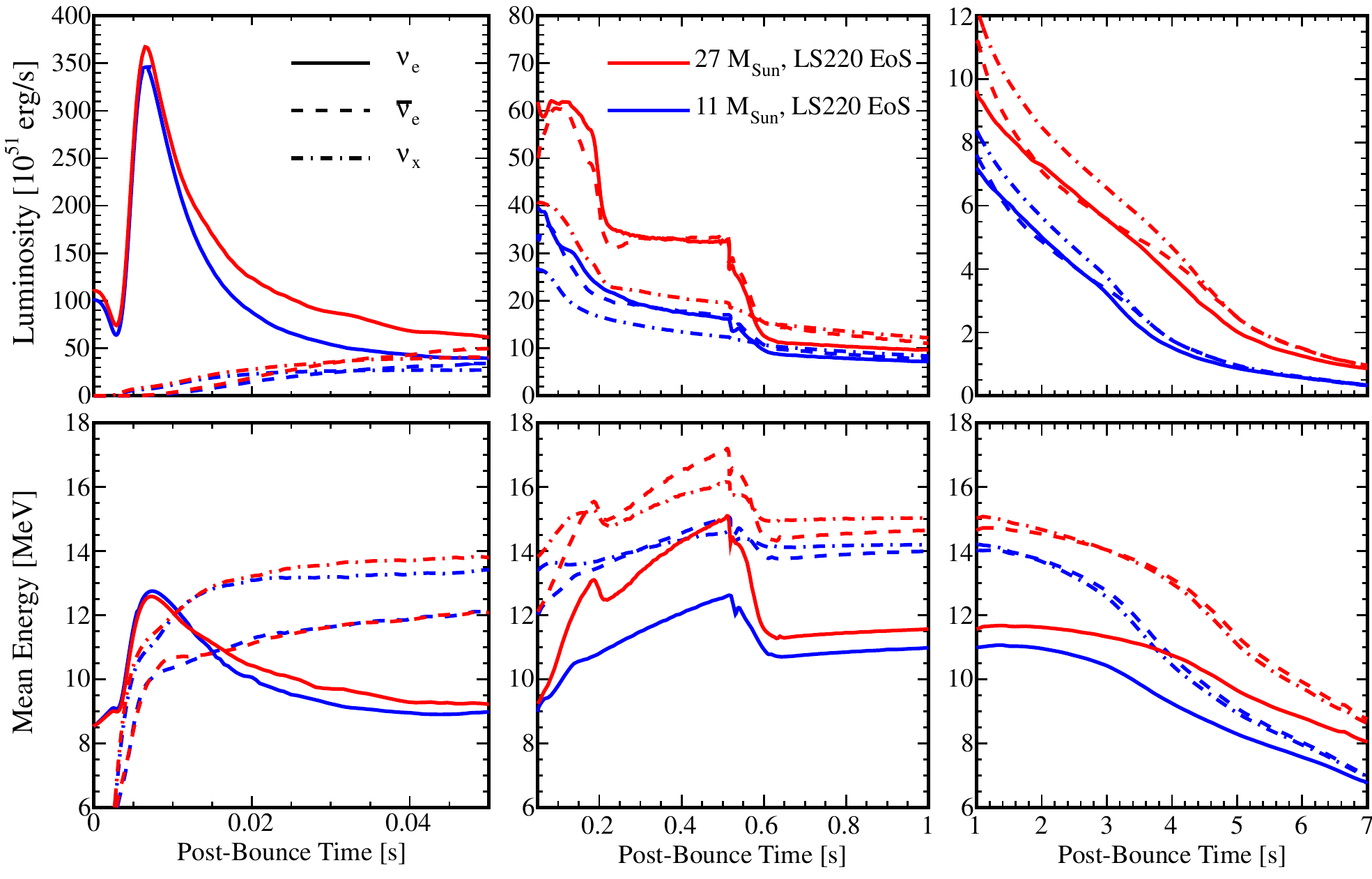}
\caption{The upper and lower panels show the neutrino luminosity~$L_{\nu_{\beta}}$ and mean energy $\langle E_{\nu_\beta} \rangle$, respectively, as a function of the post-bounce time~$t_{\rm{pb}}$ for the $11~\!\Msun$ (in blue) and $27~\!\Msun$ (in red) SN progenitors with the LS220 EoS for $\nu_e$ (continuous lines), $\bar{\nu}_e$ (dashed lines) and $\nu_x$ (dot-dashed lines). The panels on the left show the neutrino properties during the neutronization burst phase, the middle panes refer to the accretion phase, and panels on the right describe the Kelvin-Helmholtz cooling phase. The differences in the neutrino properties from different progenitors during the neutronization burst are small, but they become considerable at later times. The variation of the neutrino properties owing to a different nuclear EoS is smaller than the differences from a different progenitor mass, so for clarity, the progenitors with the Shen EoS are not shown here.}
\label{fig:LEpanels}
\end{figure*}

Figure~\ref{fig:LEpanels} displays the neutrino luminosities (top panels) and mean energies (bottom panels) of all neutrino flavours ($\nu_e$, $\bar{\nu}_e$ and $\nu_x$) as a function of the post-bounce time in the observer frame for the $27~\!\Msun$ and $11~\Msun$ SN progenitors with the LS220 EoS. The variation of the neutrino properties owing to a different nuclear EoS is smaller than the differences shown here from the different progenitor mass. The neutrino signal emitted from a SN explosion lasts for more than 10~\!s but, as Fig.~\ref{fig:LEpanels} demonstrates, the luminosity drops considerably after a few seconds. Therefore, in this paper, we focus on the initial 7~\!s of the neutrino signal after the core bounce.

The left, middle, and right panels of Fig.~\ref{fig:LEpanels} show the three main phases of the SN neutrino signal: the neutronization burst, the accretion phase, and the Kelvin-Helmholtz cooling phase, respectively. The neutronization burst originates while the shock wave is moving outwards through the iron core. Free protons and neutrons are released as the shock wave dissociates iron nuclei. Consequently, rapid electron capture by nuclei and free protons produces a large~$\nu_e$ burst. As evident from the top left panel in Fig.~\ref{fig:LEpanels}, the width and amplitude of the~$\nu_e$ luminosity during the neutronization burst are approximately independent of the SN progenitor mass and EoS~\cite{Kachelriess:2004ds,Serpico:2011ir}. Generally, the~$\nu_x$ luminosity rises more quickly than that of~$\bar{\nu}_e$ during the first 10--20~ms of the signal due to the high abundance of $\nu_e$ and electrons, which suppress the rapid production of $\bar{\nu}_e$.

The accretion phase is shown in the middle panels of Fig.~\ref{fig:LEpanels}. During this phase, the SN shock loses energy while moving outward and dissociating iron nuclei until it stalls at a radius of about 100--200~\!km. According to the delayed-neutrino SN explosion mechanism~\cite{Bethe:1984ux,Bethe:1990mw}, neutrinos provide additional energy to the shock to revive it after tens to hundreds of milliseconds and finally trigger the explosion. As the in-falling material accretes onto the core, it is heated, and the subsequent $e^+e^-$ annihilation produces neutrinos of all flavours. Due to the high abundance of $\nu_e$ during the neutronization burst, the production of $\bar{\nu}_e$ and $\nu_x$ is initially suppressed. The production of $\bar{\nu}_e$ increases as the capture of electrons and positrons on free nucleons starts to become more efficient. The non-electron neutrinos remain less abundant as they can only be produced via neutral-current interactions. 

The explosion of 1D SN simulations may require an artificial initiation, especially for more massive progenitors. In the simulation  shown in Fig.~\ref{fig:LEpanels}, the explosion was triggered at $t_{\rm{pb}}\simeq0.5$~\!s. After this, the Kelvin-Helmholtz cooling phase of the newly born neutron star begins. As shown in the right panel of Fig.~\ref{fig:LEpanels}, the neutrino luminosities gradually decrease as the proto-neutron star cools and de-leptonizes. As the explosion is artificially triggered in these simulations, the exact transition time from the accretion to the Kelvin-Helmholtz cooling phase should be taken with caution. The neutrino signal during this phase is sensitive to the progenitor mass and the EoS. In fact, while the differences among the neutrino properties from different progenitors during the neutronization burst are small, at later times they become considerable. 

\subsection{Neutrino flavour conversion}

The neutrino transport in SN hydrodynamical simulations is solved within the weak-interaction basis for all three neutrino flavours. Neutrinos oscillate while they are propagating through the stellar envelope as well as on their way to Earth. This affects the neutrino flavour distribution detected on Earth. In particular, neutrinos undergo the Mikheev-Smirnov-Wolfenstein (MSW) effect~\cite{wolf,Wolfenstein:1977ue,Mikheev:1986if}, which affects the survival probability of each neutrino flavour according to the adiabaticity of the matter profile. The MSW effect could be modified by turbulence or significant stochastic fluctuations in the stellar matter density (see e.g.~\cite{Borriello:2013tha,Sawyer:1990tw,Fogli:2006xy,Friedland:2006ta}). In addition, neutrino--neutrino interactions are believed to be important and can affect the neutrino flavour evolution and therefore the expected energy distribution~\cite{Duan:2010bg,Mirizzi:2015eza,Chakraborty:2016yeg}.

For our purpose, however, the details of the oscillation physics are not important. This is because \CENS\ is sensitive to all neutrino flavours and the total neutrino flux is conserved. Hence, the same total flux produced at the SN core will reach the detector on Earth.

Non-standard physics may lead to situations where the total flux is not conserved, such as a scenario with light sterile neutrinos~\cite{Esmaili:2014gya,Wu:2014kaa,Tamborra:2011is,Pllumbi:2014saa}, non-standard neutrino interactions~\cite{EstebanPretel:2007yu,Nunokawa:1996tg,Stapleford:2016jgz}, or light dark matter particles~\cite{Fuller:2009zz,Hidaka:2007se,Davis:2016dqh}. All of these cases affect the heating of the star, implying that the total neutrino flux reaching the Earth could be different from the total neutrino flux at the neutrinosphere. In this paper, we do not consider these scenarios further but focus on the Standard Model scenario.

\section{Supernova neutrino scattering with dual-phase xenon detectors} \label{sec:xenondet}

With the launch of the XENON1T experiment~\cite{Aprile:2015uzo}, which contains two tonnes of instrumented xenon, direct detection dark matter searches have entered the era of tonne-scale targets. The detection principle of this experiment is similar to smaller predecessors, including LUX~\cite{Akerib:2012ys}, PandaX~\cite{Cao:2014jsa}, XENON100~\cite{Aprile:2011dd}, XENON10~\cite{Aprile:2010bt}, and the three ZEPLIN experiments~\cite{Alner:2005pa, Alner:2007ja,Akimov:2006qw}. Future experiments using the same technology include XENONnT~\cite{Aprile:2014zvw} and LZ~\cite{Akerib:2015cja} with each planning for approximately seven tonnes of instrumented xenon. The DARWIN consortium~\cite{Baudis:2012bc,Schumann:2015cpa,Aalbers:2016jon} is investigating an even larger experiment to succeed XENONnT and LZ with approximately 40~tonnes of instrumented xenon. In principle, the technology can be extended to even larger detectors at comparatively modest cost.

These experiments consist of a dual-phase cylindrical time projection chamber (TPC) filled primarily with liquid xenon and a gaseous xenon phase on top. The energy deposited by an incident particle in the instrumented volume produces two measurable signals, called the~S1 and~S2 signals, respectively, from which the energy deposition can be reconstructed. An energy deposition in the liquid xenon creates excited and ionized xenon atoms, and the prompt de-excitation of excited molecular states yields the S1 (or prompt scintillation) signal. An electric drift field of size $\mathcal{O} (1)$~\!kV/cm draws the ionization electrons to the liquid-gas interface. A second electric field of size $\mathcal{O} (10)$~\!kV/cm extracts the ionization electrons from the liquid to the gas. Within the gas phase, these extracted electrons collide with xenon atoms to produce the S2 (or proportional scintillation) signal. The S1 and S2 signals are observed with two arrays of photomultiplier tubes (PMTs) situated at the top and bottom of the TPC. A measurement of both the S1 and S2 signals allows for a full 3D reconstruction of the position of the energy deposition in the TPC. In typical dark matter searches, only an inner volume of the xenon target is used to search for dark matter (the ``fiducial volume"), but the background rate for the duration of the SN signal is sufficiently small such that all of the instrumented xenon can be used to search for SN neutrino scattering (see section~\ref{sec:S2only} for further discussion). In the following, we will thus always refer to the instrumented volume.

The general expression for the differential scattering rate $dR$ in terms of the observable~S1 and~S2 signals for a perfectly efficient detector is
\begin{equation}
\label{eq:rateS1S2}
\frac{d^2R}{d \Sone d \Stwo}= \int\! d\tpb d \Ed \,\mathrm{pdf} \left( \Sone,\Stwo | \Ed \right) \frac{d^2R}{d \Ed {d \tpb}}.
\end{equation}
The differential rate is an integral over the time-period of the SN neutrino signal, expressed in terms of the post-bounce time~$\tpb$, and an integral over the recoil energy~$\Er$ of the xenon nucleus. The differential scattering rate in terms of~$\Er$ is convolved with the probability density function $(\mathrm{pdf})$ to obtain S1 and S2 signals for a given energy deposition~$\Er$. In subsections~\ref{sec:subEr} and~\ref{sec:S1S2}, we describe the procedure to calculate~$d^2R/d\Er d \tpb$ and~$\mathrm{pdf}(\Sone,\Stwo | \Ed)$ respectively. Subsequently, in subsection~\ref{sec:obrates}, we present the expected neutrino-induced scattering rates in terms of the~S1 and~S2 observable quantities.

Before moving on, we briefly comment on single-phase xenon experiments, such as XMASS~\cite{Abe:2013tc}, which only have the liquid phase. The absence of the gas-phase implies that there is no S2 signal, so any generated ionization only adds to the S1 signal. Thus, the instrument is more sensitive in S1 but lacks the inherent amplification of the S2 signal using proportional scintillation. Ultimately, due to quantum efficiencies of photon detection and some sources of background, single-phase detectors have a higher energy threshold compared to dual-phase detectors. As we demonstrate in the next subsection, the recoil spectrum increases rapidly at low energies; so, dual-phase experiments are significantly more sensitive to SN neutrinos. For this reason, we do not consider single-phase detectors and refer the reader to the literature for further discussion~\cite{XMASS:2016cmy}.

\subsection{Scattering rates in terms of recoil energy \label{sec:subEr}}

The interaction of a SN neutrino with a xenon nucleus through \CENS\ causes the nucleus to recoil with energy~$\Er$. The differential scattering rate in terms of~$\Er$ is given by
\begin{equation}
\label{eq:d2RdEdt}
\frac{d^2R}{d \Er d t_{\rm{pb}}}=\sum_{\nu_{\beta}}N_\text{Xe}\int_{E_{\nu}^{\rm{min}}} dE_{\nu}\, f_{\nu_\beta}^0(E_{\nu},t_{\rm pb}) \frac{d\sigma}{d \Er} \ ,
\end{equation}
where the sum is over all six neutrino flavours, $N_{\rm{Xe}}\simeq4.60\times10^{27}$ is the number of xenon nuclei per tonne of liquid xenon, $E_\nu^{\rm{min}}\simeq\sqrt{\mN \Er/2}$ is the minimum neutrino energy required to induce a xenon recoil with energy~$\Er$, $\mN$~is the mass of the xenon nucleus, and $f_{\nu_\beta}^0(E_{\nu},t_{\rm pb})$~is as defined in Eq.~\eqref{eq:nuflux}. Finally, $d\sigma/ d \Er$ is the coherent elastic neutrino-nucleus scattering cross-section~\cite{Drukier:1983gj},
\begin{equation}
\frac{d\sigma}{d \Er}= \frac{G_F^2 \mN}{4\pi}Q_W^2\left(1-\frac{\mN \Er}{2 E_\nu^2}\right)F^2(\Er) \ ,
\label{eq:Xsection}
\end{equation}
where $G_F$ is the Fermi constant, $Q_W=N-(1-4\sin^2\theta_W)Z$ is the weak nuclear hypercharge of a nucleus with $N$ neutrons and $Z$ protons, $\sin^2\theta_W\simeq0.2386$ is the weak mixing angle at small momentum transfer~\cite{Erler:2004in}, and $F(\Er)$ is the nuclear form factor. For xenon, the Helm form factor provides an excellent parametrization for the small values of $\Er$ induced by \CENS\ with which we are concerned~\cite{Vietze:2014vsa},
\begin{equation}
F(\Er)=\frac{3 j_1 (q r_n)}{qr_n}\exp\left(-\frac{(qs)^2}{2} \right) \, ,
\end{equation}
where $q^2=2\mN \Er$ is the squared momentum transfer, $s=0.9$~fm is the nuclear skin thickness, $r^2_n=c^2+\frac{7}{3}\pi^2 a^2-5s^2$ is the nuclear radius parameter, $c=1.23A^{1/3}-0.60$~\!fm, $a=0.52$~\!fm, $A$ is the atomic number of xenon, and~$j_1(q r_n)$ is the spherical Bessel function. 

\begin{figure}[!htbp]
\centering
\includegraphics[width=0.975\columnwidth]{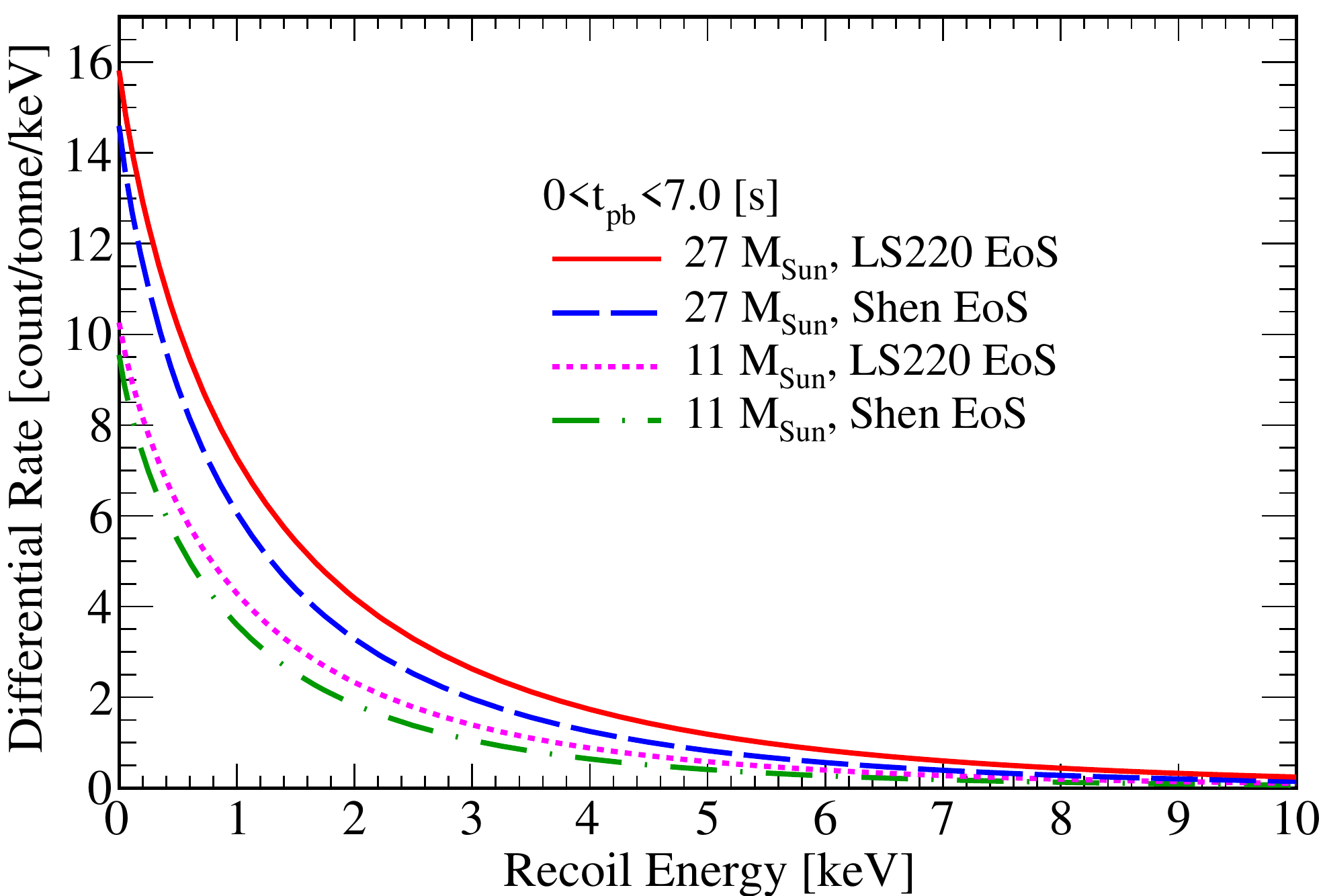}
\includegraphics[width=0.975\columnwidth]{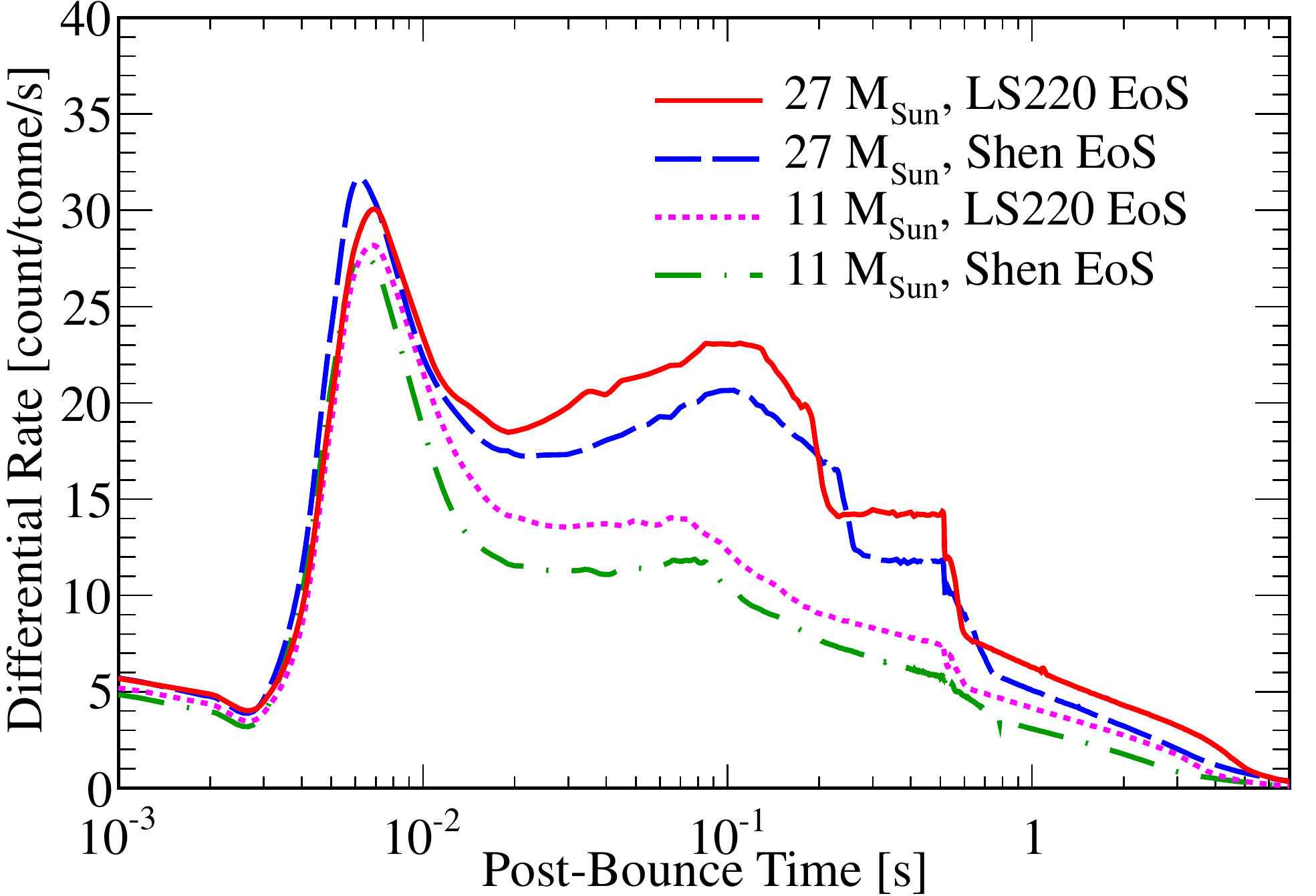}
\includegraphics[width=0.975\columnwidth]{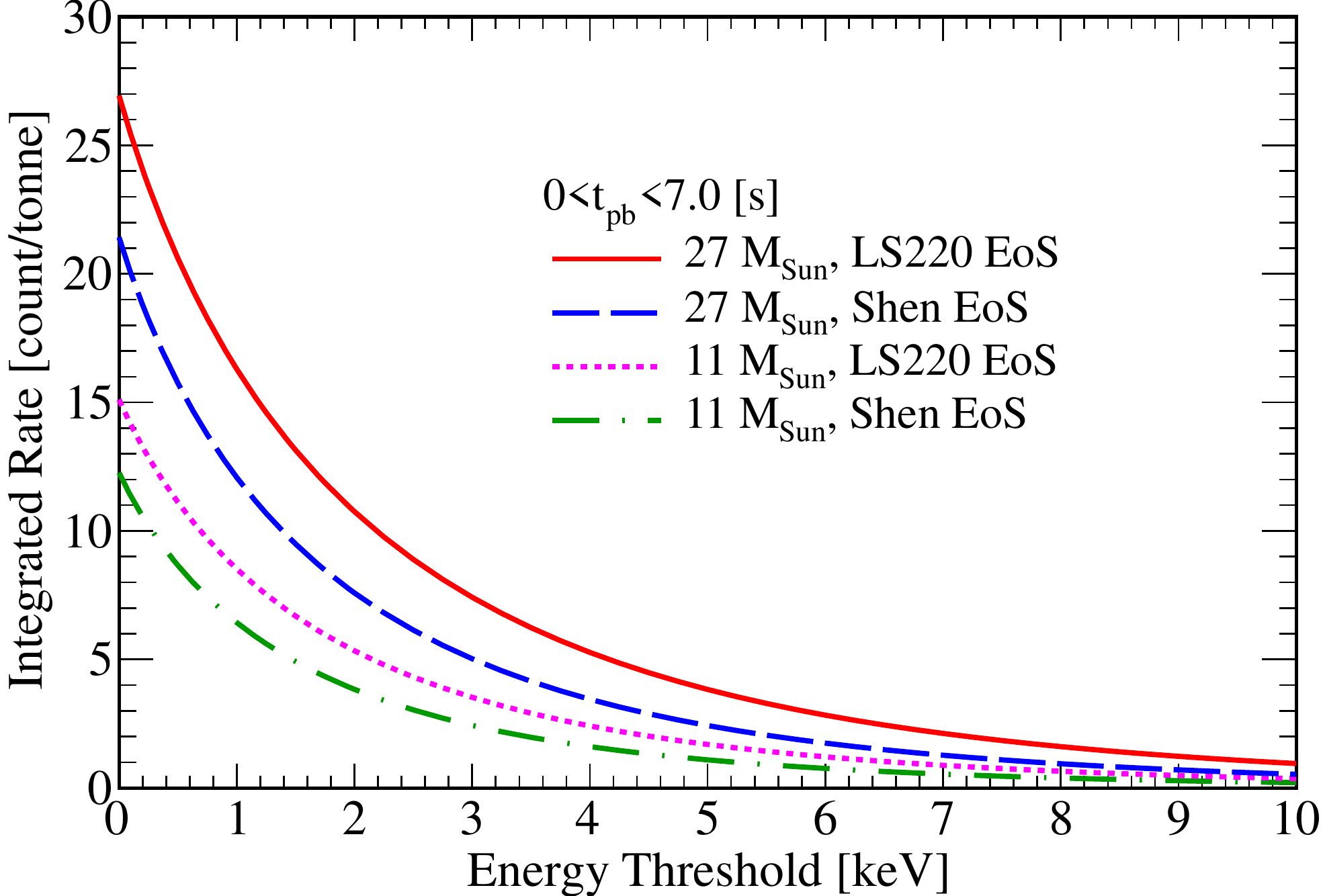}
\caption{The upper panel shows the expected differential recoil spectrum~$dR/d\Er$ as a function of the recoil energy~$\Er$. The differential rate~$dR/dt_{\rm{pb}}$ as a function of the post-bounce time~$t_{\rm{pb}}$ is plotted in the middle panel. The lower panel represents the number of observable events~$R$ as a function of the detector's energy threshold~$E_{\rm{th}}$. All panels show results for $11~\!\Msun$ and $27~\!\Msun$ progenitors with LS220 and Shen EoSs for a SN at 10~\!kpc. In the upper and lower panels, the neutrino flux is integrated over $[0,7]$~\!s after the core bounce, while the middle panel assumes~$E_{\rm{th}}=0$~\!keV. All panels show that the event rate is larger for the $27~\!\Msun$ SN progenitors while the LS220 EoS results in an $\mathcal{O}(25\%)$ larger rate than the Shen EoS. Note that these rates are not directly observable since it is S1 and S2 that is measured, rather than $\Er$.}
\label{fig:DiffRecSpectra}
\end{figure}

The differential scattering rates~$dR/d\Er$ as a function of the xenon recoil energy~$\Er$ for the four progenitor models are shown in the upper panel of Fig.~\ref{fig:DiffRecSpectra} for $t_{\rm{pb}}$ integrated over $[0,7]$~\!s. As evident in all of this figure's panels, the event rate is larger for the $27~\!\Msun$ SN progenitors, while there is a smaller difference owing to the different equations of state, with the LS220 EoS resulting in a slightly larger predicted event rate. 

The middle panel of Fig.~\ref{fig:DiffRecSpectra} shows the differential scattering rate~$dR/dt_{\rm{pb}}$ as a function of the post-bounce time~$t_{\rm{tb}}$ for the four progenitor models. In this figure, we have integrated over the recoil energy, assuming an idealistic threshold energy $E_{\rm{th}}=0$~\!keV. The qualitative behaviour is similar for other threshold energies although the rate is smaller. The differential scattering rates among the different SN progenitors are comparable for $t_{\rm pb}\lesssim 10^{-2}$~\!s, reflecting the similarities of the neutrino emission properties during the neutronization burst (cf.~left panels of Fig.~\ref{fig:LEpanels}). As the post-bounce time increases, the differences among the differential rates become larger. Most of the scattering events occur for~$t_{\rm{pb}}\lesssim1$~\!s. 

The total number of events observed by an experiment is determined by integrating the differential scattering rate above a given energy threshold $E_{\rm{th}}$ over the full time period of the SN burst. These integrated spectra are shown in the lower panel of Fig.~\ref{fig:DiffRecSpectra}. Again, we see that the number of signal events is about twice as large for the $27~\!\Msun$ SN progenitors, while there is $\mathcal{O}(25\%)$ difference owing to the different equation of state. The total number of events drops quickly as $E_{\rm{th}}$ increases, demonstrating the importance of pushing $E_{\rm{th}}$ as low as possible. However, as $\Er$ is not directly measurable, these rates are not directly observable. Therefore, a careful treatment is needed to discuss the rates in terms of~S1 and~S2 signals instead.

Besides scattering off xenon nuclei, neutrinos can also scatter off electrons in the xenon atom. We neglect the latter interaction as the rate of electron recoils is very small, approximately $10^{-5}~\mathrm{counts}/\mathrm{tonne}$, compared to the rate of approximately $10~\mathrm{counts}/\mathrm{tonne}$ for recoils with a xenon nucleus. 

\subsection{Generation of the observable S1 and S2 signals \label{sec:S1S2}}

To convert the nuclear recoil energy~$\Ed$ induced by a SN neutrino into the S1 and S2 signals, we perform a Monte Carlo simulation of a xenon TPC following the method employed by the XENON1T collaboration~\cite{Aprile:2015uzo}. In this subsection, we discuss the technical details of our Monte Carlo simulation.

The S1 and S2 signals are directly proportional to the number of scintillation photons $N_{\rm{ph}}$ and ionization electrons $N_{\rm{el}}$, respectively. The mean numbers of photons and electrons are modeled as
\begin{align}
\langle N_{\rm{ph}} \rangle&=\Ed\ L_{y}(\Ed) \ , \\
\langle N_{\rm{el}} \rangle&=\Ed\  Q_y(\Ed) \label{eq:QYdef}\ ,
\end{align}
where both the photon yield, $L_{y}$, and electron yield, $Q_y$, are functions of~$\Ed$. We use the emission model developed by the LUX collaboration with data from an {\it in situ} nuclear recoil calibration~\cite{Akerib:2015rjg}. We ignore the small effects that may arise from having different drift fields~\cite{Szydagis:2011tk,Mock:2013ila,Lenardo:2014cva} across the various detectors. The quantities~$Q_y$ and~$L_{y}$ have been directly measured down to an energy of 0.7~\!keV and 1.1~\!keV, respectively~\cite{Akerib:2015rjg}. Unless otherwise stated, we assume that~$Q_y$ and~$L_{y}$ are zero below 0.7~\!keV. Hence, our rate predictions tend to be conservative, and we discuss the impact of this assumption in section~\ref{sec:discussion}.

In a realistic detector, we must account for quantum and statistical fluctuations. Part of the nuclear recoil energy is lost to heat dissipation. Therefore, the number of scintillation photons and ionization electrons that are produced by the xenon target, $N_{\rm{Q}}^{\rm{NR}}= N_{\rm{ph}} + N_{\rm{el}}$, is only a fraction of the total number of quanta produced by an electronic recoil at the same energy. In electronic recoils, the energy lost to heat is negligible and $\langle N_{\rm{Q}} \rangle =\Ed/13.7$~\!eV is the total number of quanta available~\cite{Szydagis:20111006,Szydagis:2013sih}. We model the intrinsic fluctuation in $N_{\rm{Q}}^{\rm{NR}}$ with a Binomial distribution characterized by a trial factor $\langle N_{\rm{Q}} \rangle$ and probability $f_{\rm{NR}}=\langle N_{\rm{Q}}^{\rm{NR}} \rangle/ \langle N_{\rm{Q}} \rangle$,
\begin{equation}
N_{\rm{Q}}^{\rm{NR}}=\text{Binomial}(\langle N_{\rm{Q}}\rangle,f_{\rm{NR}}) \ .
\end{equation}
In addition to the intrinsic fluctuation in $N_{\rm{Q}}^{\rm{NR}}$, the fraction of quanta that is emitted as scintillation photons also fluctuates. We model this with a second Binomial distribution with trial factor $N_{\rm{Q}}^{\rm{NR}}$ and probability $f_{\rm{ph}}=\langle N_{\rm{ph}} \rangle/\langle N_{\rm{Q}}^{\rm{NR}} \rangle$:
\begin{equation}
N_{\rm{ph}}=\text{Binomial}(N_{\rm{Q}}^{\rm{NR}},f_{\rm{ph}}) \ .
\end{equation}
By conservation of the number of quanta, we have that the number of electrons is simply $N_{\rm{el}}=N_{\rm{Q}}^{\rm{NR}}-N_{\rm{ph}}$.

Next, we consider detector-specific fluctuations as we convert the number of generated scintillation photons and ionization electrons into observed S1 and S2 signals. Both S1 and S2 are measured in photoelectrons (PE). For the S1 signal, the number of detected photoelectrons~is
\begin{equation}
N_{\rm{PE}}=\text{Binomial}(N_{\rm{ph}},f_{\rm{PE}}) \ ,
\end{equation}
where $f_{\rm{PE}}$ is the photon detection efficiency (also referred to as $g_1$ or $\epsilon_1$ in other studies~\cite{Schumann:2015cpa,Akerib:2015rjg}). LUX calibration measurements indicate that $f_{\rm{PE}}\simeq0.12$~\cite{Akerib:2015rjg} and simulations of the XENON1T detector predict a similar value~\cite{ Aprile:2015uzo}; we assume that $f_{\rm{PE}}\simeq0.12$ for all other detectors, as well. This efficiency may be optimistic for detectors that are much larger than XENON1T since the geometry of larger detectors generally means that~$f_{\rm{PE}}$ decreases. However, as we discuss in section~\ref{sec:S2only}, the~S1 signal is less important than the~S2 signal such that this assumption does not affect our conclusions.

Finally, for the S1 signal, we must account for the response of a PMT, which is modeled with a Gaussian distribution
\begin{equation}
\Sone=\text{Gauss}(N_{\rm{PE}},0.4\sqrt{N_{\rm{PE}}}) \ .
\end{equation}

\begin{figure*}[!htbp]
\centering
\includegraphics[width=0.97\columnwidth]{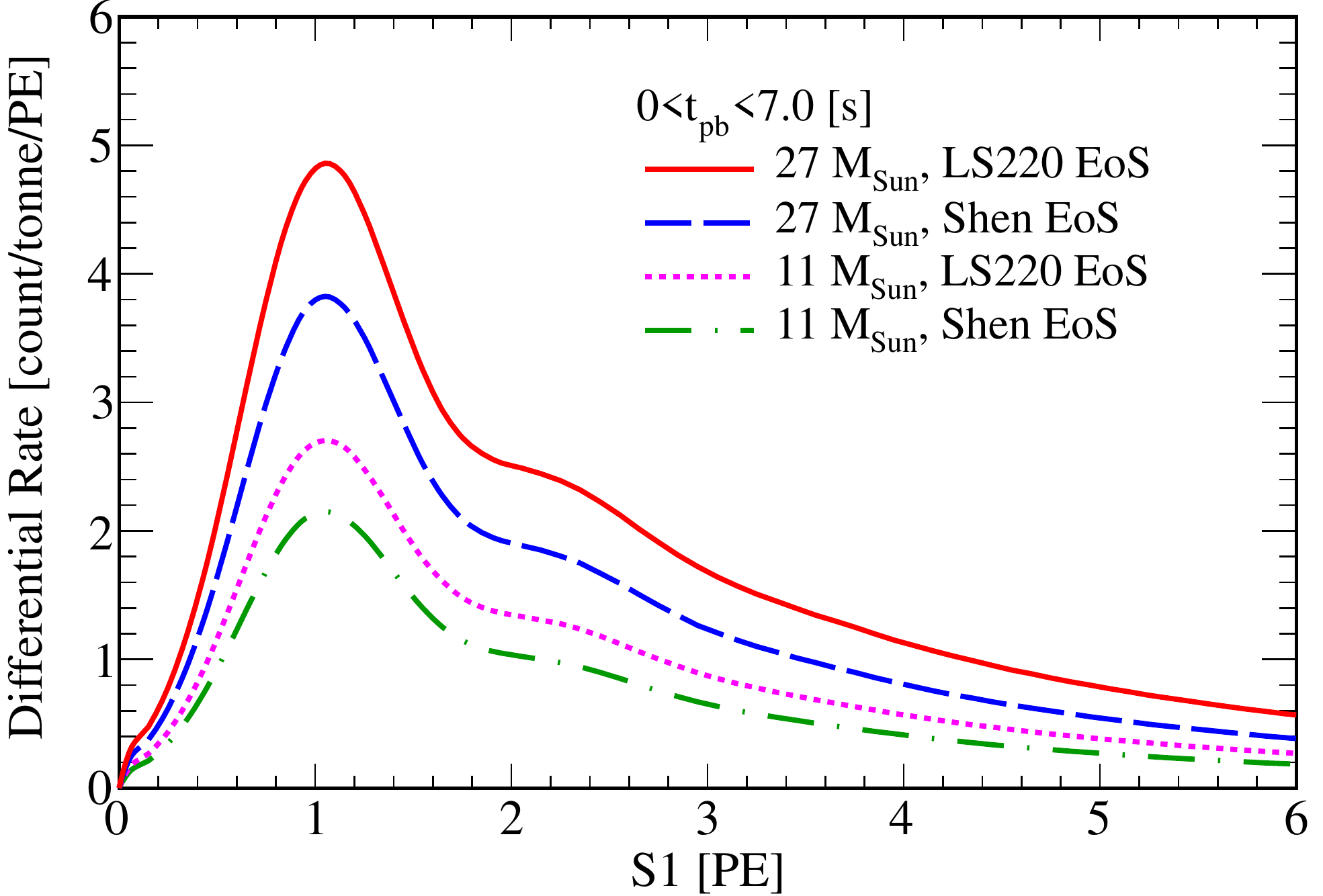} \hspace{1mm}
\includegraphics[width=0.97\columnwidth]{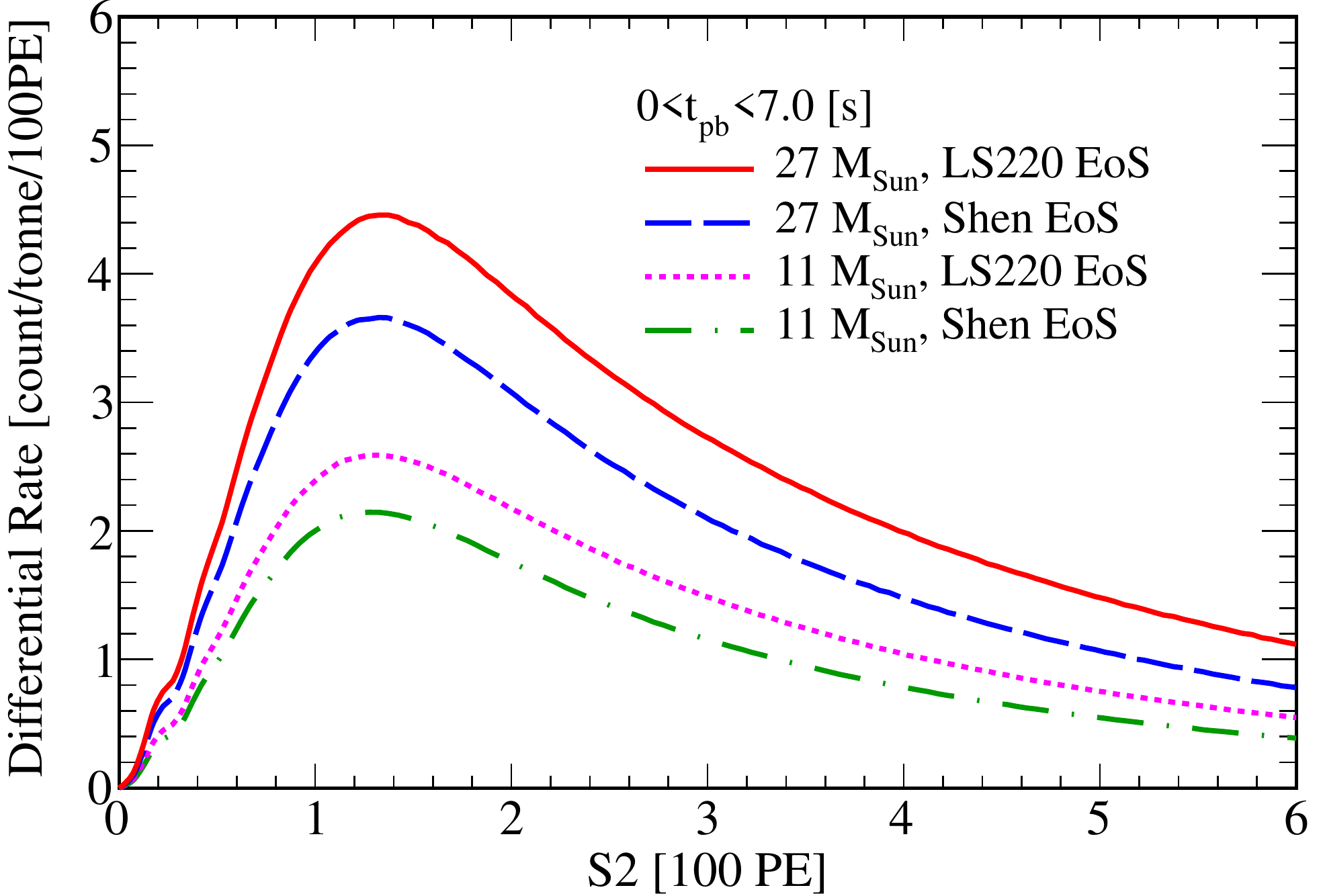}
\caption{The left and right panels show the differential rates of the four SN progenitors in terms of the observable~S1 and~S2 signals, respectively. We have integrated the neutrino flux over the first 7~\!s after the core bounce and assumed that the SN burst occurs 10~\!kpc from Earth. Note that the axes in the right panel have units of 100~PE compared to~PE in the left panel meaning that the~S2 signal is generally larger than the~S1 signal. The light and charge yields,~$L_y$ and~$Q_y,$ respectively, have been set to zero below recoil energies of 0.7~\!keV. Integrated rates are given in Table~\ref{table:S1S2numbers}.}
\label{fig:S2S1space}
\end{figure*}

To obtain the S2 signal, we must account for the loss of ionization electrons due to electronegative impurities in the liquid as they drift toward the liquid-gas interface. This attenuation is represented by the electron survival probability
\begin{equation}
p_{\rm{sur}}=\exp\left[-\frac{\Delta z}{v_{\rm{d}}\tau} \right] \ ,
\end{equation}
where $\Delta z$ is the distance an electron traverses, $\tau$ is the so-called electron lifetime in the liquid and~$v_{\rm{d}}$ is the electron drift velocity. The value measured in XENON100 was $v_{\rm{d}}\simeq1.7$~\!mm/$\mu$s~\cite{Aprile:2012vw} and we assume this value for all other detectors (cf.~\cite{Yoo:2015yza}). The electron lifetime in the liquid is in general a detector-dependent parameter. For the purpose of this study, we therefore assume that $\Delta z/\tau$ is distributed uniformly over $[0,2/3]$~\!mm/$\mu$s and that it holds for all of the detector sizes that we consider. This condition means that, as detectors become larger, their purity increases proportionally to meet or exceed this requirement. For XENON1T (LZ), the maximum drift length $\Delta z_\text{\rm{max}}\simeq967~\! (1300)$~\!mm implies that the electron lifetime is at least $\tau\simeq1450~\! (1950)$~\!$\mu$s, which is straightforward to achieve~\cite{Akerib:2015cja}.

The number of ionization electrons that reach the liquid-gas interface is
\begin{equation}
\tilde{N}_{\rm{el}}=\text{Binomial}(N_{\rm{el}},p_{\rm{sur}}) \ ,
\end{equation}
and we assume that the extraction efficiency from the liquid into gas is 100\%. Finally, the extracted electrons are accelerated by a strong electric field in the gas phase and induce an S2 signal that is modeled with a Gaussian distribution:
\begin{equation}
\label{eq:S2gas}
\Stwo=\text{Gauss}(20\ \tilde{N}_{\rm{el}},7\ \sqrt{\tilde{N}_{\rm{el}}}) \ .
\end{equation}
We have conservatively assumed that the average yield for each extracted electron is 20~PE, but the yield could be higher. For example, the LZ design goal is 50~PE per electron~\cite{Akerib:2015cja}.

\subsection{Observable scattering rates \label{sec:obrates}}

Now that we have given expressions for $dR/dE_{\rm{R}}$ and detailed our procedure for generating the~S1 and~S2 signals, it is straightforward to use Eq.~\eqref{eq:rateS1S2} to calculate the differential rates $dR/d\Sone$ and~$dR/d\Stwo$. These rates are shown in the left and right panels of Fig.~\ref{fig:S2S1space}, respectively, where we have integrated $t_{\rm{pb}}$ over $[0,7]$~\!s and assumed that~$L_y$ and~$Q_y$ are zero below 0.7~\!keV, as discussed in the previous subsection. In the left panel, we have integrated over all~S2 values by assuming an~S2 threshold of zero, while the~S1 signal has been integrated with an~S1 threshold of zero in the right panel. Similar to the previous figures, the differential rates are highest for the $27~\!\Msun$ SN progenitors and for the LS220~EoS. Comparing the two panels, it is apparent that the~S2 signal is generally larger than the~S1 signal (note that the axes in the right panel have units 100~PE compared to~PE in the left panel). The left panel shows that the differential rate has peaks at integer multiples of 1~PE. A similar behaviour is also present in the right panel, although the effect is smaller and the peaks appear at multiples of 20~PE, the average number of photoelectrons generated for each extracted electron, cf.~Eq.~\eqref{eq:S2gas} (see also Fig.~\ref{fig:QyLyvary4} where the effect is more apparent). The roll-off in~$dR/d\mathrm{S2}$ below approximately $100$~PE (right panel) is a result of the assumption that~$Q_y$ is zero below 0.7~\!keV. In section~\ref{sec:discussion}, we show that this assumption does not have a significant impact on our results.

\begin{table}
\caption{Expected number of SN neutrino events per tonne of xenon target above various~S1 and~S2 thresholds. The SN burst occurs at 10~\!kpc from Earth and the neutrino flux has been integrated over the first 7~\!s after the core bounce. The light and charge yields,~$L_y$ and~$Q_y$, respectively, have been set to zero below recoil energies of 0.7~\!keV. The number of events, for the case in which the threshold includes $0$~PE (`$\geq0$') and when it does not (`$>0$'), have been separated to show that many of the events have an~S1 or~S2 signal that is exactly zero. The symbol $(\star)$ indicates the most likely threshold values (see discussion in sections~\ref{sec:S2only} and~\ref{sec:discussion} for details). An S2-only search for \CENS\ from SN neutrinos is optimal as it results in a higher number of detected events. \label{table:S1S2numbers}}
\centering
\begin{ruledtabular}
\begin{tabular}{@{}c c c c c c @{}} 
& & \multicolumn{2}{c}{$27\,\Msun$} & \multicolumn{2}{c}{$11\,\Msun$} \\ \cline{3-4}\cline{5-6}
& & LS220 & Shen & LS220 & Shen \\ \colrule
$\mathrm{S1}_{\mathrm{th}}$ [PE] & $\langle N_{\rm{ph}} \rangle$ & & & \\\colrule
$\geq 0$ & 0 & 26.9 & 21.4 & 15.1 & 12.3 \\
$> 0$ & 0 & 13.3 & 9.8 & 6.9 & 5.2 \\
1 & 8.3 & 11.0 & 8.0 & 5.6 & 4.1 \\
2 & 16.7 & 7.3 & 5.1 & 3.6 & 2.6 \\
3 $(\star)$  &25  & 5.2 & 3.5 & 2.4 & 1.7 \\ \colrule
$\mathrm{S2}_{\mathrm{th}}$ [PE] & $\langle N_{\rm{el}} \rangle$ & & & \\ \colrule
$\geq 0$ & 0 & 26.9 & 21.4 & 15.1 & 12.3 \\
$> 0$ & 0 &  18.5 & 14.0 & 9.9 & 7.6 \\
20 & 1.2 & 18.4 & 14.0 & 9.8 & 7.6 \\
40 & 2.4 & 18.1 & 13.7 & 9.7 & 7.4 \\
60 $(\star)$ & 3.6 & 17.6 & 13.3 & 9.4 & 7.2 \\
80 & 4.8 & 17.0 & 12.8 & 9.0 & 6.9 \\
100 & 6.0 &16.3 & 12.2 & 8.6 & 6.5 \\ 
\end{tabular} 
\end{ruledtabular}
\end{table}

Table~\ref{table:S1S2numbers} lists the total number of expected events per tonne of xenon target for various values of the~S1 and~S2 thresholds. For the listed S1 thresholds, we have integrated over all~S2 values, and vice versa for the listed S2 thresholds. The S2 thresholds are given as multiples of 20~PE, the average number of detected photoelectrons for each extracted electron. The second column lists the mean number of primary photons and electrons required to produce an~S1 and~S2 signal for the listed thresholds, calculated with the relations $\langle \mathrm{S1} \rangle= f_{\rm{PE}} \langle N_{\rm{ph}} \rangle$ and $\langle \mathrm{S2} \rangle= 20 \langle p_{\rm{sur}} \rangle \langle N_{\rm{el}} \rangle$ (we quote the raw~S2 value, rather than the position corrected value). The number of events is reported for our four SN progenitor models located 10~kpc from Earth and for~$t_{\rm pb}$ integrated in the range $[0,7]$~\!s. We have separated the number of events for the case in which the threshold includes 0~PE and when it does not to show that approximately~50\% and~30\% of the events have an~S1 or~S2 signal that is exactly zero, respectively, and are therefore not observable even in an ideal detector. Generally, the number of~S2 events is much higher than the number of~S1 events, and the event rate drops more slowly as the~S2 threshold is increased, compared to an increase in the~S1 threshold. This trend reflects the fact that the~S2 signal from low-energy depositions is easier to detect in a dual-phase xenon TPC due to the amplification that is inherent to the process of proportional scintillation. For example, the mean~S1 signal of a 1~\!keV energy deposition is $\langle \mathrm{S1} \rangle\simeq0.5$~PE, while the mean number of electrons and mean~S2 signal are $\langle \mathrm{N_{\rm{el}}} \rangle\simeq7.4$ and $\langle \mathrm{S2} \rangle\simeq150$~PE, respectively. Since dual-phase xenon detectors are sensitive to single electrons~\cite{Edwards:2007nj,Aprile:2013blg}, even very small energy depositions result in detectable~S2 signals.

On the basis of these preliminary results, we show in the next section that an S2-only analysis is the optimal channel for detecting \CENS\ from SN neutrinos. We discuss realistic values of the S2~threshold and show that an S2-only search is not limited by background events. In section~\ref{sec:discussion}, we also show the signal uncertainty is not a limitation.

\section{S2-only analysis} \label{sec:S2only}

The canonical dark matter search in a dual-phase xenon experiment requires the presence of both an~S1 and an~S2 signal. This stipulation reduces the background rate by two primary means. Firstly, measuring both S1 and S2 enables discriminated between the dominant electronic recoil backgrounds and the expected nuclear recoil signal, based on the ratio S2/S1 at a given value of S1. Secondly, the S1 and S2 signals allow for a 3D reconstruction of the interaction vertex, based on the time difference between the S1 and S2 signal events and the PMT hit pattern. The latter means that events can be selected from the central region of the detector, where the background rate is lowest. In these canonical dark matter searches, which utilize data collected over $\mathcal{O}(100)$~days, the~S1 threshold is typically 2~PE or 3~PE, while the S2 threshold is typically $\sim150$~PE (see e.g.~\cite{Aprile:2012nq,Akerib:2015rjg,Tan:2016diz}). 

For SN neutrinos though, the brevity of the $\mathcal{O}(10)$~\!s burst enables the signal to be discrimination from background based on the timing information rather than the charge-to-light ratio. Although the requirement of detecting both an~S1 and an~S2 signal has the effect of further reducing the background rate, it also significantly reduces the signal rate, especially for processes such as SN neutrino scattering where the nuclear recoil energy is small~\cite{Hagmann:2004uv,Angle:2011th,Frandsen:2013cna,Essig:2011nj,Santos:2011ju,Essig:2012yx}. For example, for $\mathrm{S2}_{\rm{th}}=60$~PE and any value of~S1 (including no S1 signal), the number of SN neutrino events for the $27~\!\Msun$ SN progenitor with the LS220 EoS is 17.6~events/tonne. However, when additionally requiring an~S1 signal with $\mathrm{S1}_{\rm{th}}=2$~PE, the number of events drops to only 7.2~events/tonne. Requiring both an~S1 and an~S2 signal therefore significantly reduces the rate of \CENS\ compared to an S2-only analysis.

We now show that, for a SN burst, the expected background rate in a tonne-scale detector is small enough such that an S2-only analysis does not require the additional discrimination capabilities otherwise afforded by the S1 signal. Although the low-energy S2 background in dual-phase xenon experiments is not yet fully understood, the dominant contribution is believed to arise from photoionization of impurities in the liquid xenon and the metal surfaces in the TPC~\cite{Aprile:2013blg}, caused by the relatively high energy of the 7-\!eV xenon scintillation photons. Another background contribution may be from delayed extraction of electrons from the liquid to gas-phase~\cite{Edwards:2007nj}. Such processes create clusters of single-electron~S2 signals and, occasionally, these single-electron signals overlap and appear as a single~S2 signal from multiple electrons. The resultant low-energy background~S2 signals are very similar to those expected in the case of a SN neutrino interaction. The background rate for these lone-S2 events has been characterized by XENON10~\cite{Angle:2011th,Essig:2012yx} and XENON100~\cite{Aprile:2016wwo}, which found background rates of approximately $2.3 \times 10^{-2}$ and $1.4 \times 10^{-2}$~events/tonne/s, respectively. These rates are consistent with the general expectation that the S2-only background rate is independent of the detector size. Based on these measurements, we therefore assume that the average background rate in XENON1T and future detectors will lie in the range $(1.4-2.3) \times 10^{-2}$~events/tonne/s. This background rate corresponds to $0.1-0.2$~events/tonne during the initial~7~\!s of the~SN signal, which is at least a factor of~40 smaller than the signal rate from the $11~\!\Msun$ with Shen EoS progenitor, the smallest rate in Table~\ref{table:S1S2numbers} (assuming $\mathrm{S2}_{\rm{th}}=60$~PE). Additionally, it is worth recalling that the background signal grows linearly in time, whereas the SN neutrino signal does not, resulting in an even better signal-to-background ratio in the early times of the SN burst.

Finally, we motivate an appropriate choice of the~S2 threshold. This threshold is largely determined by two factors. The first is the `trigger-efficiency' for an experiment to detect an S2 signal. For XENON10, the trigger-efficiency was 50\% for $\mathrm{S2}\simeq20$~PE and reached 100\% for $\mathrm{S2}\simeq30$~PE~\cite{Essig:2012yx}, while for XENON100, it was 50\% for $\mathrm{S2}\simeq60$~PE and reached 100\% for $\mathrm{S2}\simeq140$~PE~\cite{Aprile:2012vw}. Values have not yet been reported for LUX. The trigger system for XENON1T has been significantly upgraded relative to XENON100 and is expected to lead to an improvement in the trigger-efficiency. Therefore, while the trigger-efficiency does vary between different experiments, here we assume a benchmark value of $\mathrm{S2}_{\rm{th}}=60$~PE and make the simplifying assumption that the trigger-efficiency is 100\% above this value. This benchmark value is consistent with the threshold in the sensitivity studies of LZ, where it was assumed that the S2-only threshold is 2.5 extracted electrons~\cite{Akerib:2015cja}, corresponding to $\mathrm{S2}_{\rm{th}}=50$~PE with an average of 20~PE per extracted electron (and ignoring the small loss owing to the finite electron lifetime).

The second consideration when deciding $\mathrm{S2}_{\rm{th}}$ is the signal uncertainty induced by the choice of the electron yield $Q_y$ (cf.~Eq.~\eqref{eq:QYdef} for where it enters our analysis). We postpone a full discussion of this uncertainty until section~\ref{sec:discussion} and, for now, simply state that the signal uncertainty from $Q_y$ is smaller than~10\% when $\mathrm{S2}_{\rm{th}}=60$~PE. This is appreciably smaller than the $\sim25\%$ variation for the LS220 and Shen EoS for the same progenitor mass as well as the approximate factor-of-two difference for different progenitor masses; so, this uncertainty should only have a small effect on our results.

\begin{figure}
\centering
\includegraphics[width=0.95\columnwidth]{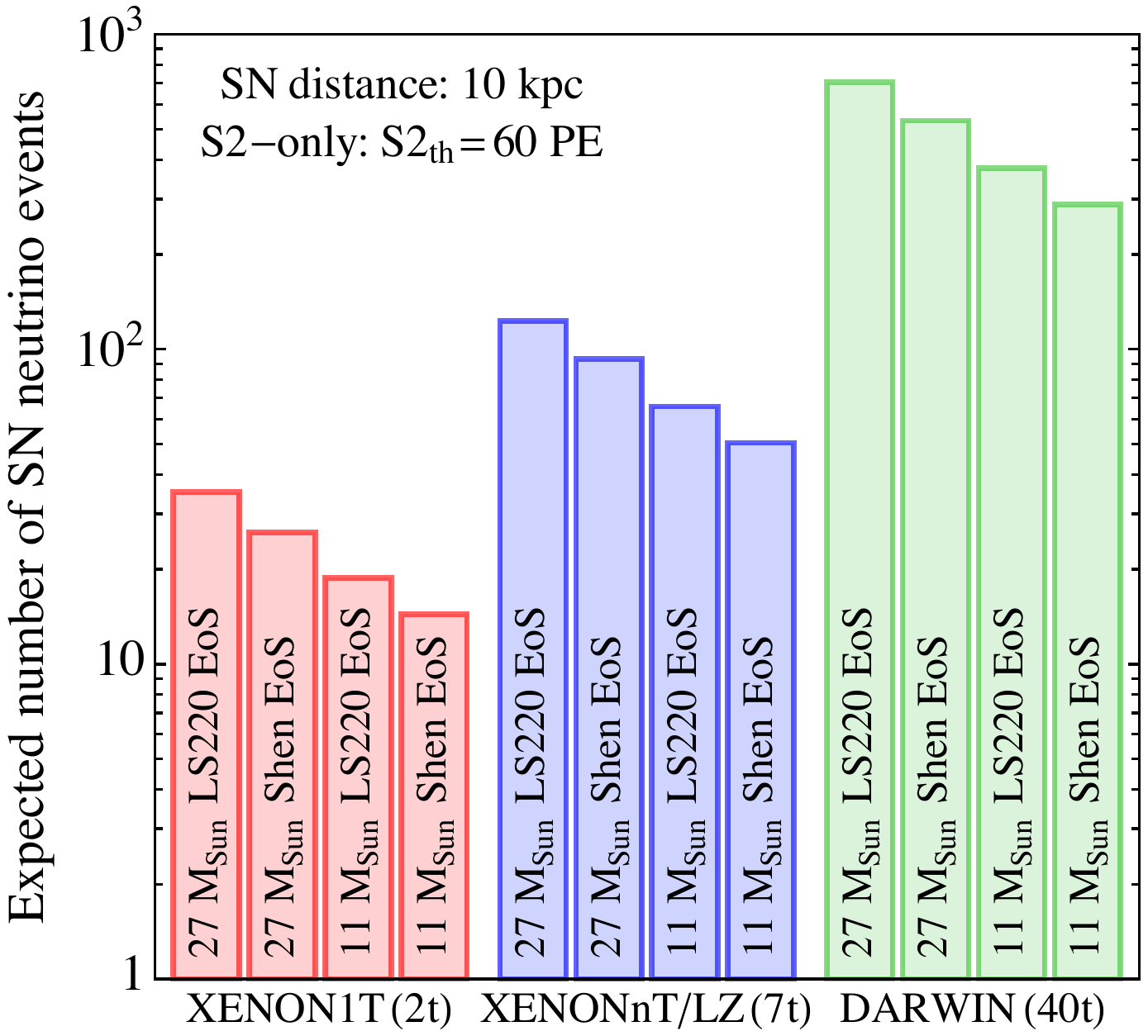}
\caption{The expected number of SN neutrino events from an S2-only analysis with a threshold of 60~PE for a SN burst at 10~\!kpc from Earth. The different colours refer to XENON1T (red), XENONnT and LZ (blue), and DARWIN (green) detectors. For each detector, the number of events is shown for the four SN progenitors that we consider in this study.}
\label{fig:Nevents}
\end{figure}

For all of the reasons outlined above, our main results have been obtained by adopting an S2-only analysis with $\mathrm{S2}_{\rm{th}}=60$~PE. Figure~\ref{fig:Nevents} displays the expected number of SN neutrino events from an S2-only analysis with this threshold for the three detectors and four SN progenitors that we consider in this study. Finally, since the background rate is significantly smaller than the signal rate, we ignore it in section~\ref{sec:results} unless stated otherwise.

\section{Supernova neutrino detection} \label{sec:results}

In this section, we calculate the discovery potential of an S2-only search for SN neutrinos as a function of the SN distance and discuss the discrimination power of xenon detectors with respect to the SN progenitor. We then show that it is possible to reconstruct the SN neutrino light curve and, therefore, to discriminate among the different phases of the neutrino signal. Furthermore, we demonstrate that xenon detectors can reconstruct both the neutrino differential spectrum and the total energy emitted by the SN into all flavours of neutrinos. Finally, we present a concise comparison of the performance of xenon detectors with dedicated neutrino detectors.

\subsection{Detection significance}
\begin{figure}[!tbp]
\centering
\includegraphics[width=0.95\columnwidth]{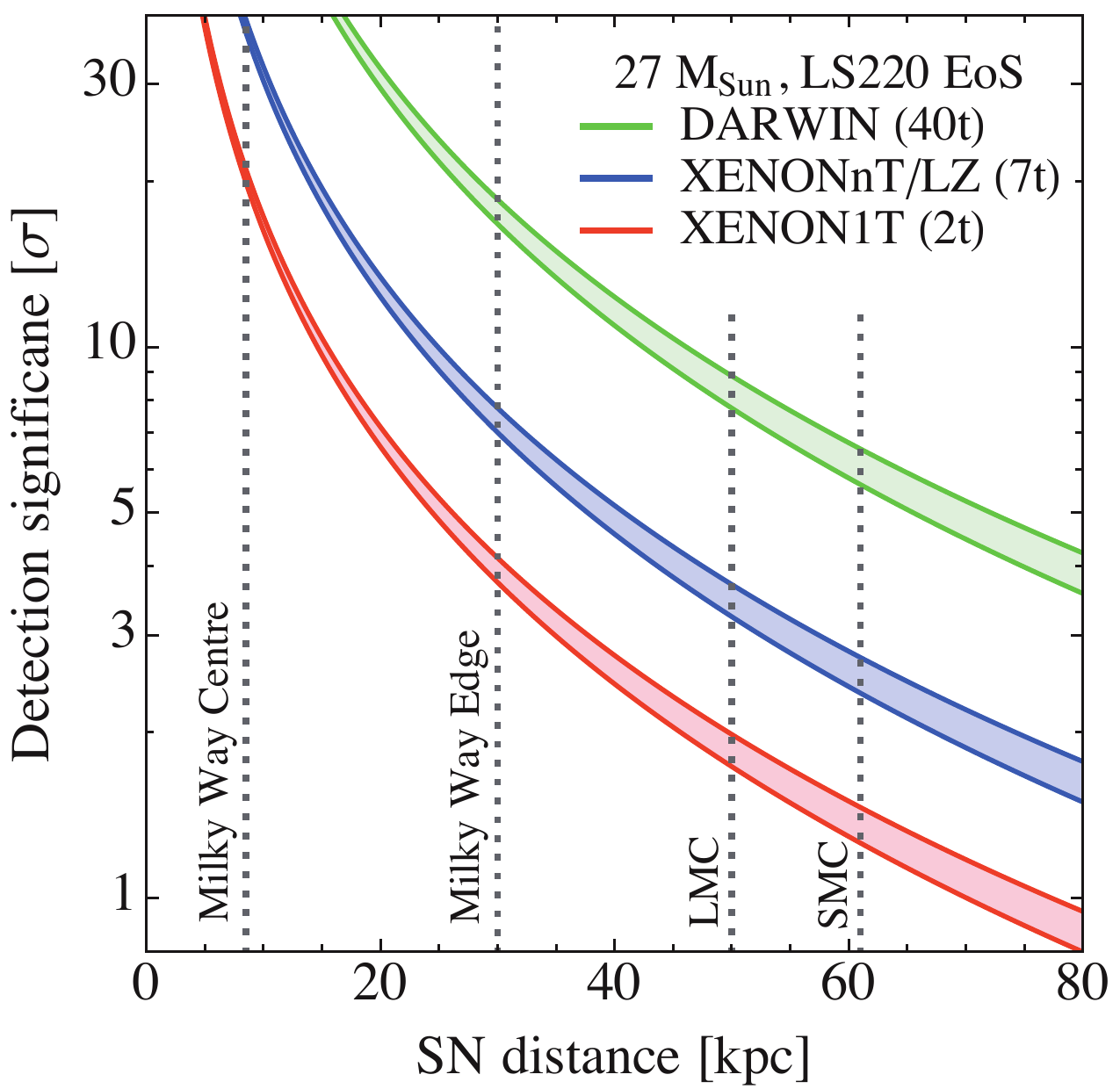}
\caption{The detection significance is given as a function of the SN distance for a 27~\!$\Msun$ progenitor with LS220 EoS. The SN signal has been integrated over $[0,7]$~\!s. The different bands refer to XENON1T (red), XENONnT and LZ (blue), and DARWIN (green). The band width reflects uncertainties from our estimates for the background rate, discussed in section~\ref{sec:S2only}. The vertical dotted lines mark the centre and edge of the Milky Way as well as the Large and Small Magellanic Clouds (LMC and SMC, respectively). For this SN progenitor, XENONnT/LZ could make at least a~$5\sigma$ discovery of the neutrinos from a SN explosion anywhere in the Milky Way. DARWIN extends the sensitivity beyond the SMC.}
\label{fig:milkyway}
\end{figure}

We first investigate the sensitivity of present and upcoming xenon detectors to a SN burst as a function of the SN distance from Earth. Figure~\ref{fig:milkyway} shows the SN burst detection significance as a function of the SN distance from Earth for the 27~\!$\Msun$ progenitor with LS220 EoS. We see that XENON1T will be able to detect this SN burst at more than~$5\sigma$ significance up to 25~kpc from Earth, while XENONnT and LZ will make at least a~$5\sigma$ discovery anywhere in the Milky Way. DARWIN's much larger target mass will extend the sensitivity to a 5$\sigma$ discovery past the Large Magellanic Cloud (LMC) and the Small Magellanic Cloud~(SMC).

In this figure, the SN signal has been integrated over the first 7 seconds after the core bounce. We calculate the detection significance following the likelihood-based test for the discovery of a positive signal described in~\cite{Cowan:2010js}. Our null hypothesis is that the observed events are only due to the background processes described in section~\ref{sec:S2only}, while our alternative hypothesis is that the observed events are due to both the background processes and from SN neutrino scattering. A detection significance of~$5\sigma$ means that we reject the background-only hypothesis at this significance, which we therefore regard as a~$5\sigma$ discovery of the SN neutrino signal. The bands in Fig.~\ref{fig:milkyway} show the detection significance for a background rate spanning the range $(1.4-2.3) \times 10^{-2}$~events/tonne/s, our assumption for the background rate discussed in section~\ref{sec:S2only}, based on the measured rates in XENON10 and XENON100. 

Figure~\ref{fig:milkyway} shows the detection significance for the 27~\!$\Msun$ LS220 EoS progenitor, which gives the highest event rate among the four progenitors that we consider. However, from this figure and Table~\ref{table:S1S2numbers}, it is straightforward to calculate the detection significance for the other progenitors. The expected number of events simply scales with the inverse square of the SN distance, which implies that the distance $d_2|_{n \sigma}$ for an $n\sigma$ detection of an alternative SN progenitor is related to the distance $d_{27,\mathrm{LS220}}|_{n \sigma}$ for an $n\sigma$ detection of the 27~\!$\Msun$ LS220 EoS progenitor by $d_2|_{n \sigma}=d_{{27,\mathrm{LS220}}}|_{n \sigma} \sqrt{\mathrm{events}_2/\mathrm{events}_{27,\mathrm{LS220}}}$. Here `events' is simply the number of events calculated from the $\Stwo_{\rm{th}}=60$~PE row in Table~\ref{table:S1S2numbers} (which gives the number of events per tonne and thus must be multiplied by the detector size). With this formula, we estimate that the SN burst from the 11~\!$\Msun$, Shen EoS progenitor can be detected at $5\sigma$ significance at 16~kpc, 26~kpc and 44~kpc from Earth for XENON1T, XENONnT/LZ and DARWIN, respectively.

\subsection{Distinguishing between supernova progenitors}

Besides spotting a SN burst, we are also interested in investigating whether dual-phase xenon detectors could help us to constrain the SN progenitor physics and the neutrino properties. Given the sensitivity of xenon detectors to SN neutrinos and the expected insignificant background, detection should allow the progenitor mass to be discerned. With the neutrino flux from only four progenitor models, we cannot perform a detailed study of the precision with which the progenitor mass could be reconstructed. However, we can make some general statements on the performance of the different xenon experiments.

For a SN at 10~kpc, the expected numbers of events in XENON1T, XENONnT/LZ and DARWIN for the 27~\!$\Msun$ LS220 EoS progenitor are 35, 123 and 704, respectively, which are $3.8\sigma$, $7.1\sigma$ and $16.9\sigma$ higher than the 11~\!$\Msun$ LS220 EoS progenitor, where the expectations are 19, 66 and 376 events. This demonstrates that when the SN distance is well known, DARWIN will be able to discern between these progenitor masses with a high degree of certainty, while even XENON1T's ability will be reasonably good. 

\subsection{Reconstructing the supernova neutrino light curve}

\begin{figure*}[!tbp]
\centering
\includegraphics[width=0.99\columnwidth]{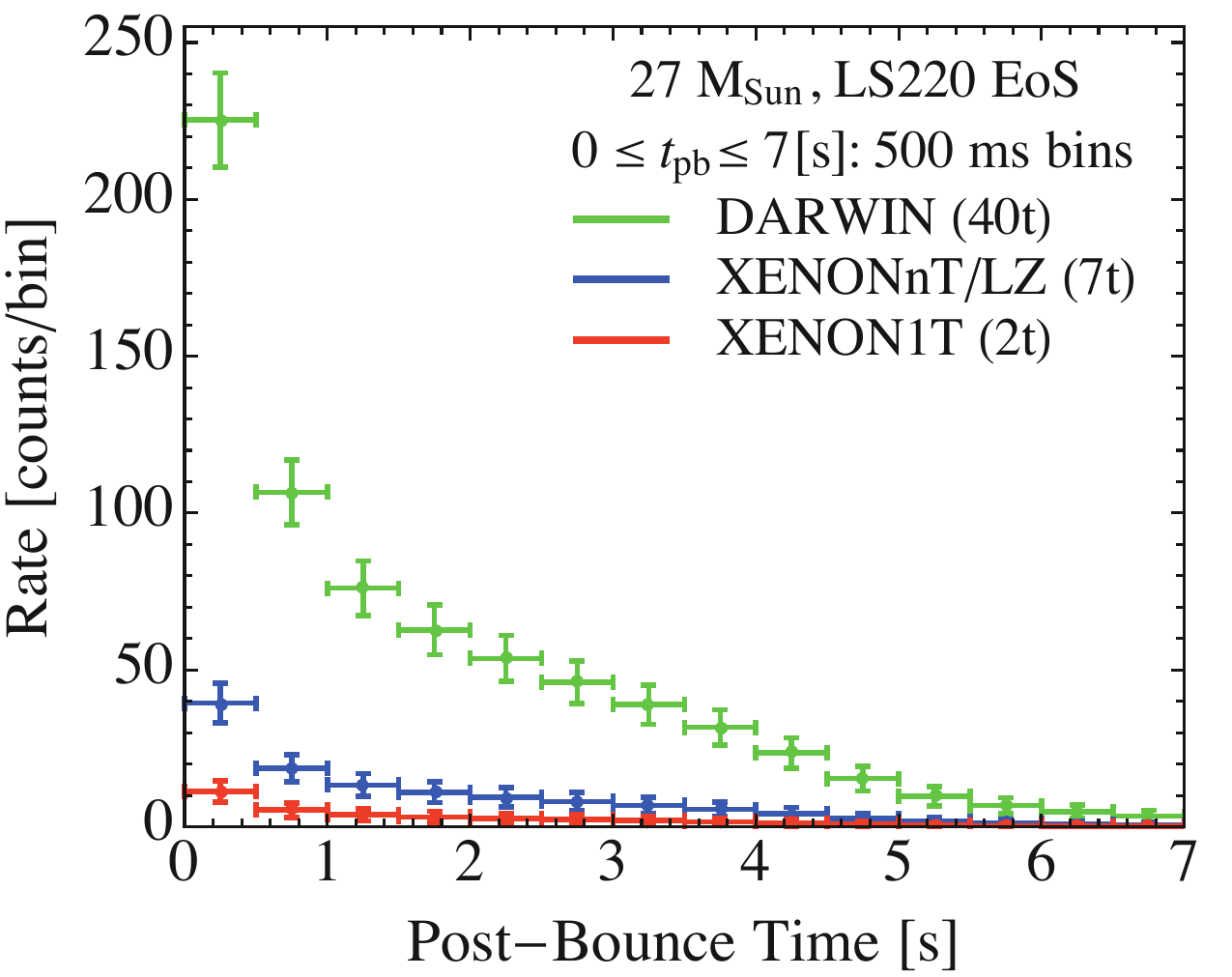}
\includegraphics[width=0.99\columnwidth]{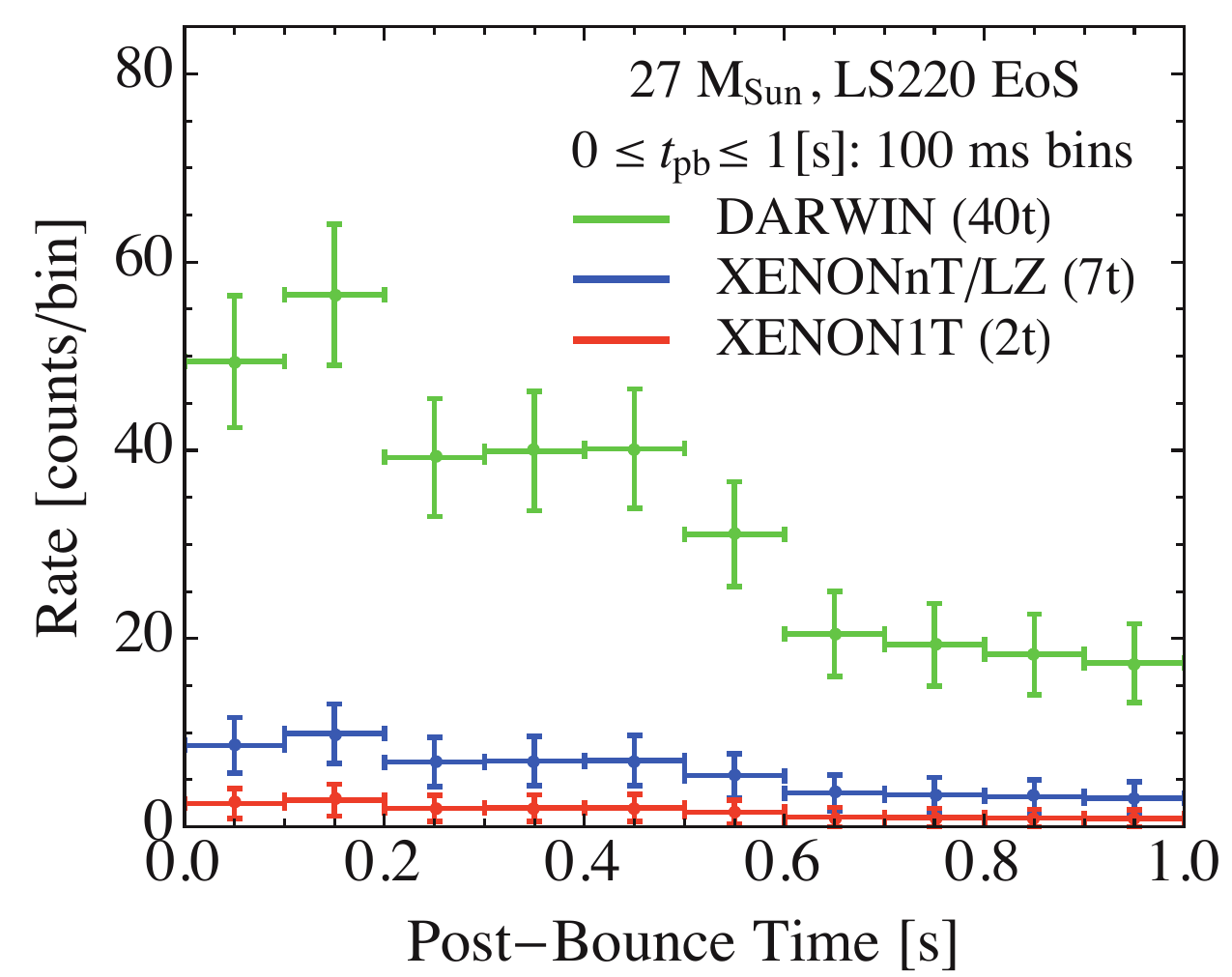}
\caption{Event rate from an S2-only analysis as a function of the post-bounce time for a SN burst at 10~kpc. The event rate is shown for a $27~\!\Msun$ SN progenitor with LS220 EoS for three target masses: 2,~7, and~40 tonnes in red, blue, and green respectively. The left panel covers the full time evolution with 500~\!ms time bins, including the Kelvin-Helmholtz cooling phase. The right panel shows the early evolution with 100~\!ms time bins and focuses on the neutronization and accretion phases.}
\label{fig:light}
\end{figure*}

We now discuss the reconstruction of the SN neutrino light curve from a Galactic SN burst. Figure~\ref{fig:light} shows the neutrino event rate for the most optimistic of the four SN progenitors ($27~\!\Msun$ with LS220 EoS) as a function of the time after the core bounce. The rate has been obtained for a SN at 10~kpc from Earth by adopting an S2-only analysis with a benchmark threshold of 60~PE for XENON1T, XENONnT/LZ and DARWIN. In this analysis, we neglect the small background rate.

The left panel of Fig.~\ref{fig:light} shows the light curve during the full time evolution of the SN burst with 500~\!ms bins. For a Galactic SN, a detector the size of DARWIN clearly shows the characteristic behaviour of the Kelvin-Helmholtz cooling phase where the event rate slowly decreases between 1 to 7~\!s, following the same neutrino luminosity trend (cf.\ Figs.~\ref{fig:LEpanels} and~\ref{fig:DiffRecSpectra}). This behaviour is also partially distinguishable with XENONnT/LZ, albeit with a smaller significance, and essentially unobservable with XENON1T where the number of events per bin is too low.

The right panel of Fig.~\ref{fig:light} focuses on the early time evolution of the SN signal during the neutronization and the accretion phases of the burst. In this panel, 100~\!ms time binning is used. A detector the size of DARWIN will be able to discern the neutronization peak in the neutrino signal shown in Fig.~\ref{fig:LEpanels}. However, the neutronization peak cannot be distinguished at a high level of significance in a seven-tonne or two-tonne detector for a SN at 10~kpc. We thus conclude that it will be necessary to have a xenon detector with $\mathcal{O}(40)$~tonnes of xenon in order to constrain the SN light curve with high precision. Such an experiment will be competitive with existing neutrino telescopes. We stress that the results shown here are for a SN exploding 10~kpc from Earth. The DARWIN results shown in Fig.~\ref{fig:light} for a SN at 10~kpc are equivalent to the XENON1T or XENONnT/LZ results for a SN at 2.2~kpc and 4.2~kpc respectively. 

As shown in Figs.~\ref{fig:LEpanels} and~\ref{fig:DiffRecSpectra}, the neutrino signal is almost independent of the progenitor mass and the nuclear EoS for $t_{\rm{pb}} \lesssim 10$~ms, while, for later times, it depends on the progenitor properties. Therefore, an accurate measurement of the later-time light curve will tell us about properties of the SN progenitor, such as its mass and~EoS. 

The single-flavour light curve, which depends on the neutrino mass ordering and on flavour oscillation physics~\cite{Mirizzi:2015eza}, should be accurately reconstructed with traditional neutrino detectors. The fact that xenon-based direct detection dark matter experiments are flavour insensitive will allow for the possibility of combining the all-flavour light curve with results from detectors sensitive to a single-flavour light curve. This complementarity between xenon detectors and traditional neutrino experiments should, therefore, allow for tests of oscillation physics as well as the possible existence of non-standard physics scenarios~(e.g.~\cite{Keranen:2004rg}). We leave a detailed study of this feature for future work.

\subsection{Neutrino differential flux}\label{sec:Erec}

\begin{figure*}[!tbp]
\centering
\includegraphics[width=0.9\columnwidth]{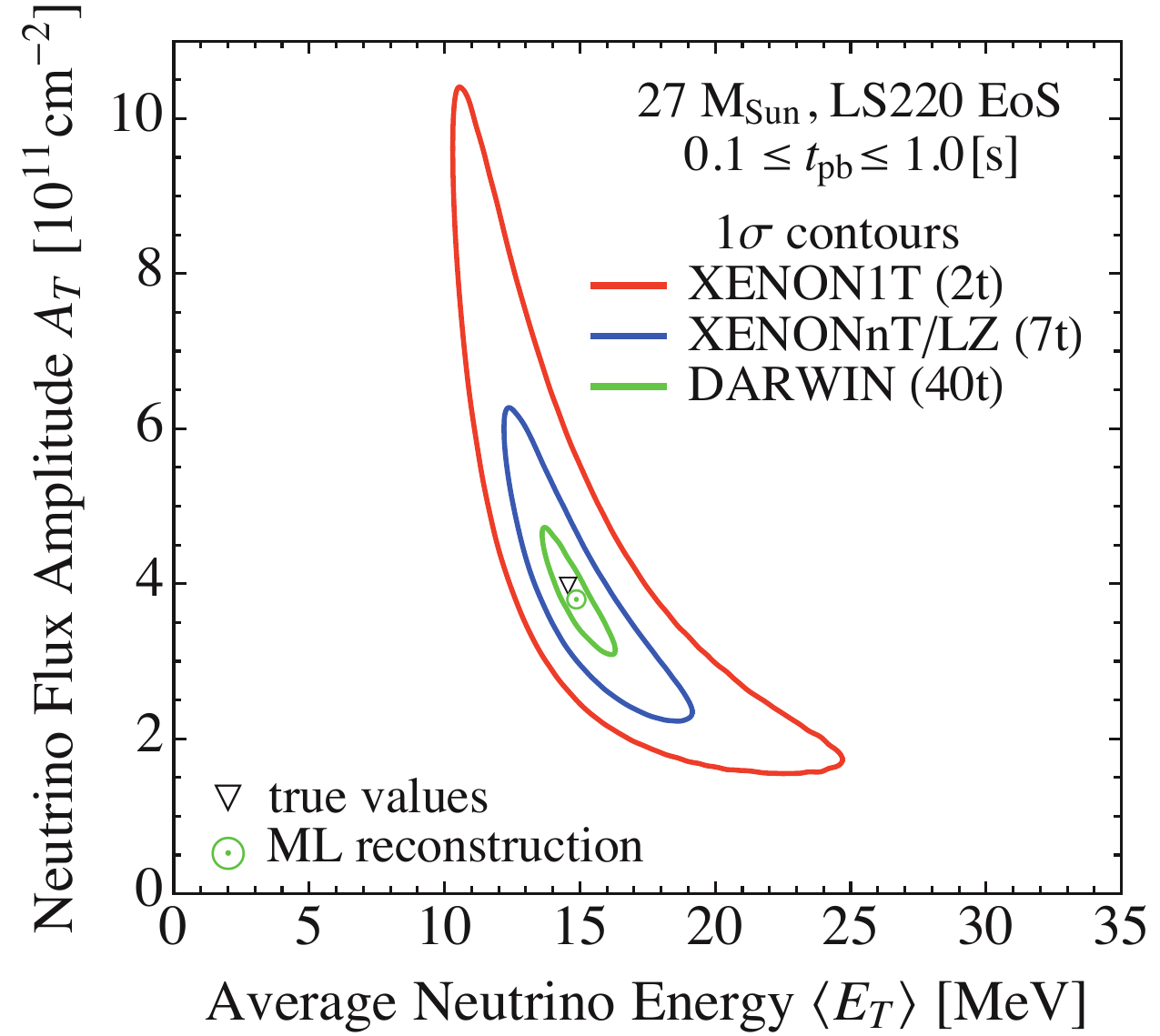}\hspace{2mm}
\includegraphics[width=0.9\columnwidth]{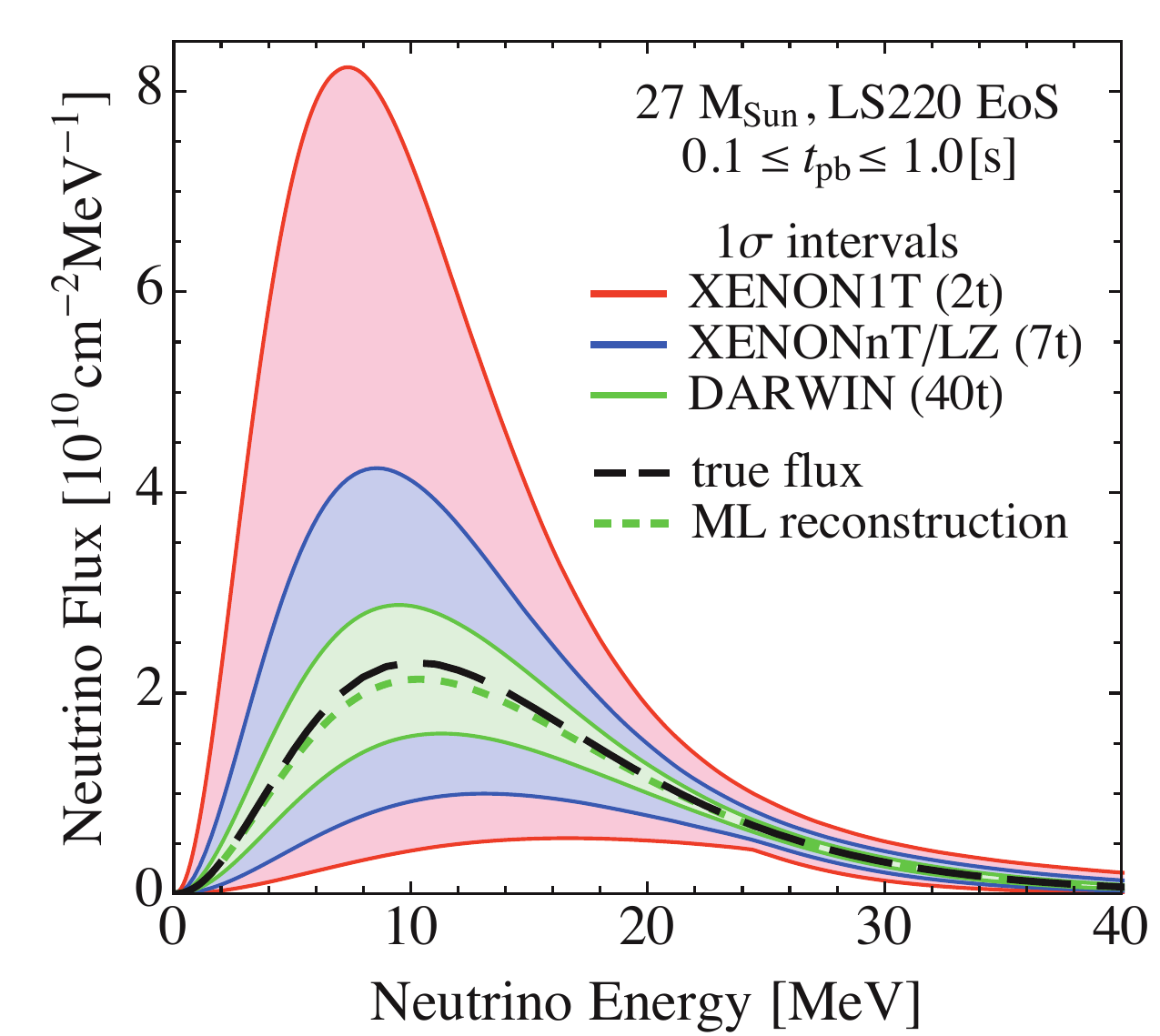}
\caption{The left panel shows the reconstructed average neutrino energy $\langle E_T \rangle$ and amplitude parameter $A_T$ (green dot-circle) compared to the true value for the $27~\!\Msun$ LS220 EoS progenitor at 10~kpc from Earth integrated between 0.1~\!s and 1~\!s (black triangle). Also shown are the $1\sigma$ contours from 2-, 7- and 40-tonne mock experiments following our maximum likelihood~(ML) analysis. The right panel shows the reconstructed neutrino flux as a function of the neutrino energy. The dashed green line represents the differential flux obtained with the best fit ML estimators and is compared with the true flux, shown by the dashed black line. Also shown are the $1\sigma$ intervals from our 2-, 7- and 40-tonne mock experiments.}
\label{fig:S2Reconstruction2}
\end{figure*}

Up to this point, we have extracted information using the event rate integrated over the S2 range for a given threshold $\mathrm{S2}_{\rm{th}}$. However, xenon detectors are also able to accurately measure the S2 value of an individual event. For the first time, we investigate the physics that can be extracted from this spectral information. In particular, in this subsection, we demonstrate that xenon detectors can reconstruct the all-flavour neutrino differential flux as a function of the energy and, in the next subsection, that the total SN energy emitted into all flavours of neutrinos can be reconstructed.

The neutrino differential flux, $f_{\nu_\beta}^0(E_{\nu},t_{\rm pb})$, defined in Eq.~\eqref{eq:nuflux}, enters the calculation for the rate of events in a xenon detector in Eq.~\eqref{eq:d2RdEdt}. From this equation, we see that it is the time-integrated differential flux summed over all neutrino flavours that determines the number of SN neutrino scattering events in a detector. This flux is typically dominated by the $\nu_x$ flavours, simply because it contributes four flavours ($\bar{\nu}_{\mu}$, $\nu_{\mu}$, $\bar{\nu}_{\tau}$ and $\nu_{\tau}$) out of the six flavours that comprise the total flux. We now show that this flux may be reconstructed. It depends on the SN progenitor and the time window of the observation and we would like to reconstruct it by making as few assumptions as possible about the initial SN progenitor. We thus make the following ansatz for the time-integrated differential flux summed over all neutrino flavours:
\begin{equation}
\label{eq:ATETdef}
\begin{split}
&\sum_{\nu_{\beta}} \int_{t_1}^{t_2} dt_{\rm{pb}}\,f_{\nu_\beta}^0(E_{\nu},t_{\rm pb}) \\
&\qquad\equiv A_T\, \xi_T\left(\frac{E_{\nu}}{\langle E_T \rangle} \right)^{\alpha_T} \exp{\left(\frac{-(1+\alpha_T)E_{\nu}}{\langle E_T \rangle} \right)}\;.
\end{split}
\end{equation}
With this ansatz, we assume that the time-integrated differential flux can be parametrized with three free parameters: an amplitude~$A_T$, an average energy~$\langle E_T \rangle$, and a shape parameter~$\alpha_T$. Here,~$ \xi_T$ is a normalization parameter defined such that
\begin{equation}
\int d E_{\nu}\,\xi_T\left(\frac{E_{\nu}}{\langle E_T \rangle} \right)^{\alpha_T} \exp{\left(\frac{-(1+\alpha_T)E_{\nu}}{\langle E_T \rangle} \right)}=1\;.
\end{equation}
With these definitions, $A_T$ has units of area$^{-1}$, $\alpha_T$ is dimensionless, and $\langle E_T \rangle$ has units of energy. As suggested by our notation, $\langle E_T \rangle$ is the average neutrino energy of the time-integrated flux summed over all flavours.

In practice, however, the shape parameter $\alpha_T$ is difficult to constrain experimentally since it is degenerate with $\langle E_T \rangle$, which also controls the shape of the observed recoil spectrum i.e.\ $d R/d\mathrm{S2}$. We therefore make a simplifying assumption motivated by the observation from SN simulations that the differential neutrino flux can be approximated by a Fermi-Dirac distribution with zero chemical potential~\cite{1989A&A...224...49J,Keil:2002in,Tamborra:2012ac}. For this distribution, the relation $\langle E^2 \rangle/\langle E \rangle^2\simeq1.3$ holds~\cite{Keil:2002in}, and from Eq.~\eqref{eq:earelate}, implies $\alpha_T\simeq2.3$. This value of~$\alpha_T$ is fixed in the subsequent results. This should be a reasonably good approximation everywhere, except during the very short neutronization burst phase $(t_{\rm{pb}}\lesssim10~\!\mathrm{ms})$, where the spectrum is significantly pinched with respect to a Fermi-Dirac distribution~\cite{Tamborra:2012ac}.

We first consider the reconstruction of the time-integrated differential flux summed over all neutrino flavours in the time window $[t_1,t_2]=[0.1,1]$~\!s (cf.\ Eq.~\eqref{eq:ATETdef}) for the $27~\!\Msun$ LS220 EoS progenitor at 10~kpc from Earth. This window corresponds to the accretion phase of the~SN burst. To reconstruct $A_T$ and $\langle E_T \rangle$, we perform a maximum likelihood (ML) analysis and maximize the extended likelihood function
\begin{equation}
\ln \mathcal{L}(A_T,\langle E_T \rangle)= N \ln \mu -\mu +\sum_{i=1}^{N}\ln f \left( \mathrm{S2}_i;\langle E_T \rangle \right) \;,
\end{equation}
where $\mu=\mu(A_T,\langle E_T \rangle)$ is the mean number of expected events, $N$ is the observed number of events, and $f(\mathrm{S2}_i;\langle E_T\rangle)$ is the probability density function evaluated at the~S2 value of the $i$th event.

Figure~\ref{fig:S2Reconstruction2} shows the ML estimators for~$A_T$ and~$\langle E_T \rangle$ for XENON1T, XENONnT/LZ and DARWIN mock experiments, where, by construction, each mock experiment has the same ML estimators for~$A_T$ and~$\langle E_T \rangle$. The~$N$ observed events are randomly drawn from the~$d R/d\mathrm{S2}$ spectrum of the $27~\!\Msun$ LS220 EoS progenitor at 10~kpc, integrated from 0.1~\!s to 1~\!s. We consider all events in the~S2 range from $\mathrm{S2}_{\rm{th}}=60$~PE to $\mathrm{S2}_{\rm{max}}=2000$~PE. For this progenitor and this time window, the mean number of expected events is 7.0~events/tonne, so $N$ is drawn from Poisson distributions with means of 14, 49 and 280 events for XENON1T, XENONnT/LZ and DARWIN, respectively.

The left panel of Fig.~\ref{fig:S2Reconstruction2} shows the best fit ML estimators (green dot-circle) together with the $1\sigma$ contours for XENON1T, XENONnT/LZ and DARWIN. These contours are obtained from the ML by $\ln \mathcal{L}=\ln \mathcal{L}_{\rm{max}}-2.3/2$. The black triangle shows the values of these parameters from our input SN progenitor. The DARWIN reconstruction of the parameters is excellent, while the XENON1T reconstruction has a significantly larger uncertainty. Besides reconstructing $A_T$ and $\langle E_T \rangle$, an estimation of the expected $\nu_x$ average energy should be also possible analogously to what was proposed for neutrino--proton elastic scattering~\cite{Beacom:2002hs,Dasgupta:2011wg}.

The dashed green line in the right panel of Fig.~\ref{fig:S2Reconstruction2} shows the differential flux obtained with the best-fit ML estimators substituted into Eq.~\eqref{eq:ATETdef}. This can be compared with the true flux from the $27~\!\Msun$ LS220 EoS progenitor, which is shown by the dashed black line. The ML reconstruction is in very good agreement with the true flux. Also shown are the $1\sigma$ intervals. At each value of the neutrino energy, the intervals were obtained by propagating all points in the $1\sigma$ regions for~$A_T$ and~$\langle E_T \rangle$ through Eq.~\eqref{eq:ATETdef} and selecting the maximum and minimum values of the neutrino flux. The right panel demonstrates that a DARWIN-sized experiment will be capable of accurately reconstructing the neutrino flux. The errors from XENON1T however are substantial, owing to the fact that with XENON1T one would observe only~14 events during this time window, compared to~280 with DARWIN.

\subsection{Total energy emitted into neutrinos}

Finally, we show that it is possible to reconstruct the total energy emitted by neutrinos, which is simply the luminosity integrated over the duration of the SN burst (here taken as the first 7~\!s) and summed over all neutrino flavours. This is related to the free parameters in our ansatz by (see Eqs.~\eqref{eq:nuflux} and \eqref{eq:ATETdef})
\begin{align}
E_{\rm{tot}}=\sum_{\nu_{\beta}} \int_{0~\!\!s}^{7~\!\!s} d t_{\rm{pb}}\, L_{\nu_{\beta}}(t_{\rm{pb}})
 =4 \pi d^2 A_T \langle E_T \rangle \label{eq:Etot}\;.
\end{align}
This relation follows from noting that 
\begin{eqnarray}
A_T \langle E_T \rangle=\int d E_{\nu} \,E_{\nu} \sum_{\nu_{\beta}} \int d t_{\rm{pb}} f_{\nu_\beta}^0(E_{\nu},t_{\rm pb})\ ,
\end{eqnarray}
and using Eq.~\eqref{eq:nuflux} to express $f_{\nu_\beta}^0(E_{\nu},t_{\rm pb})$ in terms of~$L_{\nu_{\beta}}(t_{\rm{pb}})$.

\begin{figure}[!tbp]
\centering
\includegraphics[width=0.9\columnwidth]{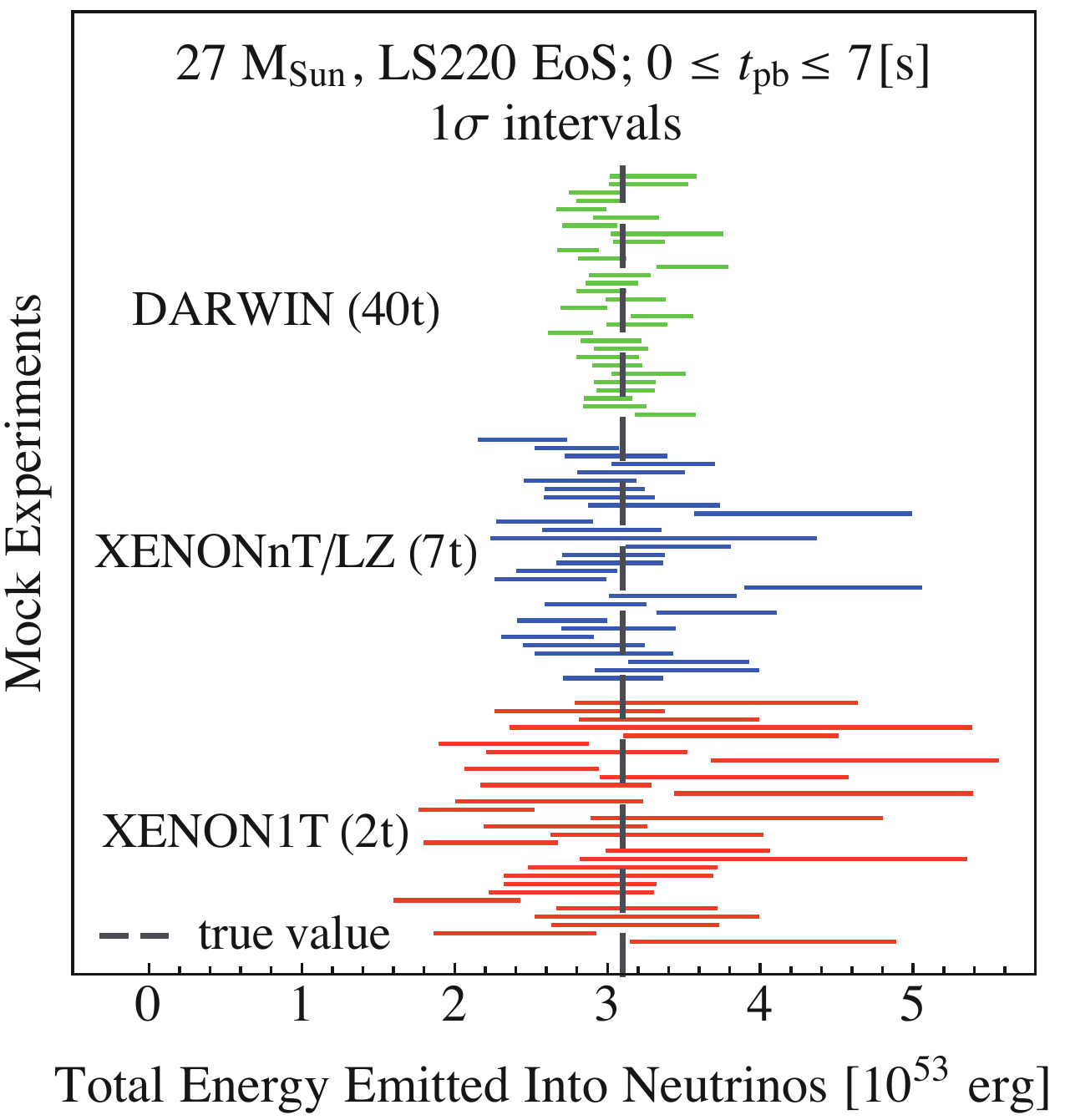}
\caption{The reconstructed $1\sigma$ band of the total energy emitted into neutrinos in~30 mock experiments for each of XENON1T (red), XENONnT/LZ (blue) and DARWIN (green). The true value for the $27~\!\Msun$ LS220 EoS progenitor integrated over the total time of the SN burst (taken as the first 7~\!s) is shown by the dashed vertical line.}
\label{fig:EReconstruction}
\end{figure}

Figure~\ref{fig:EReconstruction} shows the~$1\sigma$ range of the reconstructed total energy emitted into neutrinos in~30 mock experiments for each of XENON1T, XENONnT/LZ and DARWIN. As in the previous subsection, we use the ML method to find the estimators of the parameters~$A_T$ and~$\langle E_T \rangle$ for the signal integrated over the first 7~\!s of a $27~\!\Msun$ LS220 EoS progenitor at 10~kpc from Earth. Then, we calculate~$E_{\rm{tot}}$ from Eq.~\eqref{eq:Etot} and the $1\sigma$ range using the propagation of errors as described in~\cite{Agashe:2014kda}. We do not include any uncertainty on the distance~$d$ in our reconstruction.

\begin{table}[t!]
\caption{The typical precision of the reconstructed total energy emitted in neutrinos over the first 7~\!s, assuming our four SN progenitors situated 10~kpc from Earth, for different current and upcoming detectors.\label{table:Ttot}}
\centering
 \begin{ruledtabular}
\begin{tabular}{@{}c c c c c @{}} 
& \multicolumn{2}{c}{$27\,\Msun$} & \multicolumn{2}{c}{$11\,\Msun$} \\ \cline{2-3}\cline{4-5}
 & LS220 & Shen & LS220 & Shen \\ \colrule
XENON1T (2t) & $20\%$ & $25\%$ & $30\%$ & $36\%$ \\
XENONnT/LZ (7t) & $11\%$ & $13\%$ & $16\%$ & $20\%$ \\
DARWIN (40t) & $5\%$ & $6\%$ & $7\%$ & $9\%$ \\
\end{tabular} 
 \end{ruledtabular}
\end{table}

The dashed vertical line shows the total energy from the SN simulation of the $27~\!\Msun$ LS220 EoS progenitor. As we would expect, each mock experiment results in a different mean and variance with the property that the~$1\sigma$ region covers the true value in approximately 68\% of the mock experiments. The typical uncertainty on the reconstructed energy for all four SN progenitors is given in Table~\ref{table:Ttot}. This number is the average of the ratio of the $1\sigma$ error over the mean for~250 mock experiments. The uncertainty is smallest for the $27~\!\Msun$ LS220 EoS progenitor since it results in the highest number of events, and is largest for the $11~\!\Msun$ Shen EoS progenitor, which gives the lowest number of events. Unsurprisingly, the errors decrease substantially as the target mass is increased from~2~tonnes in XENON1T to~40~tonnes in DARWIN. However, even XENON1T can give a reasonably precise estimate of the total energy emitted into neutrinos for a~SN at 10~\!kpc.

\subsection{Comparison with dedicated neutrino detectors}

We briefly compare the expected number of events of the forthcoming xenon detectors with existing or future neutrino detectors (see also Table~1 of~\cite{Mirizzi:2015eza} for an overview). For a SN burst at 10~\!kpc, XENON1T and XENONnT/LZ will measure approximately~35 and~120 events in total. This is similar to the projected number of events from neutrino-proton elastic scattering at scintillator detectors~\cite{Dasgupta:2011wg}. However, it is one order of magnitude less than DUNE, which is expected to measure approximately $\mathcal{O}(10^3)$ events mostly in the~$\nu_e$ channel with a 40-tonne liquid argon detector (see Fig.~5.5 of~\cite{Acciarri:2015uup}). In the $\bar{\nu}_e$ channel, larger event rates are expected from IceCube, which should see approximately $10^6$ events (see Fig.~52 of~\cite{Abe:2011ts}), Hyper-Kamiokande, which is expected to measure approximately 10$^5$ events (see Fig.~54 of~\cite{Abe:2011ts}), and JUNO, which should detect about 6000 events (see Figs.~4-7 of~\cite{An:2015jdp}). The proposed DARWIN direct detection dark matter detector, with 40~tonnes of liquid xenon, will measure approximately 700~events for all six flavours, and is thus starting to be competitive in terms of the event rate with these dedicated neutrino detectors. Of course, the quoted numbers depend on different assumptions for the adopted SN model and therefore have to be viewed only as rough estimations of the expected number of events. 

For what concerns the reconstruction of the SN neutrino light curve, IceCube, Hyper-Kamiokande and JUNO will all measure many more events [$\mathcal{O}(10^4-10^5)$~events/s] compared to DARWIN, which will see approximately 330 events during the first second and 370 events in the remainder of the SN burst. Even though the number of events is smaller for DARWIN, it is important to remember that it is sensitive to all six neutrino flavours, while the existing and planned neutrino detectors are primarily sensitive to a single flavour. Moreover, as discussed in previous sections, dual-phase xenon detectors will provide us with all-flavour information about the energetics of the explosion that should be compared with the flavour-dependent energy spectra possibly reconstructed, e.g., in JUNO or  Hyper-Kamiokande with high resolution. In this sense, in the event of a SN burst, a global analysis of the burst with events from all experiments will benefit from the inclusion of DARWIN data to better constrain the properties of neutrinos and the SN progenitor. 

\section{Experimental factors}\label{sec:discussion}

In this section, we discuss the uncertainties related to~$Q_y$, the detector performance during calibration periods, and the eventual pile-up of events that could prevent a clean identification of individual~S2 signals if a SN burst occurred too close to the Earth.

\subsection{Signal uncertainty from $Q_y$}\label{subsec:signalQy}

\begin{figure}[!t]
\centering
\includegraphics[width=0.97\columnwidth]{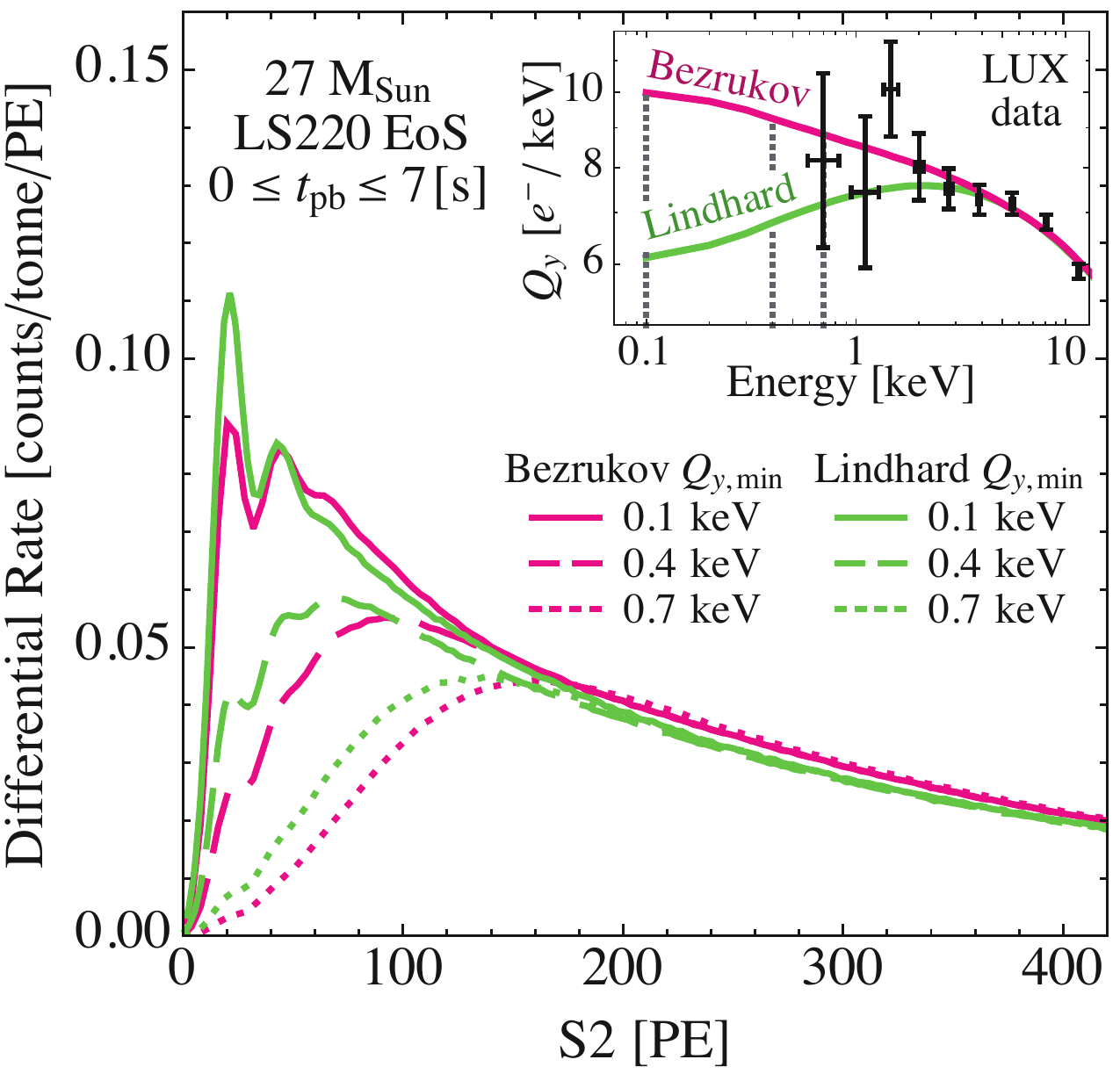}
\caption{Variations in the $dR/d\Stwo$ differential spectrum under different assumptions for $Q_y$ for the $27\,\Msun$ LS220 EoS progenitor at~10~\!kpc and integrated over the first 7~\!s. The quantity $Q_{y,\rm{min}}$ is the energy below which $Q_y=0$ and the solid, dashed and dotted lines correspond to $Q_{y,\rm{min}}$ values of 0.1~keV, 0.4~keV and 0.7 keV. The inset shows the Lindhard and Bezrukov $Q_y$ models together with the LUX measurements. The differences in~$dR/d\Stwo$ between the Lindhard and Bezrukov models are reasonably small compared to the larger differences from varying $Q_{y,\rm{min}}$.}
\label{fig:QyLyvary4}
\end{figure}

An accurate prediction of the~S2 signal relies on knowledge of~$Q_y$, the charge yield in liquid xenon, at sub-keV energies. The LUX collaboration has provided the most accurate measurement of~$Q_y$ and has measured it down to a nuclear recoil energy of 0.7~\!keV. The LUX data points from~\cite{Akerib:2015rjg} are reproduced in the inset of Fig.~\ref{fig:QyLyvary4}. The Lindhard model~\cite{Lindhard:1961zz} provides a good fit to the measurements and, following the LUX collaboration, is the default parametrization that we have assumed for energies above 0.7~\!keV. For energies below this value, we have conservatively assumed that~$Q_y=0$. In this subsection, we investigate the uncertainty that this assumption introduces on the number of events observed with a dual-phase xenon experiment.

In order to extrapolate $Q_y$ to the lowest energies, we use either the Lindhard model or the alternative model by Bezrukov~et.~al.~\cite{Bezrukov:2010qa}. As can be seen in the inset of Fig.~\ref{fig:QyLyvary4}, both the Lindhard and Bezrukov models fit the data well. We then set $Q_y$ to zero below various values~$Q_{y,\rm{min}}$. The case of $Q_{y,\rm{min}}=0.7$~keV can be seen as the minimum predicted signal. At approximately 0.1~\!keV or below, the Lindhard model is expected to break down due to atomic effects~\cite{Sorensen:2014sla}. We thus also test $Q_{y,\rm{min}}=0.1$~keV and 0.4~\!keV, as an intermediate example.

The main panel of Fig.~\ref{fig:QyLyvary4} shows different realizations of the $dR/d\Stwo$ spectrum for the $27~\!\Msun$ LS220 EoS progenitor at 10~kpc integrated over the first 7~\!s. The spectra are obtained for the Lindhard and Bezrukov models of $Q_y$ with three values of~$Q_{y,\rm{min}}$. As expected, the lower the assumed $Q_{y,\rm{min}}$ value, the greater the number of signal electrons that can be detected from low-energy nuclear recoils. The differences between the Lindhard and Bezrukov models for $Q_y$ are much smaller than the differences from varying $Q_{y,\rm{min}}$. For a given $Q_{y,\rm{min}}$, the Lindhard model gives a signal that is shifted to lower S2 values, which follows from the lower energy yield for given recoil energy, as seen in the inset of Fig.~\ref{fig:QyLyvary4}.

\begin{table}[!t]
\caption{The expected number of neutrino events per tonne for various~S2 thresholds under different assumptions for~$Q_y$. We compare the Lindhard and Bezrukov models and assume that $Q_y=0$ for energies below~$Q_{y,\rm{min}}$. The results are for the $27\,\Msun$ LS220 EoS progenitor at~10~\!kpc and integrated over the first 7~\!s. Similar results hold for other progenitor models. The signal uncertainty in each row is $(\mathrm{S2}_{\rm{max}}-\mathrm{S2}_{\rm{min}})/(\mathrm{S2}_{\rm{max}}+\mathrm{S2}_{\rm{min}})$. The Lindhard model with $Q_{y,\rm{min}}=0.7$~keV gives the smallest number of events per tonne and is the benchmark assumption that we have made in this paper. 
\label{table:QyLyvary2}}
\centering
 \begin{ruledtabular}
\begin{tabular}{@{}c c c c c c @{}} 
& \multicolumn{5}{c}{$27\,\Msun$ LS220 EoS} \\ \cline{2-6}
& \multicolumn{2}{c}{Lindhard $Q_{y,\rm{min}}$} & \multicolumn{2}{c}{Bezrukov $Q_{y,\rm{min}}$} & Signal \\ \cline{2-3}\cline{4-5}
$\mathrm{S2}_{\mathrm{th}}$ [PE] & 0.1~keV & 0.7~keV &0.1~keV & 0.7~keV & uncertainty\\ \colrule
20 & 22.9 & 18.4 & 23.8 & 18.5 & 13\% \\
40 & 21.0 & 18.1 & 22.2 & 18.3 &10\% \\
60 $(\star)$ & 19.4 & 17.6 & 20.6 & 17.9 &8\% \\
80 & 18.1 & 17.0 & 19.2 & 17.5 &6\%\\
100 & 16.9 & 16.3 & 17.9 & 16.9 &5\% \\
\end{tabular} 
 \end{ruledtabular}
\end{table}

Table~\ref{table:QyLyvary2} shows the total number of expected events per tonne of xenon target in the various $Q_y$ scenarios considered. The number of events corresponds to the $27\,\Msun$ LS220 EoS SN progenitor at 10~kpc and the neutrino signal is integrated over 7~\!s. The final column in Table~\ref{table:QyLyvary2} gives an estimate of the signal uncertainty for each~S2 threshold, calculated in each row as $(\mathrm{S2}_{\rm{max}}-\mathrm{S2}_{\rm{min}})/(\mathrm{S2}_{\rm{max}}+\mathrm{S2}_{\rm{min}})$. In all cases, the minimum number of events per tonne is found for the Lindhard model with $Q_{y,\rm{min}}=0.7$~keV, which is the benchmark assumption that we have made in all calculations reported in this paper. The highest number of events is found for the Bezrukov model with $Q_{y,\rm{min}}=0.1$~keV. For the $27\,\Msun$ LS220 EoS progenitor and our benchmark value $\mathrm{S2}_{\rm{th}}=60$~PE, the uncertainty from the $Q_y$ parametrization is around 8\%. The signal uncertainties with this~S2 threshold for the other SN progenitors are similar, with an uncertainty of $9\%$ ($9\%$, $10\%$) for the $27\,\Msun$ Shen EoS ($11\,\Msun$ LS220 EoS, $11\,\Msun$ Shen EoS) progenitor. 

The neutrino flux amplitude and mean energy reconstruction analyses in section~\ref{sec:Erec} may be more adversely affected by the uncertainty in the charge yield $Q_y$, since they also take into account the shape of the recoil spectrum. The $Q_y$ modeling uncertainty could be straightforwardly incorporated into a ML analysis (as the $L_y$, $Q_y$ and Milky Way halo uncertainties are routinely incorporated into dark matter studies). Here, we simply test a higher~S2 threshold, $\mathrm{S2}_{\rm{th}}=120$~PE, to reduce the $Q_y$ modeling uncertainty by repeating our analysis that led to Fig.~\ref{fig:S2Reconstruction2}. In this case, we find similar results as in Fig.~\ref{fig:S2Reconstruction2}. The number of events is reduced from 7.0~events/tonne to 6.3~events/tonne, which leads to an increase in the $1\sigma$ regions of the mean energy and amplitude by only 13\% and 20\% respectively. Thus, the quantitative conclusions drawn from this analysis are only slightly affected by the present uncertainty in $Q_y$. Clearly, it would be most desirable to further reduce the $Q_y$ uncertainty by having other low-energy measurements of this quantity.

\subsection{Sources of increased background rates}

The low background rates discussed in section~\ref{sec:S2only} are applicable when the detector is in dark matter search mode. However, in contrast to dedicated SN neutrino detectors, direct detection dark matter detectors can in some cases spend half of their time taking calibration data~\cite{Aprile:2015ibr}. Various calibration sources are utilized, from external Compton or neutron calibrations to radioactive isotopes that are dissolved directly in the liquid target. The particular background rate in the S2-only channel discussed previously can vary significantly during calibration and may depend on the particular calibration source employed. However, even with an event rate during calibration two orders of magnitude above the rates during a dark matter search, the background is still smaller than the expected signal rates from a Galactic SN.

Another potential source of increased background to SN signals comes from photoionization on impurities in the liquid xenon. During the commissioning of a detector, the purity may be low, and thus the background rate may be increased. Furthermore, the diminished electron survival probability from their drift would in effect raise the S2-based energy threshold, possibly rendering the detector blind to SN events. Since such initial commissioning times are supposed to be short, we do not discuss them further here.

\subsection{Sensitivity limitation from event pile-up}

In a xenon TPC, a single SN neutrino scattering event produces a number of ionization electrons that are drifted to the gas phase, where the~S2 signal is produced from proportional scintillation. At a drift velocity of order 2~mm/$\mu$s~\cite{Yoo:2015yza}, a~1~\!m high TPC is expected to smear the arrival times of the electrons by about 250~$\mu$s. This aspect limits the timing resolution of this detection channel.

To get a better estimate of the maximum number of SN neutrino events ($N_{\rm{pile-up}}$) before pile-up becomes an issue, we perform a Monte Carlo simulation of the events in the TPC. Once the electrons are extracted from the liquid, the observed~S2 signal is a pulse with a width of~$\mathcal{O}(1)~\!\mu$s~\cite{Aprile:2012vw}. To resolve individual events without the need to use information from the PMT hit pattern, one~S2 pulse should not overlap another S2~pulse. This will limit the sensitivity of the detector once pile-up becomes significant. Motivated by Fig.~8 in~\cite{Aprile:2012vw}, we define events to be well separated if the spacing from the start of one~S2 pulse to the start of the next pulse is more than $10~\!\mu$s. We randomly distribute events in time according to the differential time distribution $dR/d \tpb$. We test both the 27~\!$\Msun$ LS220 EoS and 11~\!$\Msun$ Shen EoS progenitors to get an idea about the impact of these models on our conclusions. As the event rate is highest at the start of the SN burst (cf.~Figs.~\ref{fig:LEpanels} and~\ref{fig:DiffRecSpectra}), we focus on the first second after the explosion. We then distribute the events uniformly throughout the TPC and take into account the time delay as the ionization electrons drift from the interaction site to the liquid-gas interface, assuming the XENON100 drift velocity~$v_{\mathrm{d}}=1.7~\!\mathrm{mm}/\mu\mathrm{s}$~\cite{Aprile:2012vw}. In each mock (and real) experiment, the vertex sites, the number, and the time distribution of the events vary. We thus use a statistical procedure and define $N_{\rm{pile-up}}$ to be the number of events at which~90\% of mock experiments observe at least~5\% of events with a spacing of less than~$10~\!\mu$s.

We have performed our calculation for three TPC sizes, 967~\!mm, 1450~\!mm and 2600~\!mm, corresponding to the expected sizes of the XENON1T~\cite{Aprile:2015uzo}, XENONnT/LZ~\cite{Akerib:2015cja} and DARWIN TPCs~\cite{Schumann:2015cpa}. We find that~$N_{\rm{pile-up}}$ is approximately independent of these three TPC sizes, varying by less than 0.2\%. For the 27~\!$\Msun$ LS220 EoS and 11~\!$\Msun$ Shen EoS progenitors, $N_{\rm{pile-up}}=4810$~events and $N_{\rm{pile-up}}=4780$~events respectively, consistent with our simple estimate of approximately 250~$\mu$s for the timing resolution.

The maximum number of SN neutrino events, $N_{\rm{pile-up}}$, can be converted into a minimum progenitor distance from Earth so that pile-up is not an issue. Given that 8.3~events per tonne and 4.1~events per tonne are expected during the first second for the 27~\!$\Msun$ LS220 EoS and 11~\!$\Msun$ Shen EoS progenitors at 10~\!kpc, we find minimum distances of $\{0.6,1.1,2.6\}$~\!kpc and $\{0.4,0.8,1.8\}$~\!kpc for \{XENON1T, XENONnT/LZ, DARWIN\} for the 27~\!$\Msun$ LS220 EoS and 11~\!$\Msun$ Shen EoS progenitors, respectively. A SN explosion that is much closer than these distances will still be detected by a xenon detector, but precision studies of the SN neutrino light curve or neutrino flux parameters will become degraded as it becomes difficult to distinguish between individual events.

\section{Conclusions} \label{sec:conclusions}

With the launch of XENON1T with 2~tonnes of xenon target, and given the plans for larger experiments employing the same technology such as XENONnT and LZ with 7~tonnes and DARWIN with 40~tonnes, we here revisited the possibility of detecting a Galactic supernova (SN) through coherent elastic neutrino-nucleus scattering~(\CENS) with such dual-phase xenon direct detection dark matter experiments. In order to gauge the astrophysical variability of the expected signal, we studied the neutrino signal from four hydrodynamical SN simulations, differing in the progenitor mass and nuclear equation of state. For the first time, we have performed a realistic detector simulation of SN neutrino scattering, expressing the scattering rates in terms of the observed signals S1 (prompt scintillation) and S2 (proportional scintillation).

We have shown that focusing on the S2 channel maximizes the number of events that can be detected, thanks to the lower energy threshold. We have discussed appropriate values of the S2 threshold and proved that the background rate is negligible compared to the expected signal. Hence, high-significance discoveries can be expected. As a concrete example, we have shown that for a~$27~\!\Msun$ SN progenitor, the XENON1T experiment will be able to detect a SN burst with more than $5\sigma$ significance up to 25~\!kpc from Earth. Furthermore, the XENONnT and LZ experiments will extend this sensitivity beyond the edge of the Milky Way, and the DARWIN experiment will be sensitive to SN bursts in the Large and Small Magellanic Clouds. Due to the low background rate, these experiments should even be able to actively contribute to the Supernova Early Warning System (SNEWS)~\cite{Antonioli:2004zb,Scholberg:2008fa}.

For a~SN burst at 10~\!kpc, features of the neutrino signal such as the neutronization burst, accretion phase, and Kelvin-Helmholtz cooling phase will be distinguishable with the DARWIN experiment. In addition, with DARWIN it will be possible to make a high-precision reconstruction of the average neutrino energy and differential neutrino flux. Since \CENS\ is insensitive to the neutrino flavour, the signal in dual-phase xenon detectors is unaffected by uncertainties from neutrino oscillation physics. A high-precision measurement of \CENS\ from~SN neutrinos will therefore offer a unique way of testing our understanding of the~SN explosion mechanism. The sensitivity to all neutrino flavours also means that it is straightforward to reconstruct the total energy emitted into neutrinos. We have shown that even XENON1T could provide a reasonably good reconstruction of this energy.

It has already been discussed that a large multi-tonne xenon detector such as DARWIN would be able to measure solar neutrino physics~\cite{Baudis:2013qla} and exploit novel dark matter signals~\cite{Baudis:2013bba,Schumann:2015cpa,McCabe:2015eia}. Here, we have illustrated that DARWIN will also be able to reconstruct many properties of~SN progenitors and their neutrinos with high precision. Large dual-phase xenon detectors are expected to be less expensive and more compact than future-generation dedicated neutrino telescopes, encouraging the construction of liquid xenon experiments as~SN neutrino detectors. Specifically, DARWIN will allow for all-flavour event statistics that are competitive with next-generation liquid argon or scintillation neutrino detectors~\cite{Acciarri:2015uup,An:2015jdp}, which are sensitive to only some of the neutrino flavours. At the same time, being flavour blind, dual-phase xenon detectors will provide complementary information on the SN neutrino signal that is not obtainable with existing or planned neutrino telescopes.

\acknowledgments
We thank John Beacom and Sebastian Liem for discussions as well as Alec Habig and Georg Raffelt for comments on the manuscript. RFL and SR are supported by Grant No.~\#PHYS-1412965 from the National Science Foundation (NSF). CM acknowledges support from the Foundation for Fundamental Research on Matter (FOM), which is part of the Netherlands Organisation for Scientific Research (NWO). MS thanks the Istituto Nazionale di Fisica Nucleare (INFN). IT~acknowledges support from the Knud H\o jgaard Foundation and from the Danish National Research Foundation (DNRF91).


\begin{thebibliography}{111}
\expandafter\ifx\csname natexlab\endcsname\relax\def\natexlab#1{#1}\fi
\expandafter\ifx\csname bibnamefont\endcsname\relax
  \def\bibnamefont#1{#1}\fi
\expandafter\ifx\csname bibfnamefont\endcsname\relax
  \def\bibfnamefont#1{#1}\fi
\expandafter\ifx\csname citenamefont\endcsname\relax
  \def\citenamefont#1{#1}\fi
\expandafter\ifx\csname url\endcsname\relax
  \def\url#1{\texttt{#1}}\fi
\expandafter\ifx\csname urlprefix\endcsname\relax\def\urlprefix{URL }\fi
\providecommand{\bibinfo}[2]{#2}
\providecommand{\eprint}[2][]{\url{#2}}

\bibitem[{\citenamefont{Janka et~al.}(2016)\citenamefont{Janka, Melson, and
  Summa}}]{Janka:2016fox}
\bibinfo{author}{\bibfnamefont{H.~T.} \bibnamefont{Janka}},
  \bibinfo{author}{\bibfnamefont{T.}~\bibnamefont{Melson}}, \bibnamefont{and}
  \bibinfo{author}{\bibfnamefont{A.}~\bibnamefont{Summa}}
  (\bibinfo{year}{2016}), \eprint{1602.05576}.

\bibitem[{\citenamefont{Janka}(2012)}]{Janka:2012wk}
\bibinfo{author}{\bibfnamefont{H.-T.} \bibnamefont{Janka}},
  \bibinfo{journal}{Ann. Rev. Nucl. Part. Sci.} \textbf{\bibinfo{volume}{62}},
  \bibinfo{pages}{407} (\bibinfo{year}{2012}), \eprint{1206.2503}.

\bibitem[{\citenamefont{Mirizzi et~al.}(2016)\citenamefont{Mirizzi, Tamborra,
  Janka, Saviano, Scholberg, Bollig, H{\"u}depohl, and
  Chakraborty}}]{Mirizzi:2015eza}
\bibinfo{author}{\bibfnamefont{A.}~\bibnamefont{Mirizzi}},
  \bibinfo{author}{\bibfnamefont{I.}~\bibnamefont{Tamborra}},
  \bibinfo{author}{\bibfnamefont{H.-T.} \bibnamefont{Janka}},
  \bibinfo{author}{\bibfnamefont{N.}~\bibnamefont{Saviano}},
  \bibinfo{author}{\bibfnamefont{K.}~\bibnamefont{Scholberg}},
  \bibinfo{author}{\bibfnamefont{R.}~\bibnamefont{Bollig}},
  \bibinfo{author}{\bibfnamefont{L.}~\bibnamefont{H{\"u}depohl}},
  \bibnamefont{and}
  \bibinfo{author}{\bibfnamefont{S.}~\bibnamefont{Chakraborty}},
  \bibinfo{journal}{Riv. Nuovo Cim.} \textbf{\bibinfo{volume}{39}},
  \bibinfo{pages}{1} (\bibinfo{year}{2016}), \eprint{1508.00785}.

\bibitem[{\citenamefont{Chakraborty et~al.}(2016)\citenamefont{Chakraborty,
  Hansen, Izaguirre, and Raffelt}}]{Chakraborty:2016yeg}
\bibinfo{author}{\bibfnamefont{S.}~\bibnamefont{Chakraborty}},
  \bibinfo{author}{\bibfnamefont{R.}~\bibnamefont{Hansen}},
  \bibinfo{author}{\bibfnamefont{I.}~\bibnamefont{Izaguirre}},
  \bibnamefont{and} \bibinfo{author}{\bibfnamefont{G.}~\bibnamefont{Raffelt}},
  \bibinfo{journal}{Nucl. Phys.} \textbf{\bibinfo{volume}{B}}
  (\bibinfo{year}{2016}), \eprint{1602.02766}.

\bibitem[{\citenamefont{Esmaili et~al.}(2014)\citenamefont{Esmaili, Peres, and
  Serpico}}]{Esmaili:2014gya}
\bibinfo{author}{\bibfnamefont{A.}~\bibnamefont{Esmaili}},
  \bibinfo{author}{\bibfnamefont{O.~L.~G.} \bibnamefont{Peres}},
  \bibnamefont{and} \bibinfo{author}{\bibfnamefont{P.~D.}
  \bibnamefont{Serpico}}, \bibinfo{journal}{Phys. Rev.}
  \textbf{\bibinfo{volume}{D90}}, \bibinfo{pages}{033013}
  (\bibinfo{year}{2014}), \eprint{1402.1453}.

\bibitem[{\citenamefont{Wu et~al.}(2015)\citenamefont{Wu, Qian,
  Mart{\'{\i}}nez-Pinedo, Fischer, and Huther}}]{Wu:2014kaa}
\bibinfo{author}{\bibfnamefont{M.-R.} \bibnamefont{Wu}},
  \bibinfo{author}{\bibfnamefont{Y.-Z.} \bibnamefont{Qian}},
  \bibinfo{author}{\bibfnamefont{G.}~\bibnamefont{Mart{\'{\i}}nez-Pinedo}},
  \bibinfo{author}{\bibfnamefont{T.}~\bibnamefont{Fischer}}, \bibnamefont{and}
  \bibinfo{author}{\bibfnamefont{L.}~\bibnamefont{Huther}},
  \bibinfo{journal}{Phys. Rev.} \textbf{\bibinfo{volume}{D91}},
  \bibinfo{pages}{065016} (\bibinfo{year}{2015}), \eprint{1412.8587}.

\bibitem[{\citenamefont{Esteban-Pretel
  et~al.}(2007)\citenamefont{Esteban-Pretel, Tom{\`a}s, and
  Valle}}]{EstebanPretel:2007yu}
\bibinfo{author}{\bibfnamefont{A.}~\bibnamefont{Esteban-Pretel}},
  \bibinfo{author}{\bibfnamefont{R.}~\bibnamefont{Tom{\`a}s}},
  \bibnamefont{and} \bibinfo{author}{\bibfnamefont{J.~W.~F.}
  \bibnamefont{Valle}}, \bibinfo{journal}{Phys. Rev.}
  \textbf{\bibinfo{volume}{D76}}, \bibinfo{pages}{053001}
  (\bibinfo{year}{2007}), \eprint{0704.0032}.

\bibitem[{\citenamefont{Stapleford et~al.}(2016)\citenamefont{Stapleford,
  V{\"a}{\"a}n{\"a}nen, Kneller, McLaughlin, and Shapiro}}]{Stapleford:2016jgz}
\bibinfo{author}{\bibfnamefont{C.~J.} \bibnamefont{Stapleford}},
  \bibinfo{author}{\bibfnamefont{D.~J.} \bibnamefont{V{\"a}{\"a}n{\"a}nen}},
  \bibinfo{author}{\bibfnamefont{J.~P.} \bibnamefont{Kneller}},
  \bibinfo{author}{\bibfnamefont{G.~C.} \bibnamefont{McLaughlin}},
  \bibnamefont{and} \bibinfo{author}{\bibfnamefont{B.~T.}
  \bibnamefont{Shapiro}} (\bibinfo{year}{2016}), \eprint{1605.04903}.

\bibitem[{\citenamefont{Hidaka and Fuller}(2007)}]{Hidaka:2007se}
\bibinfo{author}{\bibfnamefont{J.}~\bibnamefont{Hidaka}} \bibnamefont{and}
  \bibinfo{author}{\bibfnamefont{G.~M.} \bibnamefont{Fuller}},
  \bibinfo{journal}{Phys. Rev.} \textbf{\bibinfo{volume}{D76}},
  \bibinfo{pages}{083516} (\bibinfo{year}{2007}), \eprint{0706.3886}.

\bibitem[{\citenamefont{Drukier and Stodolsky}(1984)}]{Drukier:1983gj}
\bibinfo{author}{\bibfnamefont{A.}~\bibnamefont{Drukier}} \bibnamefont{and}
  \bibinfo{author}{\bibfnamefont{L.}~\bibnamefont{Stodolsky}},
  \bibinfo{journal}{Phys. Rev.} \textbf{\bibinfo{volume}{D30}},
  \bibinfo{pages}{2295} (\bibinfo{year}{1984}), \bibinfo{note}{[395(1984)]}.
  
 \bibitem[{\citenamefont{Beacom et~al.}(2002)\citenamefont{Beacom, Farr, and
  Vogel}}]{Beacom:2002hs}
\bibinfo{author}{\bibfnamefont{J.~F.} \bibnamefont{Beacom}},
  \bibinfo{author}{\bibfnamefont{W.~M.} \bibnamefont{Farr}}, \bibnamefont{and}
  \bibinfo{author}{\bibfnamefont{P.}~\bibnamefont{Vogel}},
  \bibinfo{journal}{Phys. Rev.} \textbf{\bibinfo{volume}{D66}},
  \bibinfo{pages}{033001} (\bibinfo{year}{2002}), \eprint{hep-ph/0205220}. 

\bibitem[{\citenamefont{Horowitz et~al.}(2003)\citenamefont{Horowitz, Coakley,
  and McKinsey}}]{Horowitz:2003cz}
\bibinfo{author}{\bibfnamefont{C.~J.} \bibnamefont{Horowitz}},
  \bibinfo{author}{\bibfnamefont{K.~J.} \bibnamefont{Coakley}},
  \bibnamefont{and} \bibinfo{author}{\bibfnamefont{D.~N.}
  \bibnamefont{McKinsey}}, \bibinfo{journal}{Phys. Rev.}
  \textbf{\bibinfo{volume}{D68}}, \bibinfo{pages}{023005}
  (\bibinfo{year}{2003}), \eprint{astro-ph/0302071}.

\bibitem[{\citenamefont{Scholberg}(2012)}]{Scholberg:2012id}
\bibinfo{author}{\bibfnamefont{K.}~\bibnamefont{Scholberg}},
  \bibinfo{journal}{Ann. Rev. Nucl. Part. Sci.} \textbf{\bibinfo{volume}{62}},
  \bibinfo{pages}{81} (\bibinfo{year}{2012}), \eprint{1205.6003}.

\bibitem[{\citenamefont{Abbasi et~al.}(2011)}]{Abbasi:2011ss}
\bibinfo{author}{\bibfnamefont{R.}~\bibnamefont{Abbasi}} \bibnamefont{et~al.}
  (\bibinfo{collaboration}{IceCube}), \bibinfo{journal}{Astron. Astrophys.}
  \textbf{\bibinfo{volume}{535}}, \bibinfo{pages}{A109} (\bibinfo{year}{2011}),
  \bibinfo{note}{[Erratum: Astron. Astrophys.563,C1(2014)]},
  \eprint{1108.0171}.

\bibitem[{\citenamefont{Abe et~al.}(2011{\natexlab{a}})}]{Abe:2010hy}
\bibinfo{author}{\bibfnamefont{K.}~\bibnamefont{Abe}} \bibnamefont{et~al.}
  (\bibinfo{collaboration}{Super-Kamiokande}), \bibinfo{journal}{Phys. Rev.}
  \textbf{\bibinfo{volume}{D83}}, \bibinfo{pages}{052010}
  (\bibinfo{year}{2011}{\natexlab{a}}), \eprint{1010.0118}.

\bibitem[{\citenamefont{Abe et~al.}(2016{\natexlab{a}})}]{Abe:2016waf}
\bibinfo{author}{\bibfnamefont{K.}~\bibnamefont{Abe}} \bibnamefont{et~al.}
  (\bibinfo{collaboration}{Super-Kamiokande}), \bibinfo{journal}{Astropart.
  Phys.} \textbf{\bibinfo{volume}{81}}, \bibinfo{pages}{39}
  (\bibinfo{year}{2016}{\natexlab{a}}), \eprint{1601.04778}.

\bibitem[{\citenamefont{Aartsen et~al.}(2014)}]{Aartsen:2014njl}
\bibinfo{author}{\bibfnamefont{M.~G.} \bibnamefont{Aartsen}}
  \bibnamefont{et~al.} (\bibinfo{collaboration}{IceCube})
  (\bibinfo{year}{2014}), \eprint{1412.5106}.

\bibitem[{\citenamefont{Abe et~al.}(2011{\natexlab{b}})}]{Abe:2011ts}
\bibinfo{author}{\bibfnamefont{K.}~\bibnamefont{Abe}} \bibnamefont{et~al.}
  (\bibinfo{year}{2011}{\natexlab{b}}), \eprint{1109.3262}.

\bibitem[{\citenamefont{Agafonova et~al.}(2015)}]{Agafonova:2014leu}
\bibinfo{author}{\bibfnamefont{N.~Y.} \bibnamefont{Agafonova}}
  \bibnamefont{et~al.} (\bibinfo{collaboration}{LVD}),
  \bibinfo{journal}{Astrophys. J.} \textbf{\bibinfo{volume}{802}},
  \bibinfo{pages}{47} (\bibinfo{year}{2015}), \eprint{1411.1709}.

\bibitem[{\citenamefont{Cadonati et~al.}(2002)\citenamefont{Cadonati,
  Calaprice, and Chen}}]{Cadonati:2000kq}
\bibinfo{author}{\bibfnamefont{L.}~\bibnamefont{Cadonati}},
  \bibinfo{author}{\bibfnamefont{F.~P.} \bibnamefont{Calaprice}},
  \bibnamefont{and} \bibinfo{author}{\bibfnamefont{M.~C.} \bibnamefont{Chen}},
  \bibinfo{journal}{Astropart. Phys.} \textbf{\bibinfo{volume}{16}},
  \bibinfo{pages}{361} (\bibinfo{year}{2002}), \eprint{hep-ph/0012082}.

\bibitem[{\citenamefont{An et~al.}(2016)}]{An:2015jdp}
\bibinfo{author}{\bibfnamefont{F.}~\bibnamefont{An}} \bibnamefont{et~al.}
  (\bibinfo{collaboration}{JUNO}), \bibinfo{journal}{J. Phys.}
  \textbf{\bibinfo{volume}{G43}}, \bibinfo{pages}{030401}
  (\bibinfo{year}{2016}), \eprint{1507.05613}.

\bibitem[{\citenamefont{Kim}(2015)}]{Kim:2014rfa}
\bibinfo{author}{\bibfnamefont{S.-B.} \bibnamefont{Kim}},
  \bibinfo{journal}{Nucl. Part. Phys. Proc.}
  \textbf{\bibinfo{volume}{265-266}}, \bibinfo{pages}{93}
  (\bibinfo{year}{2015}), \eprint{1412.2199}.
  
  \bibitem[{\citenamefont{Barger et~al.}(2001)\citenamefont{Barger, Marfatia, and
  Wood}}]{Barger:2000hy}
\bibinfo{author}{\bibfnamefont{V.~D.} \bibnamefont{Barger}},
  \bibinfo{author}{\bibfnamefont{D.}~\bibnamefont{Marfatia}}, \bibnamefont{and}
  \bibinfo{author}{\bibfnamefont{B.~P.} \bibnamefont{Wood}},
  \bibinfo{journal}{Phys. Lett.} \textbf{\bibinfo{volume}{B498}},
  \bibinfo{pages}{53} (\bibinfo{year}{2001}), \eprint{hep-ph/0011251}.

\bibitem[{\citenamefont{Asakura et~al.}(2016)}]{Asakura:2015bga}
\bibinfo{author}{\bibfnamefont{K.}~\bibnamefont{Asakura}} \bibnamefont{et~al.}
  (\bibinfo{collaboration}{KamLAND}), \bibinfo{journal}{Astrophys. J.}
  \textbf{\bibinfo{volume}{818}}, \bibinfo{pages}{91} (\bibinfo{year}{2016}),
  \eprint{1506.01175}.

\bibitem[{\citenamefont{Acciarri et~al.}(2015)}]{Acciarri:2015uup}
\bibinfo{author}{\bibfnamefont{R.}~\bibnamefont{Acciarri}} \bibnamefont{et~al.}
  (\bibinfo{collaboration}{DUNE}) (\bibinfo{year}{2015}), \eprint{1512.06148}.

\bibitem[{\citenamefont{Laha et~al.}(2014)\citenamefont{Laha, Beacom, and
  Agarwalla}}]{Laha:2014yua}
\bibinfo{author}{\bibfnamefont{R.}~\bibnamefont{Laha}},
  \bibinfo{author}{\bibfnamefont{J.~F.} \bibnamefont{Beacom}},
  \bibnamefont{and} \bibinfo{author}{\bibfnamefont{S.~K.}
  \bibnamefont{Agarwalla}} (\bibinfo{year}{2014}), \eprint{1412.8425}.

\bibitem[{\citenamefont{Laha and Beacom}(2014)}]{Laha:2013hva}
\bibinfo{author}{\bibfnamefont{R.}~\bibnamefont{Laha}} \bibnamefont{and}
  \bibinfo{author}{\bibfnamefont{J.~F.} \bibnamefont{Beacom}},
  \bibinfo{journal}{Phys. Rev.} \textbf{\bibinfo{volume}{D89}},
  \bibinfo{pages}{063007} (\bibinfo{year}{2014}), \eprint{1311.6407}.

\bibitem[{\citenamefont{Lu et~al.}(2016)\citenamefont{Lu, Li, and
  Zhou}}]{Jia-Shu2016}
\bibinfo{author}{\bibfnamefont{J.-S.} \bibnamefont{Lu}},
  \bibinfo{author}{\bibfnamefont{Y.-F.} \bibnamefont{Li}}, \bibnamefont{and}
  \bibinfo{author}{\bibfnamefont{S.}~\bibnamefont{Zhou}}
   \bibinfo{journal}{Phys. Rev.} \textbf{\bibinfo{volume}{D94}},
  \bibinfo{pages}{023006} (\bibinfo{year}{2016}), \eprint{1605.07803}.
 
\bibitem[{\citenamefont{Kolbe and Langanke}(2001)}]{Kolbe:2000np}
\bibinfo{author}{\bibfnamefont{E.}~\bibnamefont{Kolbe}} \bibnamefont{and}
  \bibinfo{author}{\bibfnamefont{K.}~\bibnamefont{Langanke}},
  \bibinfo{journal}{Phys. Rev.} \textbf{\bibinfo{volume}{C63}},
  \bibinfo{pages}{025802} (\bibinfo{year}{2001}), \eprint{nucl-th/0003060}.

\bibitem[{\citenamefont{Volpe et~al.}(2002)\citenamefont{Volpe, Auerbach,
  Col{\`o}, and Van~Giai}}]{Volpe:2001gy}
\bibinfo{author}{\bibfnamefont{C.}~\bibnamefont{Volpe}},
  \bibinfo{author}{\bibfnamefont{N.}~\bibnamefont{Auerbach}},
  \bibinfo{author}{\bibfnamefont{G.}~\bibnamefont{Col{\`o}}}, \bibnamefont{and}
  \bibinfo{author}{\bibfnamefont{N.}~\bibnamefont{Van~Giai}},
  \bibinfo{journal}{Phys. Rev.} \textbf{\bibinfo{volume}{C65}},
  \bibinfo{pages}{044603} (\bibinfo{year}{2002}), \eprint{nucl-th/0103039}.

\bibitem[{\citenamefont{Shantz}(2010)}]{Shantz:2010th}
\bibinfo{author}{\bibfnamefont{T.}~\bibnamefont{Shantz}}, Master's thesis,
  \bibinfo{school}{Laurentian University}, \bibinfo{address}{Canada}
  (\bibinfo{year}{2010}).

  
  
 \bibitem[{\citenamefont{Dasgupta and Beacom}(2011)}]{Dasgupta:2011wg}
\bibinfo{author}{\bibfnamefont{B.} \bibnamefont{Dasgupta}},
  \bibinfo{author}{\bibfnamefont{J.F.}~\bibnamefont{Beacom}},
  \bibinfo{journal}{Phys. Rev.} \textbf{\bibinfo{volume}{D83}},
  \bibinfo{pages}{113006} (\bibinfo{year}{2011}), \eprint{1103.2768}. 
  



\bibitem[{\citenamefont{Freedman et~al.}(1977)\citenamefont{Freedman, Schramm,
  and Tubbs}}]{Freedman:1977xn}
\bibinfo{author}{\bibfnamefont{D.~Z.} \bibnamefont{Freedman}},
  \bibinfo{author}{\bibfnamefont{D.~N.} \bibnamefont{Schramm}},
  \bibnamefont{and} \bibinfo{author}{\bibfnamefont{D.~L.} \bibnamefont{Tubbs}},
  \bibinfo{journal}{Ann. Rev. Nucl. Part. Sci.} \textbf{\bibinfo{volume}{27}},
  \bibinfo{pages}{167} (\bibinfo{year}{1977}).

\bibitem[{\citenamefont{Marrod{\'a}n~Undagoitia and
  Rauch}(2016)}]{Undagoitia:2015gya}
\bibinfo{author}{\bibfnamefont{T.}~\bibnamefont{Marrod{\'a}n~Undagoitia}}
  \bibnamefont{and} \bibinfo{author}{\bibfnamefont{L.}~\bibnamefont{Rauch}},
  \bibinfo{journal}{J. Phys.} \textbf{\bibinfo{volume}{G43}},
  \bibinfo{pages}{013001} (\bibinfo{year}{2016}), \eprint{1509.08767}.

\bibitem[{\citenamefont{Monroe and Fisher}(2007)}]{Monroe:2007xp}
\bibinfo{author}{\bibfnamefont{J.}~\bibnamefont{Monroe}} \bibnamefont{and}
  \bibinfo{author}{\bibfnamefont{P.}~\bibnamefont{Fisher}},
  \bibinfo{journal}{Phys. Rev.} \textbf{\bibinfo{volume}{D76}},
  \bibinfo{pages}{033007} (\bibinfo{year}{2007}), \eprint{0706.3019}.

\bibitem[{\citenamefont{Strigari}(2009)}]{Strigari:2009bq}
\bibinfo{author}{\bibfnamefont{L.~E.} \bibnamefont{Strigari}},
  \bibinfo{journal}{New J. Phys.} \textbf{\bibinfo{volume}{11}},
  \bibinfo{pages}{105011} (\bibinfo{year}{2009}), \eprint{0903.3630}.

\bibitem[{\citenamefont{Abe et~al.}(2016{\natexlab{b}})}]{XMASS:2016cmy}
\bibinfo{author}{\bibfnamefont{K.}~\bibnamefont{Abe}} \bibnamefont{et~al.}
  (\bibinfo{collaboration}{XMASS}) (\bibinfo{year}{2016}{\natexlab{b}}),
  \eprint{1604.01218}.

\bibitem[{\citenamefont{Chakraborty et~al.}(2014)\citenamefont{Chakraborty,
  Bhattacharjee, and Kar}}]{Chakraborty:2013zua}
\bibinfo{author}{\bibfnamefont{S.}~\bibnamefont{Chakraborty}},
  \bibinfo{author}{\bibfnamefont{P.}~\bibnamefont{Bhattacharjee}},
  \bibnamefont{and} \bibinfo{author}{\bibfnamefont{K.}~\bibnamefont{Kar}},
  \bibinfo{journal}{Phys. Rev.} \textbf{\bibinfo{volume}{D89}},
  \bibinfo{pages}{013011} (\bibinfo{year}{2014}), \eprint{1309.4492}.

\bibitem[{\citenamefont{Aprile et~al.}(2016{\natexlab{a}})}]{Aprile:2015uzo}
\bibinfo{author}{\bibfnamefont{E.}~\bibnamefont{Aprile}} \bibnamefont{et~al.}
  (\bibinfo{collaboration}{XENON}), \bibinfo{journal}{JCAP}
  \textbf{\bibinfo{volume}{1604}}, \bibinfo{pages}{027}
  (\bibinfo{year}{2016}{\natexlab{a}}), \eprint{1512.07501}.

\bibitem[{\citenamefont{Akerib et~al.}(2015)}]{Akerib:2015cja}
\bibinfo{author}{\bibfnamefont{D.~S.} \bibnamefont{Akerib}}
  \bibnamefont{et~al.} (\bibinfo{collaboration}{LZ}) (\bibinfo{year}{2015}),
  \eprint{1509.02910}.

\bibitem[{\citenamefont{Aalbers et~al.}(2016)}]{Aalbers:2016jon}
\bibinfo{author}{\bibfnamefont{J.}~\bibnamefont{Aalbers}} \bibnamefont{et~al.}
  (\bibinfo{year}{2016}), \eprint{1606.07001}.

\bibitem[{\citenamefont{H{\"u}depohl}(2013)}]{Huedepohl:2013}
\bibinfo{author}{\bibfnamefont{L.}~\bibnamefont{H{\"u}depohl}}, Ph.D. thesis,
  \bibinfo{school}{Technische Universit{\"a}t M{\"u}nchen}
  (\bibinfo{year}{2013}).

\bibitem[{\citenamefont{Keil et~al.}(2003)\citenamefont{Keil, Raffelt, and
  Janka}}]{Keil:2002in}
\bibinfo{author}{\bibfnamefont{M.~T.} \bibnamefont{Keil}},
  \bibinfo{author}{\bibfnamefont{G.~G.} \bibnamefont{Raffelt}},
  \bibnamefont{and} \bibinfo{author}{\bibfnamefont{H.-T.} \bibnamefont{Janka}},
  \bibinfo{journal}{Astrophys. J.} \textbf{\bibinfo{volume}{590}},
  \bibinfo{pages}{971} (\bibinfo{year}{2003}), \eprint{astro-ph/0208035}.

\bibitem[{\citenamefont{Tamborra
  et~al.}(2012{\natexlab{a}})\citenamefont{Tamborra, M{\"u}ller, H{\"u}depohl,
  Janka, and Raffelt}}]{Tamborra:2012ac}
\bibinfo{author}{\bibfnamefont{I.}~\bibnamefont{Tamborra}},
  \bibinfo{author}{\bibfnamefont{B.}~\bibnamefont{M{\"u}ller}},
  \bibinfo{author}{\bibfnamefont{L.}~\bibnamefont{H{\"u}depohl}},
  \bibinfo{author}{\bibfnamefont{H.-T.} \bibnamefont{Janka}}, \bibnamefont{and}
  \bibinfo{author}{\bibfnamefont{G.}~\bibnamefont{Raffelt}},
  \bibinfo{journal}{Phys. Rev.} \textbf{\bibinfo{volume}{D86}},
  \bibinfo{pages}{125031} (\bibinfo{year}{2012}{\natexlab{a}}),
  \eprint{1211.3920}.

\bibitem[{sna()}]{snarchive}
\bibinfo{note}{{The SN neutrino data are available from
  \url{http://wwwmpa.mpa-garching.mpg.de/ccsnarchive/index.html}.}}

\bibitem[{\citenamefont{Lund et~al.}(2010)\citenamefont{Lund, Marek, Lunardini,
  Janka, and Raffelt}}]{Lund:2010kh}
\bibinfo{author}{\bibfnamefont{T.}~\bibnamefont{Lund}},
  \bibinfo{author}{\bibfnamefont{A.}~\bibnamefont{Marek}},
  \bibinfo{author}{\bibfnamefont{C.}~\bibnamefont{Lunardini}},
  \bibinfo{author}{\bibfnamefont{H.-T.} \bibnamefont{Janka}}, \bibnamefont{and}
  \bibinfo{author}{\bibfnamefont{G.}~\bibnamefont{Raffelt}},
  \bibinfo{journal}{Phys. Rev.} \textbf{\bibinfo{volume}{D82}},
  \bibinfo{pages}{063007} (\bibinfo{year}{2010}), \eprint{1006.1889}.

\bibitem[{\citenamefont{Tamborra et~al.}(2013)\citenamefont{Tamborra, Hanke,
  M{\"u}ller, Janka, and Raffelt}}]{Tamborra:2013laa}
\bibinfo{author}{\bibfnamefont{I.}~\bibnamefont{Tamborra}},
  \bibinfo{author}{\bibfnamefont{F.}~\bibnamefont{Hanke}},
  \bibinfo{author}{\bibfnamefont{B.}~\bibnamefont{M{\"u}ller}},
  \bibinfo{author}{\bibfnamefont{H.-T.} \bibnamefont{Janka}}, \bibnamefont{and}
  \bibinfo{author}{\bibfnamefont{G.}~\bibnamefont{Raffelt}},
  \bibinfo{journal}{Phys. Rev. Lett.} \textbf{\bibinfo{volume}{111}},
  \bibinfo{pages}{121104} (\bibinfo{year}{2013}), \eprint{1307.7936}.

\bibitem[{\citenamefont{Tamborra
  et~al.}(2014{\natexlab{a}})\citenamefont{Tamborra, Raffelt, Hanke, Janka, and
  M{\"u}ller}}]{Tamborra:2014hga}
\bibinfo{author}{\bibfnamefont{I.}~\bibnamefont{Tamborra}},
  \bibinfo{author}{\bibfnamefont{G.}~\bibnamefont{Raffelt}},
  \bibinfo{author}{\bibfnamefont{F.}~\bibnamefont{Hanke}},
  \bibinfo{author}{\bibfnamefont{H.-T.} \bibnamefont{Janka}}, \bibnamefont{and}
  \bibinfo{author}{\bibfnamefont{B.}~\bibnamefont{M{\"u}ller}},
  \bibinfo{journal}{Phys. Rev.} \textbf{\bibinfo{volume}{D90}},
  \bibinfo{pages}{045032} (\bibinfo{year}{2014}{\natexlab{a}}),
  \eprint{1406.0006}.

\bibitem[{\citenamefont{Tamborra
  et~al.}(2014{\natexlab{b}})\citenamefont{Tamborra, Hanke, Janka, M{\"u}ller,
  Raffelt, and Marek}}]{Tamborra:2014aua}
\bibinfo{author}{\bibfnamefont{I.}~\bibnamefont{Tamborra}},
  \bibinfo{author}{\bibfnamefont{F.}~\bibnamefont{Hanke}},
  \bibinfo{author}{\bibfnamefont{H.-T.} \bibnamefont{Janka}},
  \bibinfo{author}{\bibfnamefont{B.}~\bibnamefont{M{\"u}ller}},
  \bibinfo{author}{\bibfnamefont{G.~G.} \bibnamefont{Raffelt}},
  \bibnamefont{and} \bibinfo{author}{\bibfnamefont{A.}~\bibnamefont{Marek}},
  \bibinfo{journal}{Astrophys. J.} \textbf{\bibinfo{volume}{792}},
  \bibinfo{pages}{96} (\bibinfo{year}{2014}{\natexlab{b}}), \eprint{1402.5418}.

\bibitem[{\citenamefont{Lattimer and Swesty}(1991)}]{Lattimer:1991nc}
\bibinfo{author}{\bibfnamefont{J.~M.} \bibnamefont{Lattimer}} \bibnamefont{and}
  \bibinfo{author}{\bibfnamefont{F.~D.} \bibnamefont{Swesty}},
  \bibinfo{journal}{Nucl. Phys.} \textbf{\bibinfo{volume}{A535}},
  \bibinfo{pages}{331} (\bibinfo{year}{1991}).

\bibitem[{\citenamefont{Shen et~al.}(1998)\citenamefont{Shen, Toki, Oyamatsu,
  and Sumiyoshi}}]{Shen:1998gq}
\bibinfo{author}{\bibfnamefont{H.}~\bibnamefont{Shen}},
  \bibinfo{author}{\bibfnamefont{H.}~\bibnamefont{Toki}},
  \bibinfo{author}{\bibfnamefont{K.}~\bibnamefont{Oyamatsu}}, \bibnamefont{and}
  \bibinfo{author}{\bibfnamefont{K.}~\bibnamefont{Sumiyoshi}},
  \bibinfo{journal}{Nucl. Phys.} \textbf{\bibinfo{volume}{A637}},
  \bibinfo{pages}{435} (\bibinfo{year}{1998}), \eprint{nucl-th/9805035}.

\bibitem[{\citenamefont{Kachelriess et~al.}(2005)\citenamefont{Kachelriess,
  Tom{\`a}s, Buras, Janka, Marek, and Rampp}}]{Kachelriess:2004ds}
\bibinfo{author}{\bibfnamefont{M.}~\bibnamefont{Kachelriess}},
  \bibinfo{author}{\bibfnamefont{R.}~\bibnamefont{Tom{\`a}s}},
  \bibinfo{author}{\bibfnamefont{R.}~\bibnamefont{Buras}},
  \bibinfo{author}{\bibfnamefont{H.~T.} \bibnamefont{Janka}},
  \bibinfo{author}{\bibfnamefont{A.}~\bibnamefont{Marek}}, \bibnamefont{and}
  \bibinfo{author}{\bibfnamefont{M.}~\bibnamefont{Rampp}},
  \bibinfo{journal}{Phys. Rev.} \textbf{\bibinfo{volume}{D71}},
  \bibinfo{pages}{063003} (\bibinfo{year}{2005}), \eprint{astro-ph/0412082}.

\bibitem[{\citenamefont{Serpico et~al.}(2012)\citenamefont{Serpico,
  Chakraborty, Fischer, H{\"u}depohl, Janka, and Mirizzi}}]{Serpico:2011ir}
\bibinfo{author}{\bibfnamefont{P.~D.} \bibnamefont{Serpico}},
  \bibinfo{author}{\bibfnamefont{S.}~\bibnamefont{Chakraborty}},
  \bibinfo{author}{\bibfnamefont{T.}~\bibnamefont{Fischer}},
  \bibinfo{author}{\bibfnamefont{L.}~\bibnamefont{H{\"u}depohl}},
  \bibinfo{author}{\bibfnamefont{H.-T.} \bibnamefont{Janka}}, \bibnamefont{and}
  \bibinfo{author}{\bibfnamefont{A.}~\bibnamefont{Mirizzi}},
  \bibinfo{journal}{Phys. Rev.} \textbf{\bibinfo{volume}{D85}},
  \bibinfo{pages}{085031} (\bibinfo{year}{2012}), \eprint{1111.4483}.

\bibitem[{\citenamefont{Bethe and Wilson}(1985)}]{Bethe:1984ux}
\bibinfo{author}{\bibfnamefont{H.~A.} \bibnamefont{Bethe}} \bibnamefont{and}
  \bibinfo{author}{\bibfnamefont{R.}~\bibnamefont{Wilson},
  \bibfnamefont{James}}, \bibinfo{journal}{Astrophys. J.}
  \textbf{\bibinfo{volume}{295}}, \bibinfo{pages}{14} (\bibinfo{year}{1985}).

\bibitem[{\citenamefont{Bethe}(1990)}]{Bethe:1990mw}
\bibinfo{author}{\bibfnamefont{H.~A.} \bibnamefont{Bethe}},
  \bibinfo{journal}{Rev. Mod. Phys.} \textbf{\bibinfo{volume}{62}},
  \bibinfo{pages}{801} (\bibinfo{year}{1990}).

\bibitem[{\citenamefont{{Mikheyev} and {Smirnov}}(1985)}]{wolf}
\bibinfo{author}{\bibfnamefont{S.~P.} \bibnamefont{{Mikheyev}}}
  \bibnamefont{and} \bibinfo{author}{\bibfnamefont{A.~Yu.}
  \bibnamefont{{Smirnov}}}, \bibinfo{journal}{Yadernaya Fizika}
  \textbf{\bibinfo{volume}{42}}, \bibinfo{pages}{1441} (\bibinfo{year}{1985}).

\bibitem[{\citenamefont{{Wolfenstein}}(1978)}]{Wolfenstein:1977ue}
\bibinfo{author}{\bibfnamefont{L.}~\bibnamefont{{Wolfenstein}}},
  \bibinfo{journal}{\prd} \textbf{\bibinfo{volume}{17}}, \bibinfo{pages}{2369}
  (\bibinfo{year}{1978}).
  
  \bibitem[{\citenamefont{{Mikheev} and {Smirnov}} (1986)}]{Mikheev:1986if}
  \bibinfo{author}{\bibfnamefont{S.~P.} \bibnamefont{{Mikheyev}}}
  \bibnamefont{and} \bibinfo{author}{\bibfnamefont{A.~Yu.}
  \bibnamefont{{Smirnov}}},\bibinfo{journal}{Sov. Phys. JETP} \textbf{\bibinfo{volume}{64}}, \bibinfo{pages}{4-7}
  (\bibinfo{year}{1986}), \eprint{0706.0454}.


\bibitem[{\citenamefont{Borriello et~al.}(2014)\citenamefont{Borriello,
  Chakraborty, Janka, Lisi, and Mirizzi}}]{Borriello:2013tha}
\bibinfo{author}{\bibfnamefont{E.}~\bibnamefont{Borriello}},
  \bibinfo{author}{\bibfnamefont{S.}~\bibnamefont{Chakraborty}},
  \bibinfo{author}{\bibfnamefont{H.-T.} \bibnamefont{Janka}},
  \bibinfo{author}{\bibfnamefont{E.}~\bibnamefont{Lisi}}, \bibnamefont{and}
  \bibinfo{author}{\bibfnamefont{A.}~\bibnamefont{Mirizzi}},
  \bibinfo{journal}{JCAP} \textbf{\bibinfo{volume}{1411}}, \bibinfo{pages}{030}
  (\bibinfo{year}{2014}), \eprint{1310.7488}.

\bibitem[{\citenamefont{Sawyer}(1990)}]{Sawyer:1990tw}
\bibinfo{author}{\bibfnamefont{R.~F.} \bibnamefont{Sawyer}},
  \bibinfo{journal}{Phys. Rev.} \textbf{\bibinfo{volume}{D42}},
  \bibinfo{pages}{3908} (\bibinfo{year}{1990}).

\bibitem[{\citenamefont{Fogli et~al.}(2006)\citenamefont{Fogli, Lisi, Mirizzi,
  and Montanino}}]{Fogli:2006xy}
\bibinfo{author}{\bibfnamefont{G.~L.} \bibnamefont{Fogli}},
  \bibinfo{author}{\bibfnamefont{E.}~\bibnamefont{Lisi}},
  \bibinfo{author}{\bibfnamefont{A.}~\bibnamefont{Mirizzi}}, \bibnamefont{and}
  \bibinfo{author}{\bibfnamefont{D.}~\bibnamefont{Montanino}},
  \bibinfo{journal}{JCAP} \textbf{\bibinfo{volume}{0606}}, \bibinfo{pages}{012}
  (\bibinfo{year}{2006}), \eprint{hep-ph/0603033}.

\bibitem[{\citenamefont{Friedland and Gruzinov}(2006)}]{Friedland:2006ta}
\bibinfo{author}{\bibfnamefont{A.}~\bibnamefont{Friedland}} \bibnamefont{and}
  \bibinfo{author}{\bibfnamefont{A.}~\bibnamefont{Gruzinov}}
  (\bibinfo{year}{2006}), \eprint{astro-ph/0607244}.

\bibitem[{\citenamefont{Duan et~al.}(2010)\citenamefont{Duan, Fuller, and
  Qian}}]{Duan:2010bg}
\bibinfo{author}{\bibfnamefont{H.}~\bibnamefont{Duan}},
  \bibinfo{author}{\bibfnamefont{G.~M.} \bibnamefont{Fuller}},
  \bibnamefont{and} \bibinfo{author}{\bibfnamefont{Y.-Z.} \bibnamefont{Qian}},
  \bibinfo{journal}{Ann. Rev. Nucl. Part. Sci.} \textbf{\bibinfo{volume}{60}},
  \bibinfo{pages}{569} (\bibinfo{year}{2010}), \eprint{1001.2799}.

\bibitem[{\citenamefont{Tamborra
  et~al.}(2012{\natexlab{b}})\citenamefont{Tamborra, Raffelt, H{\"u}depohl, and
  Janka}}]{Tamborra:2011is}
\bibinfo{author}{\bibfnamefont{I.}~\bibnamefont{Tamborra}},
  \bibinfo{author}{\bibfnamefont{G.~G.} \bibnamefont{Raffelt}},
  \bibinfo{author}{\bibfnamefont{L.}~\bibnamefont{H{\"u}depohl}},
  \bibnamefont{and} \bibinfo{author}{\bibfnamefont{H.-T.} \bibnamefont{Janka}},
  \bibinfo{journal}{JCAP} \textbf{\bibinfo{volume}{1201}}, \bibinfo{pages}{013}
  (\bibinfo{year}{2012}{\natexlab{b}}), \eprint{1110.2104}.

\bibitem[{\citenamefont{Pllumbi et~al.}(2015)\citenamefont{Pllumbi, Tamborra,
  Wanajo, Janka, and H{\"u}depohl}}]{Pllumbi:2014saa}
\bibinfo{author}{\bibfnamefont{E.}~\bibnamefont{Pllumbi}},
  \bibinfo{author}{\bibfnamefont{I.}~\bibnamefont{Tamborra}},
  \bibinfo{author}{\bibfnamefont{S.}~\bibnamefont{Wanajo}},
  \bibinfo{author}{\bibfnamefont{H.-T.} \bibnamefont{Janka}}, \bibnamefont{and}
  \bibinfo{author}{\bibfnamefont{L.}~\bibnamefont{H{\"u}depohl}},
  \bibinfo{journal}{Astrophys. J.} \textbf{\bibinfo{volume}{808}},
  \bibinfo{pages}{188} (\bibinfo{year}{2015}), \eprint{1406.2596}.

\bibitem[{\citenamefont{Nunokawa et~al.}(1996)\citenamefont{Nunokawa, Qian,
  Rossi, and Valle}}]{Nunokawa:1996tg}
\bibinfo{author}{\bibfnamefont{H.}~\bibnamefont{Nunokawa}},
  \bibinfo{author}{\bibfnamefont{Y.~Z.} \bibnamefont{Qian}},
  \bibinfo{author}{\bibfnamefont{A.}~\bibnamefont{Rossi}}, \bibnamefont{and}
  \bibinfo{author}{\bibfnamefont{J.~W.~F.} \bibnamefont{Valle}},
  \bibinfo{journal}{Phys. Rev.} \textbf{\bibinfo{volume}{D54}},
  \bibinfo{pages}{4356} (\bibinfo{year}{1996}), \eprint{hep-ph/9605301}.

\bibitem[{\citenamefont{Fuller et~al.}(2009)\citenamefont{Fuller, Kusenko, and
  Petraki}}]{Fuller:2009zz}
\bibinfo{author}{\bibfnamefont{G.~M.} \bibnamefont{Fuller}},
  \bibinfo{author}{\bibfnamefont{A.}~\bibnamefont{Kusenko}}, \bibnamefont{and}
  \bibinfo{author}{\bibfnamefont{K.}~\bibnamefont{Petraki}},
  \bibinfo{journal}{Phys. Lett.} \textbf{\bibinfo{volume}{B670}},
  \bibinfo{pages}{281} (\bibinfo{year}{2009}), \eprint{0806.4273}.

\bibitem[{\citenamefont{Davis}(2016)}]{Davis:2016dqh}
\bibinfo{author}{\bibfnamefont{J.~H.} \bibnamefont{Davis}}
  (\bibinfo{year}{2016}), \eprint{1605.00011}.

\bibitem[{\citenamefont{Akerib et~al.}(2013)}]{Akerib:2012ys}
\bibinfo{author}{\bibfnamefont{D.~S.} \bibnamefont{Akerib}}
  \bibnamefont{et~al.} (\bibinfo{collaboration}{LUX}), \bibinfo{journal}{Nucl.
  Instrum. Meth.} \textbf{\bibinfo{volume}{A704}}, \bibinfo{pages}{111}
  (\bibinfo{year}{2013}), \eprint{1211.3788}.

\bibitem[{\citenamefont{Cao et~al.}(2014)}]{Cao:2014jsa}
\bibinfo{author}{\bibfnamefont{X.}~\bibnamefont{Cao}} \bibnamefont{et~al.}
  (\bibinfo{collaboration}{PandaX}), \bibinfo{journal}{Sci. China Phys. Mech.
  Astron.} \textbf{\bibinfo{volume}{57}}, \bibinfo{pages}{1476}
  (\bibinfo{year}{2014}), \eprint{1405.2882}.

\bibitem[{\citenamefont{Aprile et~al.}(2012{\natexlab{a}})}]{Aprile:2011dd}
\bibinfo{author}{\bibfnamefont{E.}~\bibnamefont{Aprile}} \bibnamefont{et~al.}
  (\bibinfo{collaboration}{XENON100}), \bibinfo{journal}{Astropart. Phys.}
  \textbf{\bibinfo{volume}{35}}, \bibinfo{pages}{573}
  (\bibinfo{year}{2012}{\natexlab{a}}), \eprint{1107.2155}.

\bibitem[{\citenamefont{Aprile et~al.}(2011)}]{Aprile:2010bt}
\bibinfo{author}{\bibfnamefont{E.}~\bibnamefont{Aprile}} \bibnamefont{et~al.}
  (\bibinfo{collaboration}{XENON10}), \bibinfo{journal}{Astropart. Phys.}
  \textbf{\bibinfo{volume}{34}}, \bibinfo{pages}{679} (\bibinfo{year}{2011}),
  \eprint{1001.2834}.

\bibitem[{\citenamefont{Alner et~al.}(2005)}]{Alner:2005pa}
\bibinfo{author}{\bibfnamefont{G.~J.} \bibnamefont{Alner}} \bibnamefont{et~al.}
  (\bibinfo{collaboration}{UK Dark Matter}), \bibinfo{journal}{Astropart.
  Phys.} \textbf{\bibinfo{volume}{23}}, \bibinfo{pages}{444}
  (\bibinfo{year}{2005}).

\bibitem[{\citenamefont{Alner et~al.}(2007)}]{Alner:2007ja}
\bibinfo{author}{\bibfnamefont{G.~J.} \bibnamefont{Alner}}
  \bibnamefont{et~al.}, \bibinfo{journal}{Astropart. Phys.}
  \textbf{\bibinfo{volume}{28}}, \bibinfo{pages}{287} (\bibinfo{year}{2007}),
  \eprint{astro-ph/0701858}.

\bibitem[{\citenamefont{Akimov et~al.}(2007)}]{Akimov:2006qw}
\bibinfo{author}{\bibfnamefont{D.~{\relax Yu}.} \bibnamefont{Akimov}}
  \bibnamefont{et~al.}, \bibinfo{journal}{Astropart. Phys.}
  \textbf{\bibinfo{volume}{27}}, \bibinfo{pages}{46} (\bibinfo{year}{2007}),
  \eprint{astro-ph/0605500}.

\bibitem[{\citenamefont{Aprile et~al.}(2014{\natexlab{a}})}]{Aprile:2014zvw}
\bibinfo{author}{\bibfnamefont{E.}~\bibnamefont{Aprile}} \bibnamefont{et~al.}
  (\bibinfo{collaboration}{XENON1T}), \bibinfo{journal}{JINST}
  \textbf{\bibinfo{volume}{9}}, \bibinfo{pages}{11006}
  (\bibinfo{year}{2014}{\natexlab{a}}), \eprint{1406.2374}.

\bibitem[{\citenamefont{Baudis}(2012)}]{Baudis:2012bc}
\bibinfo{author}{\bibfnamefont{L.}~\bibnamefont{Baudis}}
  (\bibinfo{collaboration}{DARWIN Consortium}), \bibinfo{journal}{J. Phys.
  Conf. Ser.} \textbf{\bibinfo{volume}{375}}, \bibinfo{pages}{012028}
  (\bibinfo{year}{2012}), \eprint{1201.2402}.

\bibitem[{\citenamefont{Schumann et~al.}(2015)\citenamefont{Schumann, Baudis,
  Butikofer, Kish, and Selvi}}]{Schumann:2015cpa}
\bibinfo{author}{\bibfnamefont{M.}~\bibnamefont{Schumann}},
  \bibinfo{author}{\bibfnamefont{L.}~\bibnamefont{Baudis}},
  \bibinfo{author}{\bibfnamefont{L.}~\bibnamefont{Butikofer}},
  \bibinfo{author}{\bibfnamefont{A.}~\bibnamefont{Kish}}, \bibnamefont{and}
  \bibinfo{author}{\bibfnamefont{M.}~\bibnamefont{Selvi}},
  \bibinfo{journal}{JCAP} \textbf{\bibinfo{volume}{1510}}, \bibinfo{pages}{016}
  (\bibinfo{year}{2015}), \eprint{1506.08309}.

\bibitem[{\citenamefont{Abe et~al.}(2013)}]{Abe:2013tc}
\bibinfo{author}{\bibfnamefont{K.}~\bibnamefont{Abe}} \bibnamefont{et~al.},
  \bibinfo{journal}{Nucl. Instrum. Meth.} \textbf{\bibinfo{volume}{A716}},
  \bibinfo{pages}{78} (\bibinfo{year}{2013}), \eprint{1301.2815}.

\bibitem[{\citenamefont{Erler and Ramsey-Musolf}(2005)}]{Erler:2004in}
\bibinfo{author}{\bibfnamefont{J.}~\bibnamefont{Erler}} \bibnamefont{and}
  \bibinfo{author}{\bibfnamefont{M.~J.} \bibnamefont{Ramsey-Musolf}},
  \bibinfo{journal}{Phys. Rev.} \textbf{\bibinfo{volume}{D72}},
  \bibinfo{pages}{073003} (\bibinfo{year}{2005}), \eprint{hep-ph/0409169}.

\bibitem[{\citenamefont{Vietze et~al.}(2015)\citenamefont{Vietze, Klos,
  Menendez, Haxton, and Schwenk}}]{Vietze:2014vsa}
\bibinfo{author}{\bibfnamefont{L.}~\bibnamefont{Vietze}},
  \bibinfo{author}{\bibfnamefont{P.}~\bibnamefont{Klos}},
  \bibinfo{author}{\bibfnamefont{J.}~\bibnamefont{Menendez}},
  \bibinfo{author}{\bibfnamefont{W.~C.} \bibnamefont{Haxton}},
  \bibnamefont{and} \bibinfo{author}{\bibfnamefont{A.}~\bibnamefont{Schwenk}},
  \bibinfo{journal}{Phys. Rev.} \textbf{\bibinfo{volume}{D91}},
  \bibinfo{pages}{043520} (\bibinfo{year}{2015}), \eprint{1412.6091}.

\bibitem[{\citenamefont{Akerib et~al.}(2016)}]{Akerib:2015rjg}
\bibinfo{author}{\bibfnamefont{D.~S.} \bibnamefont{Akerib}}
  \bibnamefont{et~al.} (\bibinfo{collaboration}{LUX}), \bibinfo{journal}{Phys.
  Rev. Lett.} \textbf{\bibinfo{volume}{116}}, \bibinfo{pages}{161301}
  (\bibinfo{year}{2016}), \eprint{1512.03506}.

\bibitem[{\citenamefont{Szydagis
  et~al.}(2011{\natexlab{a}})\citenamefont{Szydagis, Barry, Kazkaz, Mock,
  Stolp, Sweany, Tripathi, Uvarov, Walsh, and Woods}}]{Szydagis:2011tk}
\bibinfo{author}{\bibfnamefont{M.}~\bibnamefont{Szydagis}},
  \bibinfo{author}{\bibfnamefont{N.}~\bibnamefont{Barry}},
  \bibinfo{author}{\bibfnamefont{K.}~\bibnamefont{Kazkaz}},
  \bibinfo{author}{\bibfnamefont{J.}~\bibnamefont{Mock}},
  \bibinfo{author}{\bibfnamefont{D.}~\bibnamefont{Stolp}},
  \bibinfo{author}{\bibfnamefont{M.}~\bibnamefont{Sweany}},
  \bibinfo{author}{\bibfnamefont{M.}~\bibnamefont{Tripathi}},
  \bibinfo{author}{\bibfnamefont{S.}~\bibnamefont{Uvarov}},
  \bibinfo{author}{\bibfnamefont{N.}~\bibnamefont{Walsh}}, \bibnamefont{and}
  \bibinfo{author}{\bibfnamefont{M.}~\bibnamefont{Woods}},
  \bibinfo{journal}{JINST} \textbf{\bibinfo{volume}{6}},
  \bibinfo{pages}{P10002} (\bibinfo{year}{2011}{\natexlab{a}}),
  \eprint{1106.1613}.

\bibitem[{\citenamefont{Mock et~al.}(2014)\citenamefont{Mock, Barry, Kazkaz,
  Szydagis, Tripathi, Uvarov, Woods, and Walsh}}]{Mock:2013ila}
\bibinfo{author}{\bibfnamefont{J.}~\bibnamefont{Mock}},
  \bibinfo{author}{\bibfnamefont{N.}~\bibnamefont{Barry}},
  \bibinfo{author}{\bibfnamefont{K.}~\bibnamefont{Kazkaz}},
  \bibinfo{author}{\bibfnamefont{M.}~\bibnamefont{Szydagis}},
  \bibinfo{author}{\bibfnamefont{M.}~\bibnamefont{Tripathi}},
  \bibinfo{author}{\bibfnamefont{S.}~\bibnamefont{Uvarov}},
  \bibinfo{author}{\bibfnamefont{M.}~\bibnamefont{Woods}}, \bibnamefont{and}
  \bibinfo{author}{\bibfnamefont{N.}~\bibnamefont{Walsh}},
  \bibinfo{journal}{JINST} \textbf{\bibinfo{volume}{9}},
  \bibinfo{pages}{T04002} (\bibinfo{year}{2014}), \eprint{1310.1117}.

\bibitem[{\citenamefont{Lenardo et~al.}(2015)\citenamefont{Lenardo, Kazkaz,
  Manalaysay, Mock, Szydagis, and Tripathi}}]{Lenardo:2014cva}
\bibinfo{author}{\bibfnamefont{B.}~\bibnamefont{Lenardo}},
  \bibinfo{author}{\bibfnamefont{K.}~\bibnamefont{Kazkaz}},
  \bibinfo{author}{\bibfnamefont{A.}~\bibnamefont{Manalaysay}},
  \bibinfo{author}{\bibfnamefont{J.}~\bibnamefont{Mock}},
  \bibinfo{author}{\bibfnamefont{M.}~\bibnamefont{Szydagis}}, \bibnamefont{and}
  \bibinfo{author}{\bibfnamefont{M.}~\bibnamefont{Tripathi}},
  \bibinfo{journal}{IEEE Trans. Nucl. Sci.} \textbf{\bibinfo{volume}{62}},
  \bibinfo{pages}{3387} (\bibinfo{year}{2015}), \eprint{1412.4417}.

\bibitem[{\citenamefont{Szydagis
  et~al.}(2011{\natexlab{b}})\citenamefont{Szydagis, Barry, Kazkaz, Mock,
  Stolp, Sweany, Tripathi, Uvarov, Walsh, and Woods}}]{Szydagis:20111006}
\bibinfo{author}{\bibfnamefont{M.}~\bibnamefont{Szydagis}},
  \bibinfo{author}{\bibfnamefont{N.}~\bibnamefont{Barry}},
  \bibinfo{author}{\bibfnamefont{K.}~\bibnamefont{Kazkaz}},
  \bibinfo{author}{\bibfnamefont{J.}~\bibnamefont{Mock}},
  \bibinfo{author}{\bibfnamefont{D.}~\bibnamefont{Stolp}},
  \bibinfo{author}{\bibfnamefont{M.}~\bibnamefont{Sweany}},
  \bibinfo{author}{\bibfnamefont{M.}~\bibnamefont{Tripathi}},
  \bibinfo{author}{\bibfnamefont{S.}~\bibnamefont{Uvarov}},
  \bibinfo{author}{\bibfnamefont{N.}~\bibnamefont{Walsh}}, \bibnamefont{and}
  \bibinfo{author}{\bibfnamefont{M.}~\bibnamefont{Woods}},
  \bibinfo{journal}{JINST} \textbf{\bibinfo{volume}{6}},
  \bibinfo{pages}{P10002} (\bibinfo{year}{2011}{\natexlab{b}}),
  \eprint{1106.1613}.

\bibitem[{\citenamefont{Szydagis et~al.}(2013)\citenamefont{Szydagis, Fyhrie,
  Thorngren, and Tripathi}}]{Szydagis:2013sih}
\bibinfo{author}{\bibfnamefont{M.}~\bibnamefont{Szydagis}},
  \bibinfo{author}{\bibfnamefont{A.}~\bibnamefont{Fyhrie}},
  \bibinfo{author}{\bibfnamefont{D.}~\bibnamefont{Thorngren}},
  \bibnamefont{and} \bibinfo{author}{\bibfnamefont{M.}~\bibnamefont{Tripathi}},
  \bibinfo{journal}{JINST} \textbf{\bibinfo{volume}{8}},
  \bibinfo{pages}{C10003} (\bibinfo{year}{2013}), \eprint{1307.6601}.

\bibitem[{\citenamefont{Aprile et~al.}(2014{\natexlab{b}})}]{Aprile:2012vw}
\bibinfo{author}{\bibfnamefont{E.}~\bibnamefont{Aprile}} \bibnamefont{et~al.}
  (\bibinfo{collaboration}{XENON100}), \bibinfo{journal}{Astropart. Phys.}
  \textbf{\bibinfo{volume}{54}}, \bibinfo{pages}{11}
  (\bibinfo{year}{2014}{\natexlab{b}}), \eprint{1207.3458}.

\bibitem[{\citenamefont{Yoo and Jaskierny}(2015)}]{Yoo:2015yza}
\bibinfo{author}{\bibfnamefont{J.}~\bibnamefont{Yoo}} \bibnamefont{and}
  \bibinfo{author}{\bibfnamefont{W.~F.} \bibnamefont{Jaskierny}},
  \bibinfo{journal}{JINST} \textbf{\bibinfo{volume}{10}},
  \bibinfo{pages}{P08011} (\bibinfo{year}{2015}), \eprint{1508.05903}.

\bibitem[{\citenamefont{Edwards et~al.}(2008)}]{Edwards:2007nj}
\bibinfo{author}{\bibfnamefont{B.}~\bibnamefont{Edwards}} \bibnamefont{et~al.},
  \bibinfo{journal}{Astropart. Phys.} \textbf{\bibinfo{volume}{30}},
  \bibinfo{pages}{54} (\bibinfo{year}{2008}), \eprint{0708.0768}.

\bibitem[{\citenamefont{Aprile et~al.}(2014{\natexlab{c}})}]{Aprile:2013blg}
\bibinfo{author}{\bibfnamefont{E.}~\bibnamefont{Aprile}} \bibnamefont{et~al.}
  (\bibinfo{collaboration}{XENON100}), \bibinfo{journal}{J. Phys.}
  \textbf{\bibinfo{volume}{G41}}, \bibinfo{pages}{035201}
  (\bibinfo{year}{2014}{\natexlab{c}}), \eprint{1311.1088}.

\bibitem[{\citenamefont{Aprile et~al.}(2012{\natexlab{b}})}]{Aprile:2012nq}
\bibinfo{author}{\bibfnamefont{E.}~\bibnamefont{Aprile}} \bibnamefont{et~al.}
  (\bibinfo{collaboration}{XENON100}), \bibinfo{journal}{Phys. Rev. Lett.}
  \textbf{\bibinfo{volume}{109}}, \bibinfo{pages}{181301}
  (\bibinfo{year}{2012}{\natexlab{b}}), \eprint{1207.5988}.

\bibitem[{\citenamefont{Tan et~al.}(2016)}]{Tan:2016diz}
\bibinfo{author}{\bibfnamefont{A.}~\bibnamefont{Tan}} \bibnamefont{et~al.}
  (\bibinfo{collaboration}{PandaX}), \bibinfo{journal}{Phys. Rev.} \textbf{\bibinfo{volume}{D93}},
  \bibinfo{pages}{122009} (\bibinfo{year}{2016}),
  \eprint{1602.06563}.

\bibitem[{\citenamefont{Hagmann and Bernstein}(2004)}]{Hagmann:2004uv}
\bibinfo{author}{\bibfnamefont{C.}~\bibnamefont{Hagmann}} \bibnamefont{and}
  \bibinfo{author}{\bibfnamefont{A.}~\bibnamefont{Bernstein}},
  \bibinfo{journal}{IEEE Trans. Nucl. Sci.} \textbf{\bibinfo{volume}{51}},
  \bibinfo{pages}{2151} (\bibinfo{year}{2004}), \eprint{nucl-ex/0411004}.

\bibitem[{\citenamefont{Angle et~al.}(2011)}]{Angle:2011th}
\bibinfo{author}{\bibfnamefont{J.}~\bibnamefont{Angle}} \bibnamefont{et~al.}
  (\bibinfo{collaboration}{XENON10}), \bibinfo{journal}{Phys. Rev. Lett.}
  \textbf{\bibinfo{volume}{107}}, \bibinfo{pages}{051301}
  (\bibinfo{year}{2011}), \bibinfo{note}{[Erratum: Phys. Rev.
  Lett.110,249901(2013)]}, \eprint{1104.3088}.

\bibitem[{\citenamefont{Frandsen et~al.}(2013)}]{Frandsen:2013cna}
\bibinfo{author}{\bibfnamefont{M.~T.} \bibnamefont{Frandsen}}
  \bibnamefont{et~al.}, \bibinfo{journal}{JCAP}
  \textbf{\bibinfo{volume}{1307}}, \bibinfo{pages}{023} (\bibinfo{year}{2013}),
  \bibinfo{note}{[Corrected XENON10 limit curve: see Erratum
  Phys.~Rev.~Lett.~110, 249901 (2013)]}, \eprint{1304.6066}.

\bibitem[{\citenamefont{Essig et~al.}(2012{\natexlab{a}})\citenamefont{Essig,
  Mardon, and Volansky}}]{Essig:2011nj}
\bibinfo{author}{\bibfnamefont{R.}~\bibnamefont{Essig}},
  \bibinfo{author}{\bibfnamefont{J.}~\bibnamefont{Mardon}}, \bibnamefont{and}
  \bibinfo{author}{\bibfnamefont{T.}~\bibnamefont{Volansky}},
  \bibinfo{journal}{Phys. Rev.} \textbf{\bibinfo{volume}{D85}},
  \bibinfo{pages}{076007} (\bibinfo{year}{2012}{\natexlab{a}}),
  \eprint{1108.5383}.

\bibitem[{\citenamefont{Santos et~al.}(2011)}]{Santos:2011ju}
\bibinfo{author}{\bibfnamefont{E.}~\bibnamefont{Santos}} \bibnamefont{et~al.}
  (\bibinfo{collaboration}{ZEPLIN-III}), \bibinfo{journal}{JHEP}
  \textbf{\bibinfo{volume}{12}}, \bibinfo{pages}{115} (\bibinfo{year}{2011}),
  \eprint{1110.3056}.

\bibitem[{\citenamefont{Essig et~al.}(2012{\natexlab{b}})\citenamefont{Essig,
  Manalaysay, Mardon, Sorensen, and Volansky}}]{Essig:2012yx}
\bibinfo{author}{\bibfnamefont{R.}~\bibnamefont{Essig}},
  \bibinfo{author}{\bibfnamefont{A.}~\bibnamefont{Manalaysay}},
  \bibinfo{author}{\bibfnamefont{J.}~\bibnamefont{Mardon}},
  \bibinfo{author}{\bibfnamefont{P.}~\bibnamefont{Sorensen}}, \bibnamefont{and}
  \bibinfo{author}{\bibfnamefont{T.}~\bibnamefont{Volansky}},
  \bibinfo{journal}{Phys. Rev. Lett.} \textbf{\bibinfo{volume}{109}},
  \bibinfo{pages}{021301} (\bibinfo{year}{2012}{\natexlab{b}}),
  \eprint{1206.2644}.

\bibitem[{\citenamefont{Aprile et~al.}(2016{\natexlab{b}})}]{Aprile:2016wwo}
\bibinfo{author}{\bibfnamefont{E.}~\bibnamefont{Aprile}} \bibnamefont{et~al.}
  (\bibinfo{collaboration}{XENON100}) (\bibinfo{year}{2016}{\natexlab{b}}),
  \eprint{1605.06262}.

\bibitem[{\citenamefont{Cowan et~al.}(2011)\citenamefont{Cowan, Cranmer, Gross,
  and Vitells}}]{Cowan:2010js}
\bibinfo{author}{\bibfnamefont{G.}~\bibnamefont{Cowan}},
  \bibinfo{author}{\bibfnamefont{K.}~\bibnamefont{Cranmer}},
  \bibinfo{author}{\bibfnamefont{E.}~\bibnamefont{Gross}}, \bibnamefont{and}
  \bibinfo{author}{\bibfnamefont{O.}~\bibnamefont{Vitells}},
  \bibinfo{journal}{Eur. Phys. J.} \textbf{\bibinfo{volume}{C71}},
  \bibinfo{pages}{1554} (\bibinfo{year}{2011}), \bibinfo{note}{[Erratum: Eur.
  Phys. J.C73,2501(2013)]}, \eprint{1007.1727}.

\bibitem[{\citenamefont{Ker{\"a}nen et~al.}(2004)\citenamefont{Ker{\"a}nen,
  Maalampi, Myyryl{\"a}inen, and Riittinen}}]{Keranen:2004rg}
\bibinfo{author}{\bibfnamefont{P.}~\bibnamefont{Ker{\"a}nen}},
  \bibinfo{author}{\bibfnamefont{J.}~\bibnamefont{Maalampi}},
  \bibinfo{author}{\bibfnamefont{M.}~\bibnamefont{Myyryl{\"a}inen}},
  \bibnamefont{and}
  \bibinfo{author}{\bibfnamefont{J.}~\bibnamefont{Riittinen}},
  \bibinfo{journal}{Phys. Lett.} \textbf{\bibinfo{volume}{B597}},
  \bibinfo{pages}{374} (\bibinfo{year}{2004}), \eprint{hep-ph/0401082}.

\bibitem[{\citenamefont{{Janka} and {Hillebrandt}}(1989)}]{1989A&A...224...49J}
\bibinfo{author}{\bibfnamefont{H.-T.} \bibnamefont{{Janka}}} \bibnamefont{and}
  \bibinfo{author}{\bibfnamefont{W.}~\bibnamefont{{Hillebrandt}}},
  \bibinfo{journal}{Astronomy and Astrophysics} \textbf{\bibinfo{volume}{224}},
  \bibinfo{pages}{49} (\bibinfo{year}{1989}).

\bibitem[{\citenamefont{Cowan}(2014)}]{Agashe:2014kda}
\bibinfo{author}{\bibfnamefont{G.}~\bibnamefont{Cowan}}
  (\bibinfo{collaboration}{Particle Data Group: Statistics (2015 update)}),
  \bibinfo{journal}{Chin.Phys.} \textbf{\bibinfo{volume}{C38}},
  \bibinfo{pages}{090001} (\bibinfo{year}{2014}).

\bibitem[{\citenamefont{Lindhard and Scharff}(1961)}]{Lindhard:1961zz}
\bibinfo{author}{\bibfnamefont{J.}~\bibnamefont{Lindhard}} \bibnamefont{and}
  \bibinfo{author}{\bibfnamefont{M.}~\bibnamefont{Scharff}},
  \bibinfo{journal}{Phys. Rev.} \textbf{\bibinfo{volume}{124}},
  \bibinfo{pages}{128} (\bibinfo{year}{1961}).

\bibitem[{\citenamefont{Bezrukov et~al.}(2011)\citenamefont{Bezrukov,
  Kahlhoefer, and Lindner}}]{Bezrukov:2010qa}
\bibinfo{author}{\bibfnamefont{F.}~\bibnamefont{Bezrukov}},
  \bibinfo{author}{\bibfnamefont{F.}~\bibnamefont{Kahlhoefer}},
  \bibnamefont{and} \bibinfo{author}{\bibfnamefont{M.}~\bibnamefont{Lindner}},
  \bibinfo{journal}{Astropart. Phys.} \textbf{\bibinfo{volume}{35}},
  \bibinfo{pages}{119} (\bibinfo{year}{2011}), \eprint{1011.3990}.

\bibitem[{\citenamefont{Sorensen}(2015)}]{Sorensen:2014sla}
\bibinfo{author}{\bibfnamefont{P.}~\bibnamefont{Sorensen}},
  \bibinfo{journal}{Phys. Rev.} \textbf{\bibinfo{volume}{D91}},
  \bibinfo{pages}{083509} (\bibinfo{year}{2015}), \eprint{1412.3028}.

\bibitem[{\citenamefont{Aprile et~al.}(2015)}]{Aprile:2015ibr}
\bibinfo{author}{\bibfnamefont{E.}~\bibnamefont{Aprile}} \bibnamefont{et~al.}
  (\bibinfo{collaboration}{XENON100}), \bibinfo{journal}{Phys. Rev. Lett.}
  \textbf{\bibinfo{volume}{115}}, \bibinfo{pages}{091302}
  (\bibinfo{year}{2015}), \eprint{1507.07748}.

\bibitem[{\citenamefont{Antonioli et~al.}(2004)}]{Antonioli:2004zb}
\bibinfo{author}{\bibfnamefont{P.}~\bibnamefont{Antonioli}}
  \bibnamefont{et~al.}, \bibinfo{journal}{New J. Phys.}
  \textbf{\bibinfo{volume}{6}}, \bibinfo{pages}{114} (\bibinfo{year}{2004}),
  \eprint{astro-ph/0406214}.

\bibitem[{\citenamefont{Scholberg}(2008)}]{Scholberg:2008fa}
\bibinfo{author}{\bibfnamefont{K.}~\bibnamefont{Scholberg}},
  \bibinfo{journal}{Astron. Nachr.} \textbf{\bibinfo{volume}{329}},
  \bibinfo{pages}{337} (\bibinfo{year}{2008}), \eprint{0803.0531}.

\bibitem[{\citenamefont{Baudis et~al.}(2014)\citenamefont{Baudis, Ferella,
  Kish, Manalaysay, Marrodan~Undagoitia, and Schumann}}]{Baudis:2013qla}
\bibinfo{author}{\bibfnamefont{L.}~\bibnamefont{Baudis}},
  \bibinfo{author}{\bibfnamefont{A.}~\bibnamefont{Ferella}},
  \bibinfo{author}{\bibfnamefont{A.}~\bibnamefont{Kish}},
  \bibinfo{author}{\bibfnamefont{A.}~\bibnamefont{Manalaysay}},
  \bibinfo{author}{\bibfnamefont{T.}~\bibnamefont{Marrodan~Undagoitia}},
  \bibnamefont{and} \bibinfo{author}{\bibfnamefont{M.}~\bibnamefont{Schumann}},
  \bibinfo{journal}{JCAP} \textbf{\bibinfo{volume}{1401}}, \bibinfo{pages}{044}
  (\bibinfo{year}{2014}), \eprint{1309.7024}.

\bibitem[{\citenamefont{Baudis et~al.}(2013)\citenamefont{Baudis, Kessler,
  Klos, Lang, Menendez, Reichard, and Schwenk}}]{Baudis:2013bba}
\bibinfo{author}{\bibfnamefont{L.}~\bibnamefont{Baudis}},
  \bibinfo{author}{\bibfnamefont{G.}~\bibnamefont{Kessler}},
  \bibinfo{author}{\bibfnamefont{P.}~\bibnamefont{Klos}},
  \bibinfo{author}{\bibfnamefont{R.~F.} \bibnamefont{Lang}},
  \bibinfo{author}{\bibfnamefont{J.}~\bibnamefont{Menendez}},
  \bibinfo{author}{\bibfnamefont{S.}~\bibnamefont{Reichard}}, \bibnamefont{and}
  \bibinfo{author}{\bibfnamefont{A.}~\bibnamefont{Schwenk}},
  \bibinfo{journal}{Phys. Rev.} \textbf{\bibinfo{volume}{D88}},
  \bibinfo{pages}{115014} (\bibinfo{year}{2013}), \eprint{1309.0825}.

\bibitem[{\citenamefont{McCabe}(2016)}]{McCabe:2015eia}
\bibinfo{author}{\bibfnamefont{C.}~\bibnamefont{McCabe}},
  \bibinfo{journal}{JCAP} \textbf{\bibinfo{volume}{1605}}, \bibinfo{pages}{033}
  (\bibinfo{year}{2016}), \eprint{1512.00460}.

\end{thebibliography}

\end{document}